%% file: metal_enrichment.tex
\documentclass[useAMS,usenatbib]{mn2e}
\usepackage{color}
\pdfoutput=1

\include{pacchetti}


\include{definizioni}

\def\msun{{\rm M}_{\odot}}
\def\zsun{{\rm Z}_{\odot}}

\def\sn{\rm sn}
\def\zcut{Z_{\rm cut}}
\def\meanz{\langle Z\rangle}
\def\zcrit{Z_{\rm crit}}

\def\massmetal{M_{\star}-Z}

\def\popii{\rm PopII}
\def\popiii{\rm PopIII}
\def\CitDub{DT8}

\def\stella{\rm s}

\def\CIV{\hbox{C~$\scriptstyle\rm IV $}} 
\def\SIV{\hbox{Si~$\scriptstyle\rm IV $}}
\def\CII{\hbox{C~$\scriptstyle\rm II $}} 
\def\SII{\hbox{Si~$\scriptstyle\rm II $}} 
\def\OI{\hbox{O~$\scriptstyle\rm I $}}
\def\HI{\hbox{H~$\scriptstyle\rm I $}}

\begin{document}
\date{}
\pagerange{\pageref{firstpage}--\pageref{lastpage}} \pubyear{2013}
\title[Cosmic metal enrichment]{Simulating cosmic metal enrichment by the first galaxies}
\author[Pallottini et al.]{A. Pallottini$^{1}$, A. Ferrara$^{1,4}$, S. Gallerani$^{1}$, S. Salvadori$^{2}$, V. D'Odorico$^{3}$\\
$^{1}$Scuola Normale Superiore, Piazza dei Cavalieri 7, I-56126 Pisa, Italy\\
$^{2}$Kapteyn Astronomical Institute, Landleven 12, 9747 AD Groningen, The Netherlands\\
$^{3}$INAF/Osservatorio Astronomico di Trieste, Via Tiepolo 11, I-34143 Trieste, Italy\\
$^4$Kavli IPMU (WPI), Todai Institutes for Advanced Study, University of Tokyo, Japan\\	
}
\maketitle
\label{firstpage}
\begin{abstract}
We study cosmic metal enrichment via AMR hydrodynamical simulations in a $(10$~Mpc~$h^{-1})^{3}$ volume following the Pop III -- Pop II transition and for different Pop III IMFs. We have analyzed the joint evolution of metal enrichment on galactic and intergalactic scales at $z=6$ and $z=4$. Galaxies account for $\lsim9\%$ of the baryonic mass; the remaining gas resides in the diffuse phases: (a)~\textit{voids}, i.e. regions with extremely low density ($\Delta\leq 1$), (b) the true \textit{intergalactic medium} (IGM, $1<\Delta\leq 10$) and (c) the \textit{circumgalactic medium} (CGM, $10<\Delta\leq 10^{2.5}$), the interface between the IGM and galaxies. By $z=6$ a galactic mass-metallicity relation is established. At $z=4$, galaxies with a stellar mass $M_{\star}\simeq10^{8.5} M_\odot$ show $\log({\rm O}\slash H)+12=8.19$, consistent with observations. The total amount of heavy elements rises from $\Omega^{\rm SFH}_{Z}=1.52\times10^{-6}$ at $z=6$ to $8.05 \times10^{-6}$ at $z=4$. Metals in galaxies make up to $\simeq0.89$ of such budget at $z=6$; this fraction increases to $\simeq0.95$ at $z=4$. At $z=6$ ($z=4$) the remaining metals are distributed in CGM/IGM/voids with the following mass fractions: $0.06/0.04/0.01$ ($0.03/0.02/0.01$). Analogously to galaxies, at $z=4$ a density-metallicity ($\Delta-Z$) relation is in place for the diffuse phases: the IGM/voids have a spatially uniform metallicity, $Z\sim10^{-3.5}\zsun$; in the CGM $Z$ steeply rises with density up to~$\simeq10^{-2}\zsun$. In all diffuse phases a considerable fraction of metals is in a warm/hot ($T\, \mu^{-1}>10^{4.5}K$) state. Due to these physical conditions, $\CIV$ absorption line experiments can probe only $\simeq$ 2\% of the total carbon present in the IGM/CGM; however, metal absorption line spectra are very effective tools to study reionization. Finally, the Pop III star formation history is almost insensitive to the chosen Pop III IMF. Pop III stars are preferentially formed in truly pristine ($Z=0$) gas pockets, well outside polluted regions created by previous star formation episodes.
\end{abstract}


\section{Introduction}

The intergalactic medium (IGM) has been extensively investigated through the study of the $\HI$ Ly$\alpha$ forest \citep{Rauch:1998} and the absorption features due to ionized metal species \citep[e.g.][]{Songaila:1996} detected in the spectra of high redshift quasars (QSO). Observations have probed metal enrichment in different intergalactic environments: damped Ly$\alpha$ absorbers (DLA), characterized by column densities $\log N_{\rm HI}\slash{\rm cm}^{-2}\gsim 20$ and metallicities $10^{-1.5}\lsim Z\slash\zsun\lsim10^{-1}$; Lyman limit ($17\lsim\log N_{\rm HI}\slash{\rm cm}^{-2}\lsim 20$) and Ly$\alpha$ forest ($14\lsim\log N_{\rm HI}\slash{\rm cm}^{-2}\lsim 17$) systems, typically enriched at $10^{-3.5}\lsim Z\slash\zsun\lsim10^{-2}$ \citep{Meiksin:2009RvMP}.

The evolution of the IGM enrichment can be studied by measuring the abundance of ionized species at different cosmic times. For example, the $\CIV$ density parameter decreases with redshift from $\Omega_{\rm CIV}\simeq 8\times10^{-8}$ at $z\simeq 0$ to $\Omega_{\rm CIV}\simeq 10^{-8}$ at $z\simeq 2.5$ \citep[e.g.][]{DOdorico:2010}, it is constant up to $\lsim5$ \citep[e.g.][]{Schaye:2003ApJ,Cooksey:2010ApJ}, and could possibly show sign of a downturn, $\Omega_{\rm CIV}\lsim0.5\times10^{-8}$, for $z\gsim 5$ \citep[e.g.][]{Ryan-Weber:2009MNRAS,Becker:2009ApJ,Simcoe:2011ApJ,DOdorico:2013MNRAS}. Similarly, $\SIV$ displays a flat behavior for $2.5\lsim z \lsim5$ \citep{Songaila:2001ApJ,Songaila:2005AJ}.

Metals are produced by stars inside galaxies featuring a cosmic star formation rate density (SFR) $\simeq 10^{-2} \msun {\rm yr}^{-1} {\rm Mpc}^{-3}$ at $z=0$, increasing up to one order of magnitude at $z\sim 3$ and decreasing by a factor $\sim10^{2}$ for $3\lsim z \lsim9$ \citep{Dunlop:2013ASSL}. IGM is metal polluted by galaxies, whose SFR can be can used to infer the total metal density parameter. \citet{Pettini:1999cezh} have shown that observations account for $\sim20\%$ of metals at $z\gsim2$, implying that the missing ones must be locked in a warm-hot ionized phase \citep[e.g.][]{Ferrara:2005ApJ}.

Galaxies with high ($M_{\star}\gsim10^{11}\msun$) and low ($M_{\star}\lsim10^{8}\msun$) stellar mass display different characteristics. The local \citep{Panter:2008MNRAS} and $z\lsim3$ \citep{Maiolino:2008A&A,Mannucci:2010MNRAS} mass-metallicity ($\massmetal$) relation shows an increasing metal abundance with $M_{\star}$ from $\sim10^{4}\msun$ \citep{Kirby:2013arXiv} up to $\sim10^{10}\msun$ and a flattening for higher stellar mass. Similar difference is present in the dark-to-visible mass ratio \citep[e.g.][]{Guo:2010MNRAS,Tolstoy:2010IAUS,McGaugh:2010ApJ}: dwarf galaxies typically show values $\gsim30$, while Milky Way size galaxies have a ratio $\lsim15$; for increasing mass the value approaches (but does not reach) the cosmological one, $\Omega_{m}\slash\Omega_{b}\sim 6$.

A theoretical framework must auto-consistently account for the history of IGM enrichment, its thermal state and the global evolution of galaxy formation. An attractive scenario consists in the so-called {\it pre-enrichment} \citep{Madau:2001ApJ,Ferrara:2008IAUS}, in which IGM pollution is mainly due to low mass ($M_{\star}\lsim10^{7}\msun$) galaxies ejecting metals via supernova (SN) explosions at high redshift ($z\gsim8$). In this picture massive galaxies are able to retain their metals, thus their evolution follows a closed box chemical model\citep[e.g.][]{Tremonti:2004ApJ}. On the other hand, low mass galaxies are prone to material ejection because of the shallower potential well and since their smaller size allows the SN to coherently drive the outflows \citep[e.g.][]{Ferrara:2000MNRAS}.

Ejection from low mass galaxies has an obvious advantage in terms of volume filling factor, given the abundance of these sources in $\Lambda$CDM models\footnote{In this work we assume a $\Lambda$CDM cosmology with total matter, vacuum and baryonic densities in units of the critical density $\Omega_{\Lambda}= 0.727$, $\Omega_{dm}= 0.228$, $\Omega_{b}= 0.045$, Hubble constant $\rm H_0=100~h~km~s^{-1}~Mpc^{-1}$ with $\rm h=0.704$, spectral index $n=0.967$, $\sigma_{8}=0.811$ \citep[][]{Larson:2011}.} \citep[i.e.][]{Press:1974}. Additionally, an early ($z\gsim5$) IGM pollution allows the shocked and enriched gas to cool down \citep[e.g.][]{Ferrara:2008IAUS}, which can explain the observed narrowness ($\sim15~{\rm km~s}^{-1}$) of the Doppler width of metal lines \citep{Meiksin:2009RvMP}.

The {\it pre-enrichment} scenario is appealing because the same low mass galaxies which start to pollute the IGM can play an important role in the firts stages ($z\gsim8$) of cosmic reionization \citep[e.g.][]{Salvadori:2013arXiv}. In particular, \citep[e.g.][]{Choudhury:2008MNRAS} low mass galaxies are the ideal hosts for first metal-free Population III (PopIII) stars, which may possibly be the responsible for an early ($z\simeq7$) reionization \citep[e.g.][]{Gallerani:2008MNRASa,Gallerani:2008MNRASb}. However, the nature of Pop III stars is still under debate, because of lack of observations. From a theoretical point of view there is no clear consensus on their formation properties \citep{Bromm:2002ApJ,Yoshida:2006ApJ,Greif:2012MNRAS,Hosokawa:2012ApJ,Meece:2013arXiv} nor their subsequent evolution \citep{Heger:2002ApJ,Nomoto:2006NuPhA}.

Thus, understanding the IGM metal enrichment from the first galaxies is fundamental to explain both the formation and the evolution of galaxies and the reionization history \citep{Barkana:2001PhR,Ciardi:2005SSR}. However, it remains to be assessed the evolution of the temperature and chemical state of the enriched IGM, the imprint the metal transport has on the formation of new star forming regions and its role in the transition from Pop III to the successive generation of Population II (PopII) stars. The steady increase of availability of high redshift IGM observations \citep[e.g.][]{DOdorico:2013MNRAS} and their interpretation can hopefully clarify the picture, by better constraining the theoretical models.

Cosmological numerical simulations have been extensively used to study the problem \citep{Aguirre:2007EAS,Johnson:2011arXiv}. However, the huge dynamical range of the underlying physical phenomena makes a true auto-consistent simulation impossible. A viable modelization can be achieved by using subgrid models. These depend both on the considered physics and code implementation. Recently, \citet{Hopkins:2013arXiv} studied the impact of different star formation criteria, \citet{Agertz:2012arXiv} and \citet{Vogelsberger:2013arXiv} analyzed the effect of including different kind of feedback, and the AQUILA project \citep[i.e.][]{aquila:2012MNRAS} compared 13 different prescriptions of the main used cosmological codes. Subgrid modelling lessens the burden of the large dynamical range, but given the currently available computational capabilities the numerical resources have to be focused toward either the small or the large scales.

Simulations of small cosmic volumes, i.e. box sizes $\lsim\, {\rm Mpc}\,h^{-1}$, concentrate the computational power and allow the usage of highly refined physical models. \citet{Greif:2010ApJ} studied the transition from Pop III to Pop II stars in a $10^{8}\msun$ galaxy at $z\sim10$ assessing the role of radiative feedback; \citet{Maio:2010MNRAS} analyzed the same transition by varying several parameters, such as the critical metallicity $\zcrit$ that distinguishes the populations, the initial mass function (IMF), the metal yields and the star formation threshold; \citet{xu:2013arXiv} focalized on pinpointing the remnant of Pop III at high redshift, by employing the same computational scheme of \citet{Wise:2012ApJ}, which analyzed the impact of radiation from first stars on metal enrichment at $z\gsim9$; at the same redshift, \citet{Biffi:2013arXiv}, using an extensive chemical network, studied the properties and the formation of first proto-galaxies.

Large scale ($\gsim5\, {\rm Mpc}\,h^{-1}$) cosmological simulations naturally allows for a fair comparison with the observations. \citet{Scannapieco:2006MNRAS} showed that observation of line of sight (l.o.s.) correlations of $\CIV$ and $\SIV$ are consistent with a patchy IGM enrichment, confined in metal bubbles of $\sim 2\,{\rm Mpc}\,h^{-1}$ at $1.5\lsim z\lsim3$; by implementing galaxy outflows driven by a wind model \citet{Oppenheimer:2006MNRAS} managed to reproduce the flatness of $\Omega_{\rm CIV}$ at $2\lsim z\lsim5$; \citet{Tornatore:2007MNRAS} found evidence of Pop III production at $z\gsim4$, hinting at the possibility of observing metal-free stars; using a galactic super wind model \citet{Cen:2011ApJ} simulated a 50 Mpc~$h^{-1}$ box finding, among other results, a good agreement with observations for $\Omega_{\rm CIV}$ and a reasonable match for $\Omega_{\rm OVI}$; by using a $(37.5\,{\rm Mpc}\,h^{-1})^{3}$ volume simulation evolved up to $z=1.5$, and considering different IMFs and feedback mechanisms, \citet{Tescari:2011MNRAS} analyzed the evolution of $\Omega_{\rm CIV}$ and statistics of \HI\, and $\CIV$ absorbers at different redshifts; simulating a box with size of $25\,{\rm Mpc}\,h^{-1}$ and including various feedback, \citet{Vogelsberger:2013arXiv} managed to match several observations, as the SFR and stellar mass density (SMD) evolution for $z\lsim 9$, the galaxy stellar mass function and mass-metallicity relation at $z=0$.

The aim of this paper is to model the IGM metal enrichment focusing on high redshift ($z\geq4$) by simulating a volume large enough to include a statistically significant ensemble of galaxies. Clearly, the trade off consists in a limitation of the resolution and small scale complexity that can be investigated. Our modelling approach is to limit the number of free parameters of the subgrid prescriptions and constrain them with first galaxies observations, namely both global SFR densities inferred from Ultraviolet (UV) luminosity functions \citep{Bouwens:2012ApJ,Zheng:2012Natur} and SMD from stellar energy density fitting \citep{Gonzalez:2011}. This method limits the uncertainty on the feedback prescriptions \citep[e.g.][]{Vogelsberger:2013arXiv}, and at the same time it allows a large scale analysis of the metal enrichment process.

The paper is structured as follows. In Sec. \ref{sec_metodo_sim}, we describe the numerical implementation of the cosmological simulations whose free parameters are then calibrated by matching SFR and SMD data in Sec. \ref{sec_sfr_smd}. Sec. \ref{sec_gal_enrichment} and Sec. \ref{sec_global_result} contain the analysis of galactic and IGM metal enrichment, respectively. We devote Sec. \ref{sec_test} to study the effects of varying the Pop III IMF, and in Sec. \ref{sec_spettri} we compute and discuss mock QSO absorption spectra, in preparation for a future detailed comparison recent of high redshift ($4\lsim z\lsim6$) absorption line data \citep{DOdorico:2013MNRAS}. Finally, in Sec. \ref{sec_conclusioni} we present our conclusions. 

\begin{table}
\centering
    \begin{tabular}{|c|c|c|c|}
            \hline
	    $\log \left(Z\slash \zsun\right)$ & $Y$ & $R$ & $\epsilon_{\popii}$\\
            \hline\hline
            $-4.0$ & 0.0160 & 0.4680 & $10^{50}$\\ 
            $-2.0$ & 0.0192 & 0.4705 & $10^{50}$\\ 
            $-1.0$ & 0.0197 & 0.4799 &$10^{50}$\\ 
            $\,0.0$ & 0.0253 & 0.4983 &$10^{50}$\\
           \hline
   \end{tabular}
\caption{Adopted IMF-averaged Pop II metal yields and gas return fractions (\citet{vandenHoek:1997A&AS} for $0.8 \le m/\msun \le 8 $ and \citet{Woosley:1995ApJS} for $8 \le m/\msun \le 40 $); explosion energies (in \mbox{erg\,}$\msun^{-1}$) are taken from \citet{Woosley:1995ApJS}.
\label{tabella_yield}}
\end{table}
\begin{table}
 \centering
  \begin{tabular}{|c|c|c|c|}
    \hline
    IMF  & $Y$ & $R$ & $\epsilon_{\popiii}\slash\epsilon_{\popii}$ \\
    \hline\hline
    SALP & 0.0105 & 0.46 & 1  \\ 
    FHN  & 0.0081 & 0.76 & 1  \\ 
    PISN & 0.1830 & 0.45 & 10 \\ 
    \hline
  \end{tabular}
 \caption{Adopted IMF-averaged Pop III metal yields and gas return fractions, and relative explosion energies. Data are taken from \citet{vandenHoek:1997A&AS} for $0.8 \le m/\msun \le 8 $ and \citet{Woosley:1995ApJS} for $8 \le m/\msun \le 40 $ for SALP, \citet{Kobayashi:2011ApJ} for FHN, and \citet{Heger:2002ApJ} for PISN.
\label{tabella_yield_popiii}}
\end{table}

\section{Cosmological simulations}\label{sec_metodo_sim}
We perform cosmological simulations using a customized version of the publicly available code {\tt RAMSES} \citep[i.e.][]{Teyssier:2002}, which is a Fully Threaded Tree (FTT) data structure in which the hydrodynamical Adaptive Mesh Refinement (AMR) scheme is coupled with a Particle Mesh (PM3) N-body solver through a Cloud-In-Cell interpolation scheme to compute the Poisson equation.

Our simulation evolves a $10\,h^{-1}$ comoving Mpc box, with $512^{3}$ dark matter (DM) particles. The DM mass unit is $2.06 \times 10^{6}\, \Omega_{dm}\,h^{-1}\msun$, and the baryon base grid spatial resolution is $19.53\,h^{-1}$~kpc. We allow for 4 additional refinement levels, with a Lagrangian mass threshold-based criterion. This enables us to reach a maximum resolution of $\Delta x_{\min}=1.22\,h^{-1}$~kpc in the densest regions. The initial conditions are generated at $z=199$ using {\tt GRAFIC} \citep[i.e.][]{Bertschinger:2001}. The gas is characterized by a mean molecular weight $\mu=0.59$ and has a primordial BBN composition ($Z=0$) at the starting point of the simulation. We refer to metallicity as the sum of all the heavy element species without differentiating among them, as instead done, e.g., by \citet[][]{Maio:2007MNRAS}.

To account for heating and cooling processes {\tt RAMSES} is coupled with {\tt ATON} \citep[i.e.][]{Aubert:2008}, a moment-based radiative transfer code including metal cooling \citep[i.e.][]{Theuns:1998MNRAS,Sutherland:1993ApJS}. The code allows the treatment of an external, redshift dependent UV background \citep[UVB, e.g.][]{Haardt:1996,Haardt:2012} produced by both stars and QSOs. In the simulation we include the UVB and we neglect the radiation from sources inside the box, as the scale of ionized bubbles becomes rapidly comparable to the size of the simulation box \citep[e.g.][]{Wyithe:2004Natur,Zahn:2011MNRAS} and therefore the derived reionization history could not be considered as reliable anyway. We have however verified that varying the UVB within observational limits\footnote{We have used a simple analytical form of the UV background \citep{Theuns:1998MNRAS} and varied its parameters.} only marginally affects the enrichment history.

\subsection{Star formation prescription}\label{sec_sfr_prescription}

\begin{figure*}
\centering
\includegraphics[width=8.7cm]{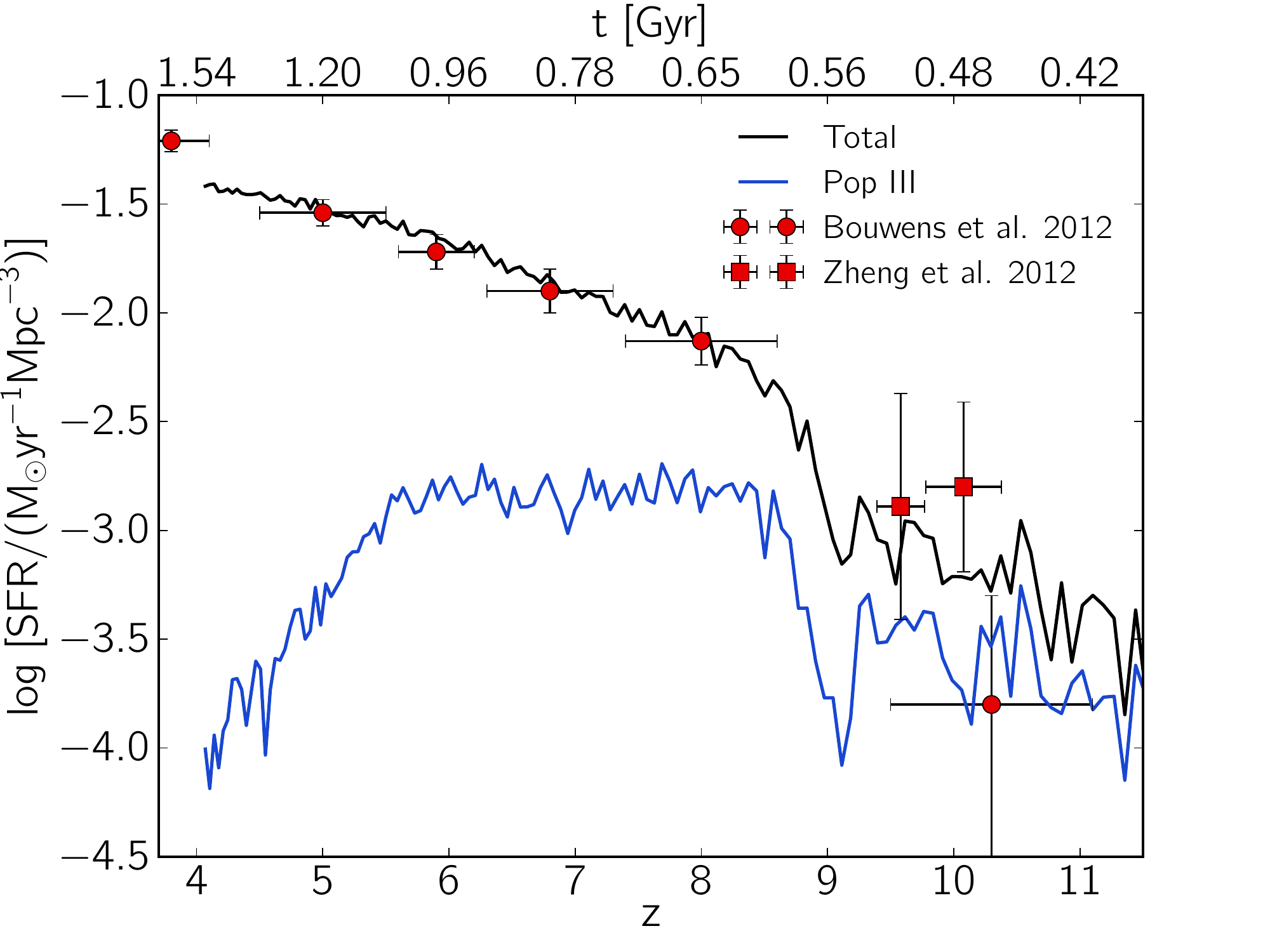}
\includegraphics[width=8.7cm]{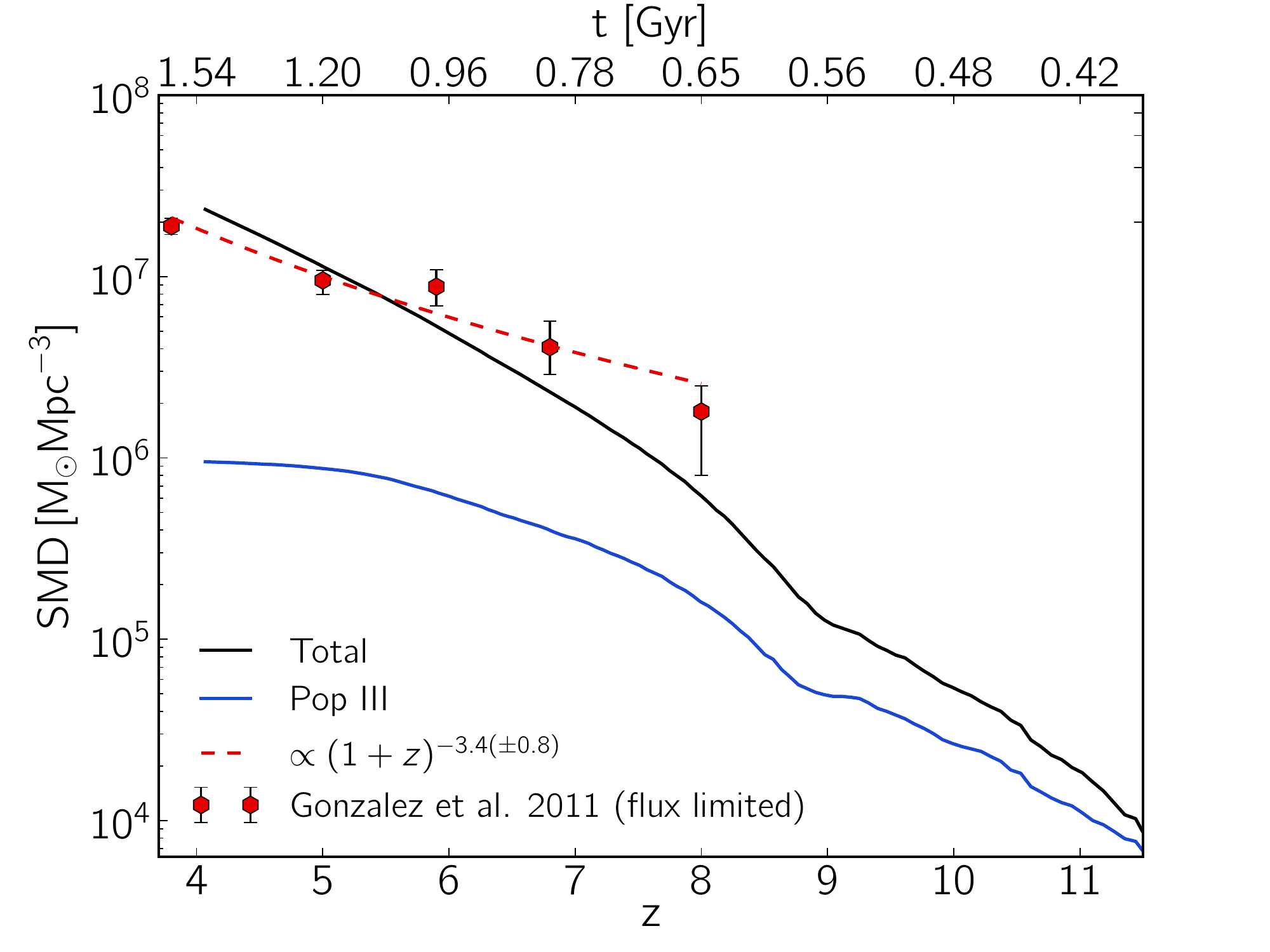}
\caption{
\textit{Left panel:} Cosmic star formation rate density (SFR) as a function of redshift (age of the Universe) for all stellar populations (black line) and for Pop III stars only (blue line). Data points (in red) are taken from \citet{Bouwens:2012ApJ} and \citet{Zheng:2012Natur}. \textit{Right:} Cosmic stellar mass density (SMD) as a function of redshift for all stellar populations (black line) and for Pop III stars only (blue line). Data points and the analytical fit (both in red) are from \citet{Gonzalez:2011}. 
\label{fig_sfr_smd}}
\end{figure*}

We adopt the star formation recipe described in \citet{Rasera:2006} and \citet{Dubois:2008}, hereafter \CitDub, whose main features are recalled below. At each coarse time step, $\Delta t$, a collisionless star particle is created in every gas cell of size $\Delta x$ which has a density $\rho$ exceeding the threshold $\rho_{\rm th}\equiv0.1\,m_H\, \mbox{cm}^{-3}$ \citep{Schaye:2004ApJ}, where $m_H$ is the proton mass. The mass of the newborn particle is $M_{\stella}=m_{\star}N$, where\footnote{Following \CitDub \, we impose that no more than half of a cell mass can be converted into stars, i.e. $M_{\stella}=\min[m_{\star}N,0.5\,\rho(\Delta x)^{3}]$. This prescription ensures the numerical stability of the code.} $m_{\star}=\rho_{\rm th}\left(\Delta x_{\min}\right)^3$ is the resolution dependent minimum mass of a star particle, and $N$ is drawn from a Poisson distribution
\begin{subequations}\label{eqs_local_sto_sfr}
\begin{equation}
	P(N)=\frac{\langle N\rangle}{N!}\exp\left(-\langle N\rangle\right)\, ,
\end{equation}
of mean $\langle N\rangle$ given by
\begin{equation}\label{eq_mean_starpt}
	\langle N\rangle=\frac{\rho\left(\Delta x\right)^{3}}{m_{\star}}\frac{\Delta t}{t_{\star}}\, ;
\end{equation}
\end{subequations}
$t_{\star}$ is the local star formation timescale and represents the first free parameter of our model. The dynamical properties of the star particle formed are inherited from the spawning cell; if the gas cell metallicity is below (above) the critical metallicity, $\zcrit\equiv10^{-4}\zsun$, we label the particle as belonging to the Pop III (Pop II) population. Such distinction is used when determining the metal yield and supernova explosion energy (Sec. \ref{sec_sub_snfeed}).

This recipe ensures that the local SFR follows a Schmidt-Kennicutt relation \citep[i.e.][]{Schmidt:1959ApJ,Kennicutt:1998ApJ}
\begin{equation}\label{eq_mimic_kennicutt}
\dot{\rho}_{\star} = \frac{\rho}{t_{\star}}\Theta\left(\rho-\rho_{\rm th}\right)\, ,
\end{equation}
where $\Theta$ is the Heaviside function\footnote{Note that the Schmidt-Kennicutt relation is written in terms of the surface density while eq. \ref{eq_mimic_kennicutt} considers the volume density. Thus we are implicitly assuming a smoothing scale which makes $t_{\star}$ resolution-dependent.}. Hereafter, we define as ``star forming'' those cells satisfying the density criterion $\rho>\rho_{\rm th}$. As highlighted in \citet{Hopkins:2013arXiv}, different definitions\footnote{Other definitions generally used to identify star forming regions are based on temperature or molecular hydrogen fraction thresholds, convergent flows requirement or Jeans instability criterion.} of star forming regions result in similar star formation histories, provided that stellar feedback is also included.

\subsection{Feedback and enrichment prescriptions}\label{sec_sub_snfeed}
The standard {\tt RAMSES} feedback prescription takes into account both thermal and momentum-driven feedback, as described in \CitDub. Momentum-driven feedback related to supernova blastwaves is important on the typical scales ($\sim \rm pc$) reached by these structures \citep{Agertz:2012arXiv}; similar arguments apply to radiative feedback from star forming regions. Since such small scales are not adequately resolved in our simulations, we decided to only include thermal feedback. Also, we do not include AGN feedback, which is thought to regulate the star formation of objects with masses $\gsim 10^{12}\msun$ \citep[e.g.][]{Teyssier:2011MNRAS,Vogelsberger:2013arXiv} that are rare or absent in our relatively small box.

Every newly-born star particle of mass $M_{\stella}$ immediately prompts a SN event. This assumption corresponds to the so-called Instantaneous Recycling Approximation \citep[i.e.][]{Tinsley:1980}, in turn consistent with the adopted coarse time step ($\Delta t\sim10~\rm Myr$). SN explosions damp a thermal energy $E_{\sn}$ 
\begin{equation}\label{eq_energy_sn}
E_{\sn}=\eta_{\sn}\epsilon_{\sn} M_{\stella}\, ,
\end{equation}
where $\epsilon_{\sn}$ is the total SN energy per stellar mass formed and $\eta_{\sn}$ is the coupling efficiency. While $\epsilon_{\sn}$ depends only on the stellar population properties, $\eta_{\sn}$ depends on the numerical implementation and resolution; in practice, we consider $\eta_{\sn}$ as the second free parameter of our model. As noted in \citet{DallaVecchia:2012MNRAS} (see also \citet[e.g.][]{Chiosi:1992ARA&A}) considering purely thermal feedback is a good assumption if the gas is able to transform the additional thermal energy into kinetic energy before cooling; this can be mimicked by a sufficiently large coupling efficiency $\eta_{\sn} \simgt 0.1$.

Every explosion returns a gas mass $R\, M_{\stella}$ and a metal mass $Y\,M_{\stella}$, that are then removed from the star particle mass. Following \citet{Salvadori:2007MNRAS, Salvadori:2008MNRAS}, and differently from \CitDub, we use metallicity dependent return fractions ($R$) and metal yields ($Y$):
\begin{subequations}\label{eqs_def_R_Y}
\begin{align}
	R=\frac{1}{\langle\Phi\rangle} &\left.\int_{m_{1}}^{m_{2}} \left(m-w\right)\Phi\, \mbox{d}m\right.\\
	Y=\frac{1}{\langle\Phi\rangle}&\left.\int_{m_{1}}^{m_{2}} m_{Z}\Phi\,\mbox{d}m\right.
\end{align}
with
\begin{equation}
\langle\Phi\rangle = \int_{ m_{1}}^{m_{2} } m\Phi\,\mbox{d}m\, ,
\end{equation}
\end{subequations}
where $w=w(m,Z)$ and $m_{Z}=m_{Z}(m,Z)$ are respectively the stellar remnant and the metal mass produced by a star of mass $m$ at a given metallicity, and $\Phi=\Phi(m)$ is the IMF. Both the IMF and the integration limits depend on the stellar population (Pop III or Pop II) considered.

\begin{figure}
\centering
\includegraphics[width=8.3cm]{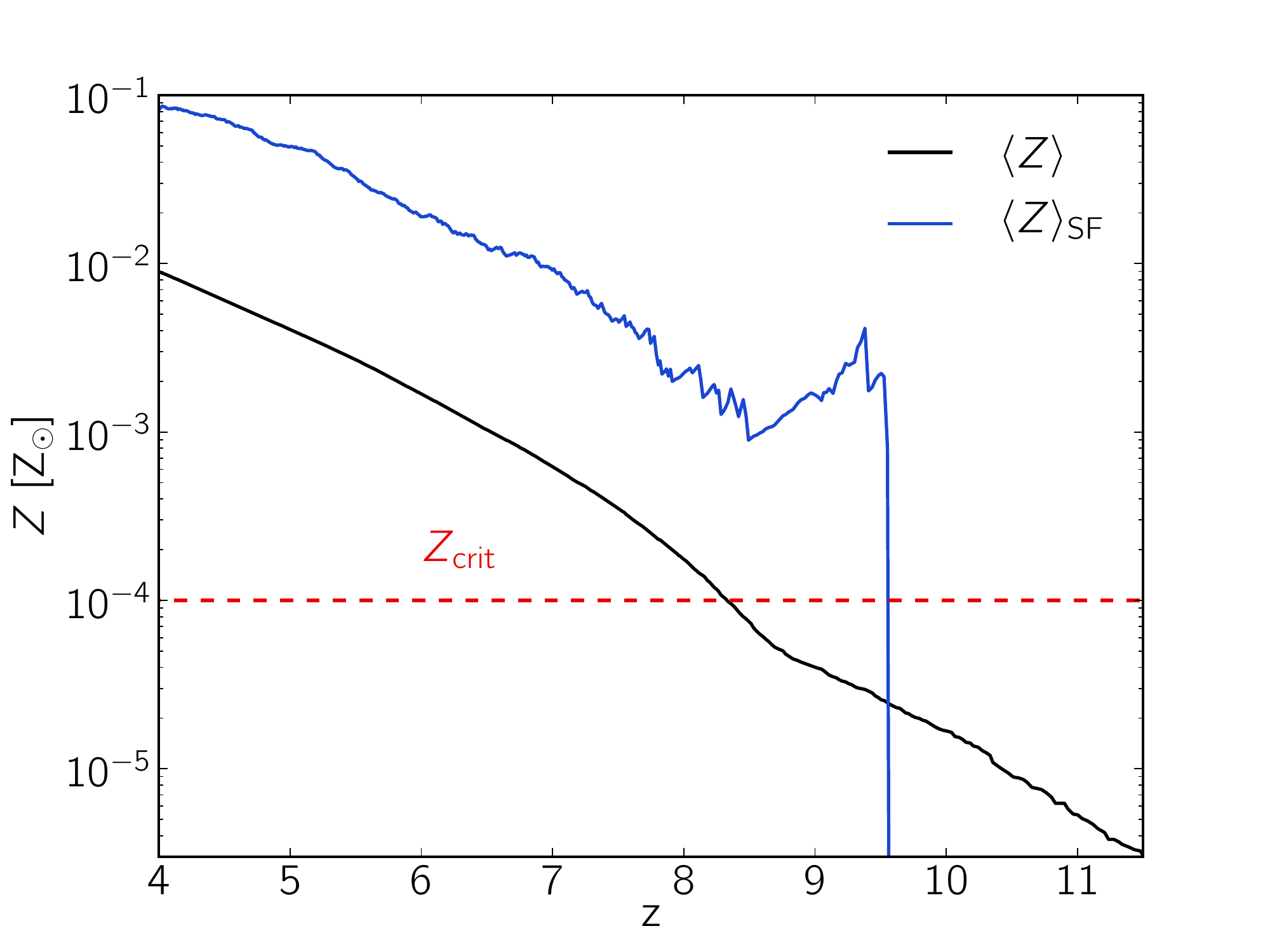}
\caption{Redshift evolution of the mean metallicity of baryons, $\meanz$, (black line) and the mean metallicity of star forming regions only, $\meanz_{\rm SF}$, (blue line). The adopted value, $\zcrit=10^{-4}\zsun$, of the critical metallicity for the Pop III - Pop II transition is indicated with a dashed red line.\label{fig_otf_anal}
} 
\end{figure}
The adopted Pop II (Pop III) metal yields, gas return fractions, and explosion energies are shown in Tab. \ref{tabella_yield} (Tab. \ref{tabella_yield_popiii}). For Pop II stars, we adopt a Larson-Salpeter IMF \citep[i.e.][]{Larson:1998MNRAS}, with\footnote{With respect to a Salpeter IMF, a Larson IMF has an exponential cut below $m=0.35\,\msun$.} $m_{1}=0.1\,\msun$, $m_{2}=100\,\msun$, following \citet{Salvadori:2007MNRAS}. The IMF of Pop III stars is very poorly constrained \citep[e.g.][]{Scannapieco:2003ApJ}. Therefore, we consider three different possibilities: (a) the same Larson-Salpeter IMF (referred to as the SALP case) as Pop II stars but with $R$, $Y$ and $\epsilon_{\sn}$ calculated for $Z=0\,$; (b) a $\delta$-function IMF, i.e. $\Phi=\delta(m-m_{0})$, with $m_{0}=25\,\msun$ appropriate for faint hypernovae (FHN) \citep{Salvadori:2012MNRAS}; (c) a top-heavy IMF with $m_{1}=100\,\msun$ and $m_{2}=500\,\msun$ allowing for pair-instability SN (PISN) in the mass range $140\,\msun\leq m\leq260\,\msun$ \citep[][]{Tornatore:2007MNRAS}.

Throughout the paper we use the SALP case as the fiducial one; Sec. \ref{sec_test} is devoted to the analysis of different prescriptions IMF.

\section{Cosmic star formation history}\label{sec_sfr_smd}
As explained above, our model contains two free parameters, namely the local star formation time, $t_{\star}$, and the supernova coupling efficiency, $\eta_{sn}$. These parameters have important effects \citep[i.e.][]{Rasera:2006} on two observable quantities, the cosmic SFR and SMD as a function of redshift. Broadly speaking, $t_{\star}$ variations shift the SFR curve with respect to the horizontal axis (i.e., redshift), while $\eta_{sn}$ sets the curve slope. Thus, in order to calibrate these unknowns, we require our simulation results to match the observed cosmic SFR \citep{Bouwens:2012ApJ,Zheng:2012Natur} and SMD \citep{Gonzalez:2011}. 

The evolution of the cosmic SFR and SMD for the best-fitting subgrid model parameters $t_{\star}=7.5~\mbox{Gyr}$ and $\eta_{\sn}=0.25$ are shown in Fig. \ref{fig_sfr_smd}, where they are also compared to the above mentioned observations. The simulated SFR reproduces the data quite accurately in the range $4\leq z\leq 8.5$; the agreement is still good at higher redshift ($9\leq z\leq10.5$), where, however, the limited mass resolution of our simulation yields a fluctuating SFR evolution. For the SMD evolution the agreement is good, although the predicted slope appears slightly steeper than the observed one (dashed red line). However, our predictions lie within 2-$\sigma$ of the data by \citet{Gonzalez:2011}. This level of agreement can be considered as satisfactory. In fact, while the SMD is inferred using Spectral Energy Distributions fitting of a flux-limited sample of galaxies, in our work the SMD is obtained simply by integrating the SFR. Our procedure then includes both stars and their remnants, regardless of the age. 

\begin{figure}
\centering
\includegraphics[width=8.3cm]{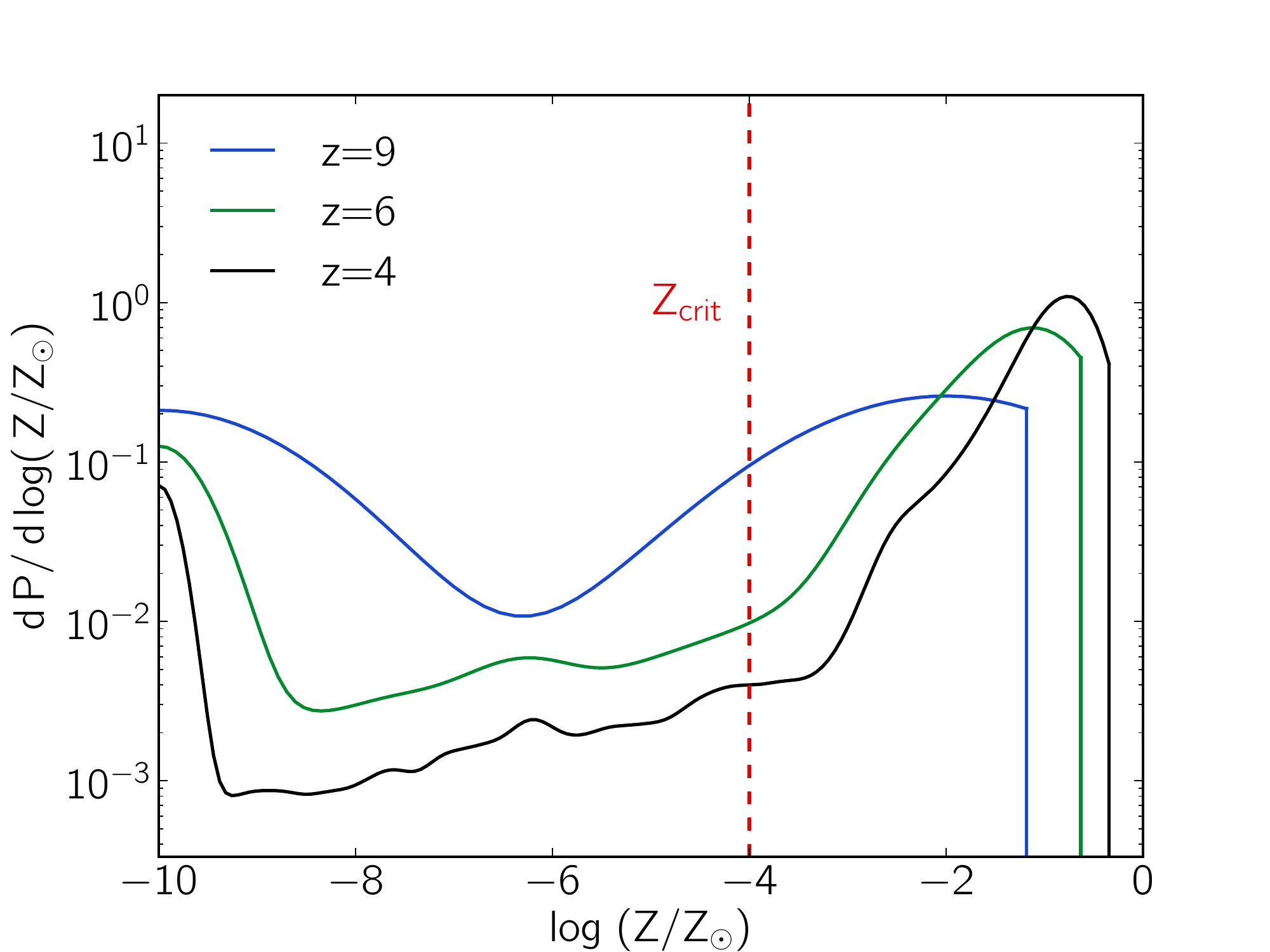}
\caption{Stellar metallicity distribution function (MDF) at $z=9$ (blue line), $z=6$ (green line) and $z=4$ (black line) for all stars in the simulation box. As in Fig. \ref{fig_otf_anal} the critical metallicity $\zcrit=10^{-4}\zsun$ is indicated with a dashed red line. The amplitude of each curve is normalized so that its integral over the available $Z$ range is equal to 1. Note that for display reasons we have set $Z=10^{-10}\zsun$ for stars with $Z<10^{-10}\zsun$\label{fig_metalpdf_star}.}
\end{figure}
In Fig. \ref{fig_sfr_smd}, we also plot separately the contribution of Pop III stars to both SFR and SMD. The Pop III SFR initially climbs to a level slightly above $10^{-3} \msun$ yr$^{-1}$ (Mpc $h^{-1})^{-3}$ at $z=9$ which is sustained until $z\simeq 6$ (which might then represent a sweet spot for experimental searches); beyond this epoch, Pop III star formation is rapidly quenched. Pop III SFR is always subdominant with respect to the Pop II one; however, at $z=9$ the Pop III rate is smaller only by a factor $\simeq 10$. This finding offers new hope for detecting these elusive stars in the near future, and particularly with JWST which is expected to be able to probe Pop III supernovae up to $z\simeq 15$ \citep[e.g.][]{Whalen:2013ApJ,deSouza:2013MNRASdec}.

Currently, observations of the $\HeII \lambda 1640$ emission line, excited by the hard UV spectra of Pop III stars, can be used to infer upper bounds on Pop III star formation. \citet{Cassata:2013A&A} estimate a SFR$_{\popiii}\lsim10^{-6}\msun$~yr$^{-1}{\rm Mpc}^{-3}$ at $z\simeq2.5$. While our simulations cannot be carried on beyond $z=4$ due to the limited volume, the upper limit is well above our simulated curve already at $z=4$. \citet{Cai:2011ApJ} used a WFC3/F130N IR narrowband filter to probe $\HeII \lambda 1640$ emission in the galaxy IOK-1 at $z = 6.96$. The detected He II flux $1.2 \pm 1.0 \times 10^{-18}$ erg s$^{-1}$ cm$^{-2}$, corresponds to a 1-$\sigma$ upper limit on Pop III SFR of $0.5~\msun$ yr$^{-1}$ in this galaxy (assuming a Salpeter IMF for this population) and representing $<~ 6$\% of the total star formation. If these figures are representative of the cosmic average, they would be in striking agreement with our results that predict at $z=6.96$ a Pop III to Pop II SFR ratio of $7.8\%$. While this extrapolation to large scales might be unwarranted, the consistency we find might hint at the fact that current experiments are on the verge of tracing the full star formation history of metal-free stars. Resolution effects might affect Pop III SFR; they are discussed in detail in Appendix \ref{sec_mass_res}.

\begin{figure}
\centering
\includegraphics[width=8.3cm]{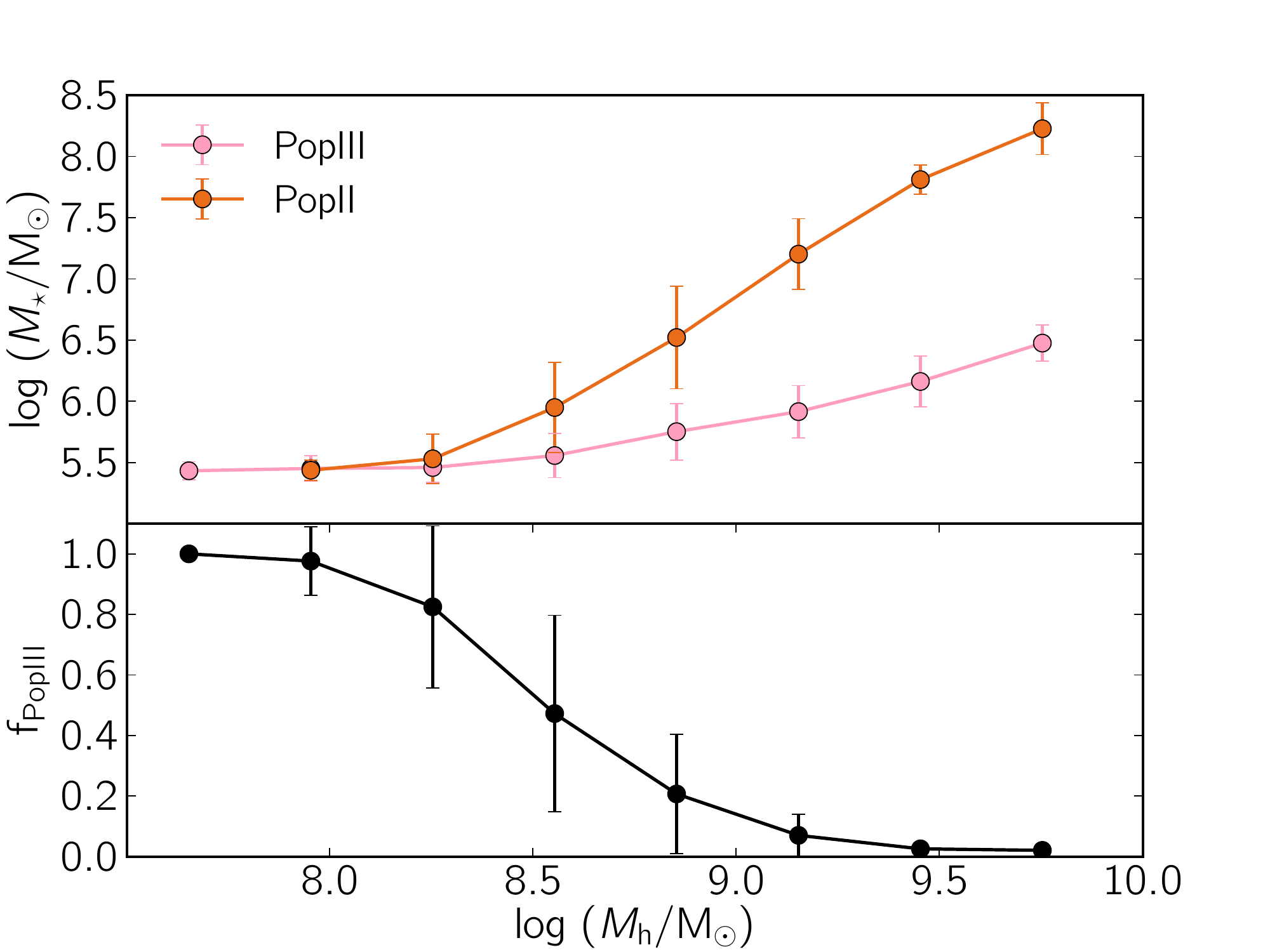}
\caption{Stellar mass content as a function of the hosting halo mass at $z=6$. {\it Upper panel}: mass in Pop II (orange line) and in Pop III (pink line) stars. {\it Lower panel}: the fraction of Pop III stars (black line), as defined in the text. The data has been binned in mass (log~$M\slash\msun \simeq 0.3$). Errorbars show the 1-$\sigma$ deviations within the bin.\label{fig_binned_halom_starm_003}}
\end{figure}

The overall Pop III evolution is not dissimilar from that found by some previous dedicated studies \citep{Tornatore:2007MNRAS, Johnson:2013MNRAS, Wise:2012ApJ}; however, we derive a Pop III SFR which is about one order of magnitude higher with respect to the results by \citet{Wise:2012ApJ} and \citet{Johnson:2013MNRAS}. At present the source of the discrepancy is unclear as it might be caused by numerical resolution, feedback treatment or limited volume effects. We will come back to this point in Sec. \ref{sec_gal_enrichment} and \ref{sec_test}.

The first Pop III stars in our simulation start to form in relatively biased regions at $z \simeq 15$. We do not attach a particular significance to this epoch as it is well known that it might depend strongly on the numerical resolution adopted. Earlier studies \citep[e.g.][]{Naoz:2006MNRAS,Trenti:2009ApJ,Gao:2010MNRASb,Fialkov:2012MNRAS,Salvadori:2013arXiv} have in fact shown that the first stars in the Universe might form as early as $z\simeq60$. Instead, it is worth noticing that in the same region after only $\sim200$~Myr the star formation mode has turned into Pop II stars. This shows that even a relatively modest Pop III star formation burst is sufficient to enrich the surrounding gas above $\zcrit$, as noted in previous works \citep[e.g.][]{Scannapieco:2003ApJ,Salvadori:2008MNRAS,Greif:2010ApJ,Wise:2012ApJ} and further elaborated in Sec. \ref{sec_halostar}.

In the following, we focus on the results specifically concerning cosmic metal enrichment. To improve the clarity of the presentation it is useful to discuss the problem separately for galaxies, the circumgalactic medium (CGM) and the intergalactic medium. The exact definitions of these components will be given in Sec. \ref{sec_global_result}. Albeit this classification might be somewhat arbitrary, there is a clear differentiation among these environments in terms of density and physical processes. We then start by analyzing the metal enrichment of galaxies.

\section{Galaxies}\label{sec_gal_enrichment}
\begin{figure}
\centering
\includegraphics[width=8.3cm]{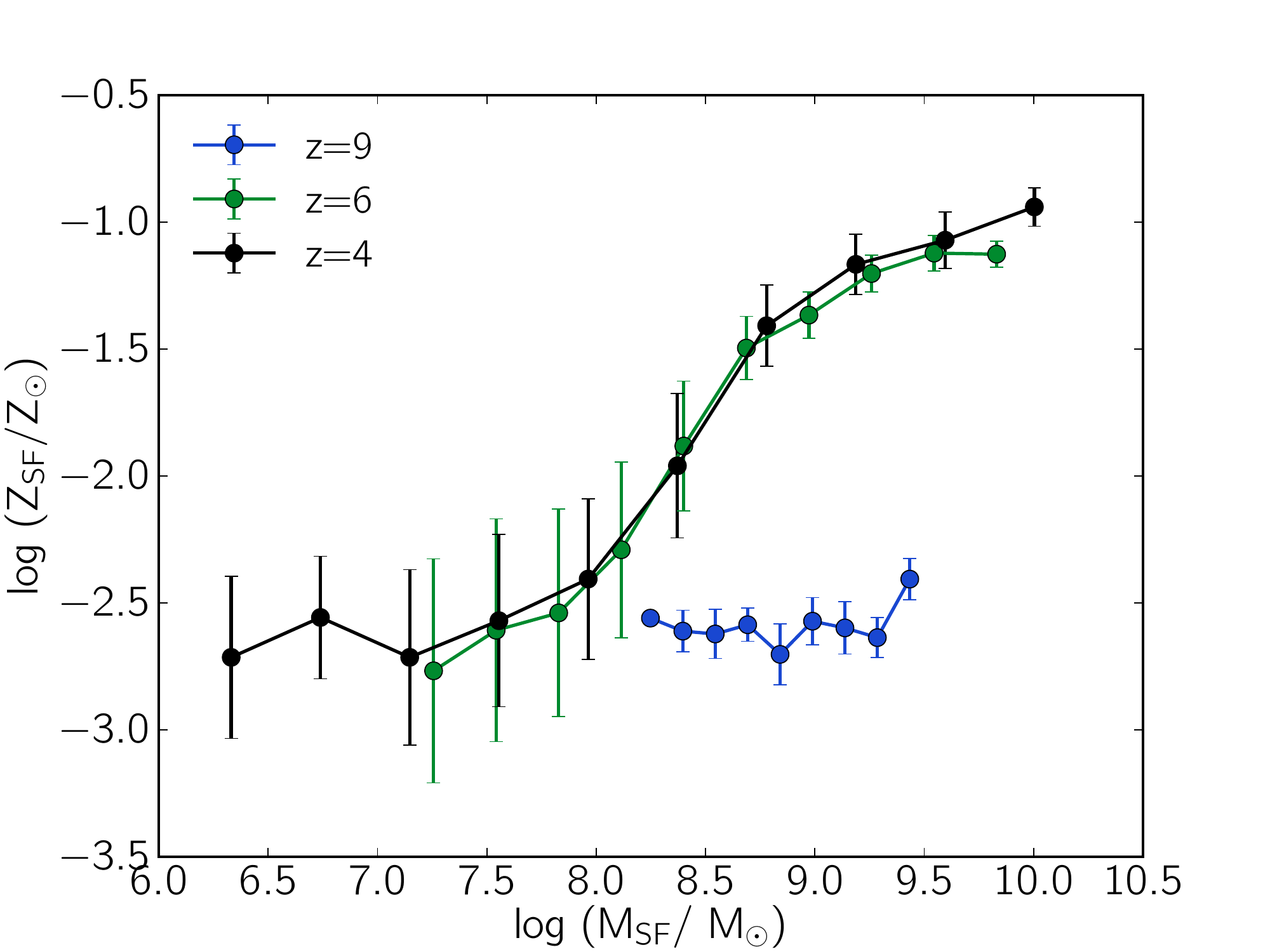}
\caption{Gas metallicity vs. total (gas+stars) baryonic mass in star forming regions at $z=9$ (blue line), $z=6$ (green line) and $z=4$ (black line). The data have been binned in mass ($\log(M/\msun)\simeq 0.2$). Errorbars show the 1-$\sigma$ deviations within the bin.\label{fig_mass_metal_sf_redshifts}}
\end{figure}
We identify galaxies and their hosting halo in each simulation snapshot with the method detailed in Appendix \ref{sec_fof_gal}. In brief, we use a Friend-of-Friend algorithm to identify both DM halos and the associated star forming regions; the latter are characterized by the condition that the gas has on overdensity $\Delta \equiv \rho/\langle\rho\rangle >\Delta_{\rm th}\equiv \rho_{\rm th} /\langle\rho\rangle$. We checked that the inferred cumulative halo mass functions are in agreement with the analytical one \citep{Press:1974,sheth:1999MNRAS} to better than $5\%$ at all masses and redshifts. This is about the maximum precision achievable from halo-finder codes \citep[e.g.][]{knebe:2013arXiv}.

As expected \citep{Christensen:2010ApJ,Wise:2012ApJ}, stars do not form in halos resolved with less then $\sim10^{2}$ particles, i.e. $M_{\rm h}\simeq10^{7.5}\msun$. However, physical arguments for star formation suppression in low mass (mini) halos have been given in several theoretical \citep{Haiman:2006ApJ, Alvarez:2012ApJ} and numerical \citep{Wise:2012ApJ,Johnson:2013MNRAS,xu:2013arXiv} works. This happens because, in absence of metals, the dominant gas cooling agent is molecular hydrogen which is easily dissociated by Lyman Werner (LW) radiation. The mass of the halo determines whether its column density is high enough to self-shield against photodissociation. Hence, for unpolluted minihalos, Pop III star formation is largely suppressed. Note that the strength of the quenching is still debated \citep{Haiman:2006ApJ} and in simulations it depends of the specific radiative transfer implementation \citep{Wolcott-Green:2011MNRAS}. 

In our simulation we do not account for the LW background nor do we apply a \HH-$\,$based star formation criterion; rather, the quenching of star formation in low mass halos is mimicked by mass resolution. This (convenient) numerical feature has also been noted by \citet{xu:2013arXiv} in comparison with the previous work by \citet{Wise:2012ApJ}. \citet{xu:2013arXiv} have a factor $\sim 16$ better resolution allowing them to resolve halos of mass $\sim10^{6}\msun$. However, a proper inclusion of the LW background leads \citet{xu:2013arXiv} to conclude that the minimum mass of star forming halos is $\simeq 3\times 10^{6}\msun$, in almost perfect agreement with findings of \citet{Wise:2012ApJ}.

The SFR history is closely connected to the evolution of metal enrichment. The total amount of metals produced by star formation rises from $\Omega^{\rm SFH}_{Z} = 1.52 \times 10^{-6}$ at $z=6$ to $8.05 \times 10^{-6}$ at $z=4$: this trend follows the growth of the cosmic stellar mass, as $\Omega^{\rm SFH}_{Z}\propto {\rm SMD}$ \citep[][]{Ferrara:2005ApJ}. The metal enrichment history has been tracked on-the-fly during the simulation. Fig. \ref{fig_otf_anal} shows the predicted redshift evolution of the mean metallicity of the gas, $\meanz$, calculated by averaging over all the baryons in the simulation volume, and the analogous one for star forming regions, i.e. $\meanz_{\rm SF}$, obtained by averaging only over cells in which $\Delta>\Delta_{\rm th}$.

As we see, $\meanz$ monotonically increases with time from $\sim10^{-6}\zsun$ at $z=11$ to $\sim10^{-2}\zsun$ at $z=4$. This result is consistent with \citet{Tornatore:2007MNRAS}, \citet{Maio:2010MNRAS} and \citet{Dave:2011MNRAS} and expected given the SFR evolution discussed previously.

Instead, if we only consider star forming regions, the mean metallicity evolution of these cells presents a slightly more complex behavior, also shown in Fig. \ref{fig_otf_anal}. It raises abruptly from very low values at $z\geq 10$ reaching a peak of $4 \times 10^{-3}\zsun$ at $z=9.3$, consistently with the metallicity level obtained by \citet{Greif:2010ApJ} in the inner part of a galaxy witnessing a single PISN explosion. After a short (100 Myr) decreasing phase, $\meanz_{\rm SF}$ begins to increase steadily. At $z\lsim 8$ star forming regions are roughly 10 times more metal-rich than the mean of all baryons. This peculiar trend is worth a more close scrutiny.

\begin{figure}
\centering
\includegraphics[width=8.3cm]{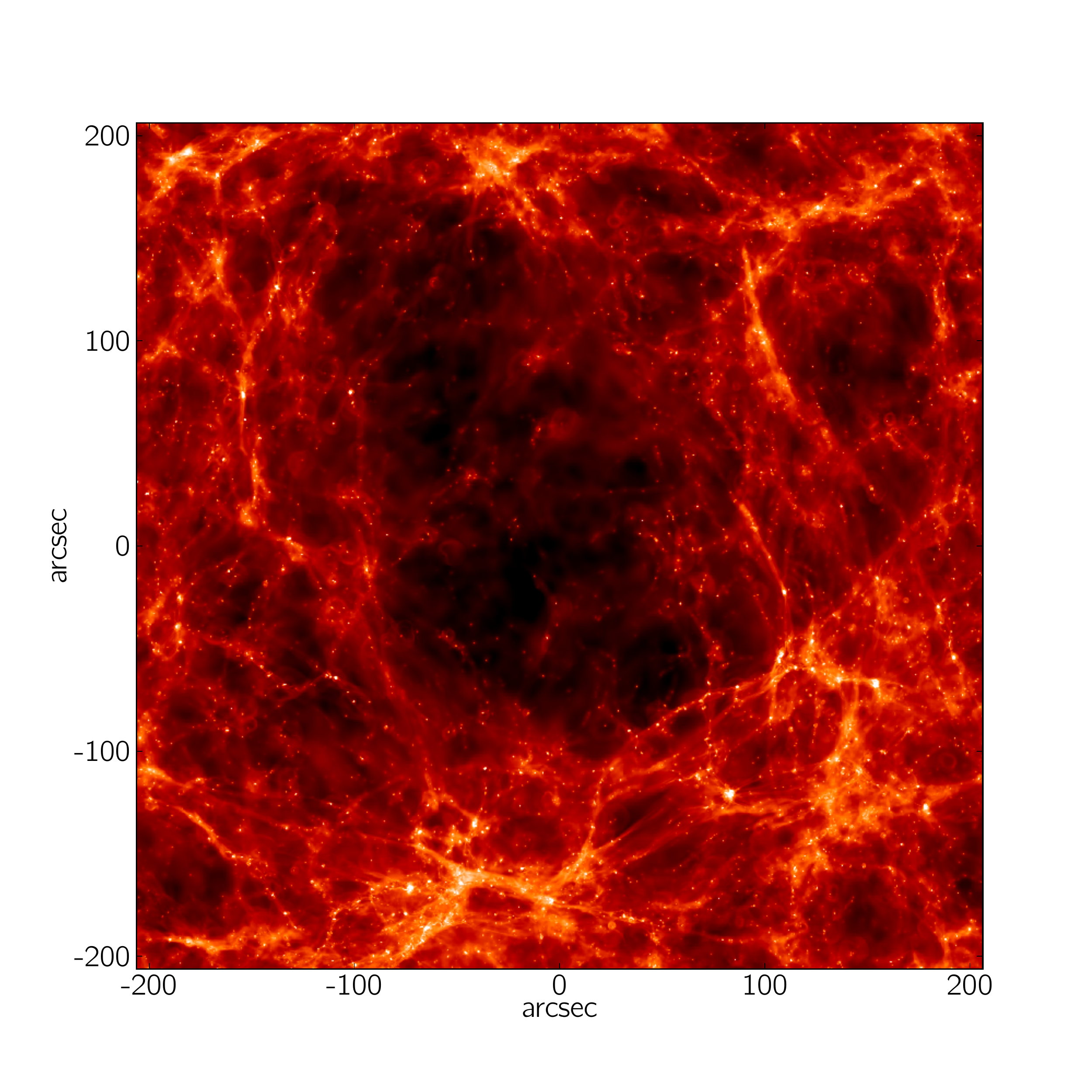}
\caption{Rendering of the gas temperature and density fields convolution at $z=4$ for the full 10 Mpc $h^{-1}$ simulation box, corresponding to about 400 arcsec. Details on the rendering technique can be found in Appendix \ref{sec_render}.\label{fig_render_density_04}}
\end{figure}

First, the drop of $\meanz_{\rm SF}$ for $z\gsim9.5$ is a selection effect introduced by our density-based definition of star forming cells ($\rho>\rho_{\rm th}$). In the first low-mass galaxies experiencing their first star formation event, SN feedback is sufficiently strong to completely disperse their gas and decrease the local gas density below $\rho_{\rm th}$. As at high $z$ this situation is almost the rule, $\meanz_{\rm SF}$ drops precipitously. This also implies that these galaxies are not able to sustain a steady star formation activity: this early flickering star formation mode is also reflected in the SFR evolution discussed in Fig. \ref{fig_sfr_smd}.

It is only when the feedback action becomes less violent and able to gently regulate the star formation activity of individual galaxies that the process stabilizes. This transition eventually occurs around $z\simeq8$, after a brief (100 Myr) phase in which metals spread from the interstellar medium (ISM) into the CGM and finally into the IGM. During this phase $\meanz_{\rm SF}$ decreases as a result of the increasing primordial gas mass into which metals are distributed. Note that as soon as a steady star formation activity can be sustained $\meanz_{\rm SF}>\zcrit$. Thus, Pop III star formation must be confined to external regions of star forming galaxies or located in the remaining pristine galaxies. We further discuss this point in the next Section and in Sec. \ref{sec_test}.

To summarize, the SFR increase occurs when galaxies (on average) enter the regulated star formation regime. The exact redshift of this transition might in principle vary with numerical resolution. Such dependence is analyzed in Sec. \ref{sec_test} and Appendix \ref{sec_mass_res}.

\subsection{Effects on stellar populations}\label{sec_halostar}
The enrichment history of a given galaxy, among other aspects, affects the evolution of its stellar populations (e.g. Pop III vs. Pop II) and it is well quantified by the evolution of $P(Z)$, the mass-weighted stellar Metallicity Distribution Function (MDF), shown in Fig. \ref{fig_metalpdf_star} for three selected epochs, $z=9, 6, 4$ and for all stars in the simulation box. Recall that we are adopting a critical metallicity for the Pop III - Pop II transition $\zcrit=10^{-4}\zsun$. Since in {\tt RAMSES} $Z$ is treated as a passively advected scalar field \citep[e.g.][]{Dubois:2007_EAS}, in the following we will consider stars with $Z\leq10^{-10}\zsun$ as primordial to exclude spurious effects due to diffusion. When calculating MDFs, we consider all stars formed before the selected redshift. 

\begin{figure*}
\centering
\includegraphics[width=5.5cm]{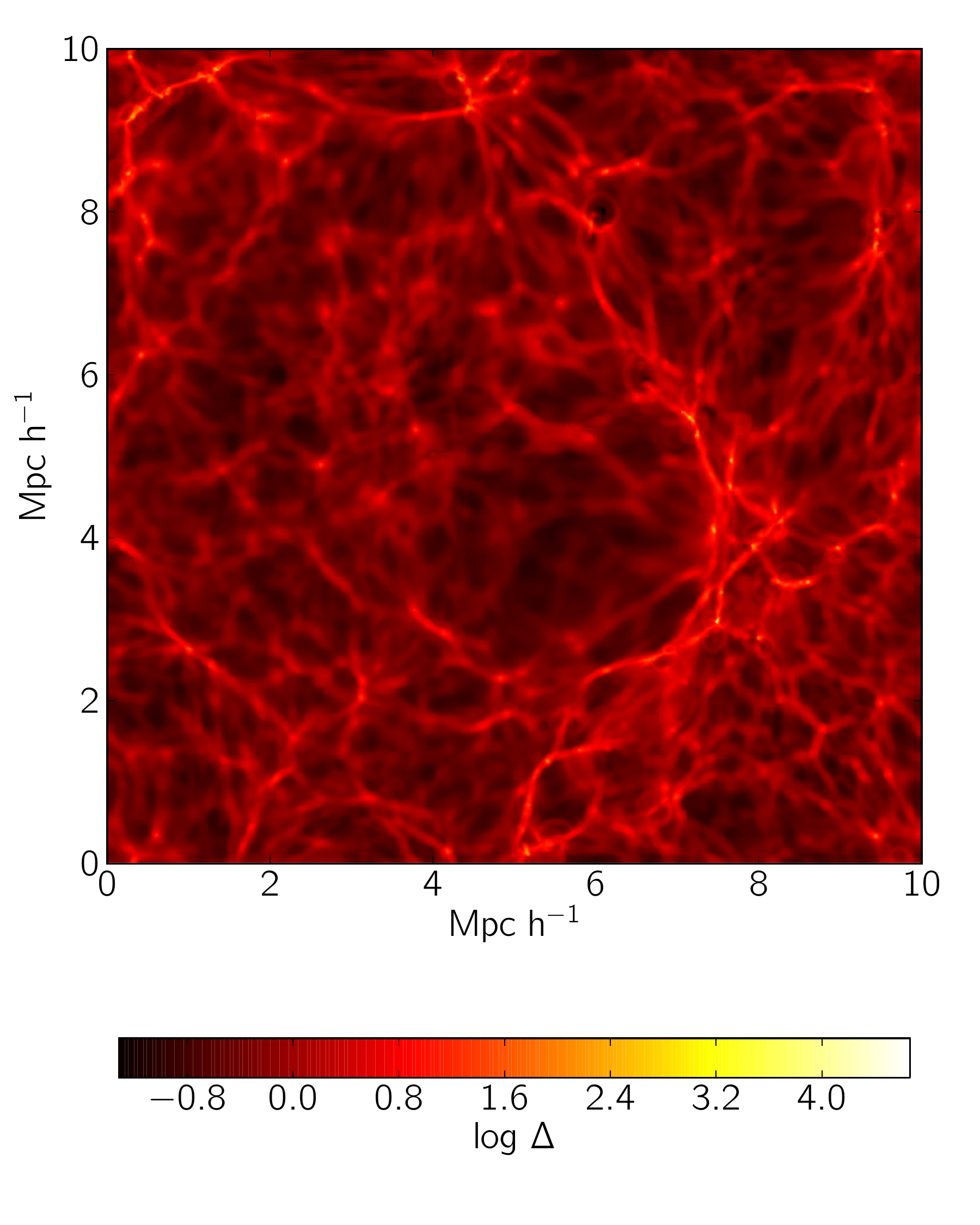}
\includegraphics[width=5.5cm]{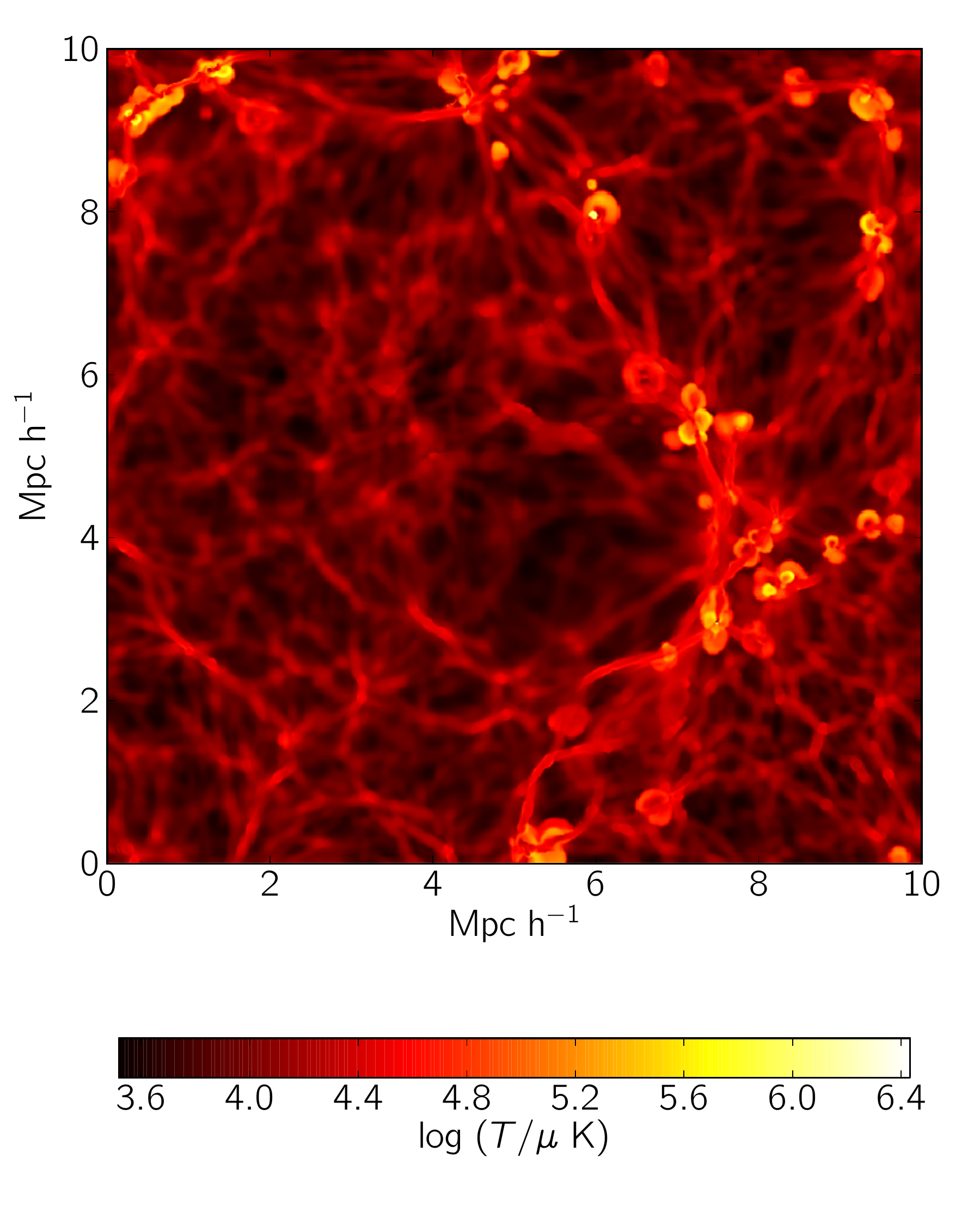}
\includegraphics[width=5.5cm]{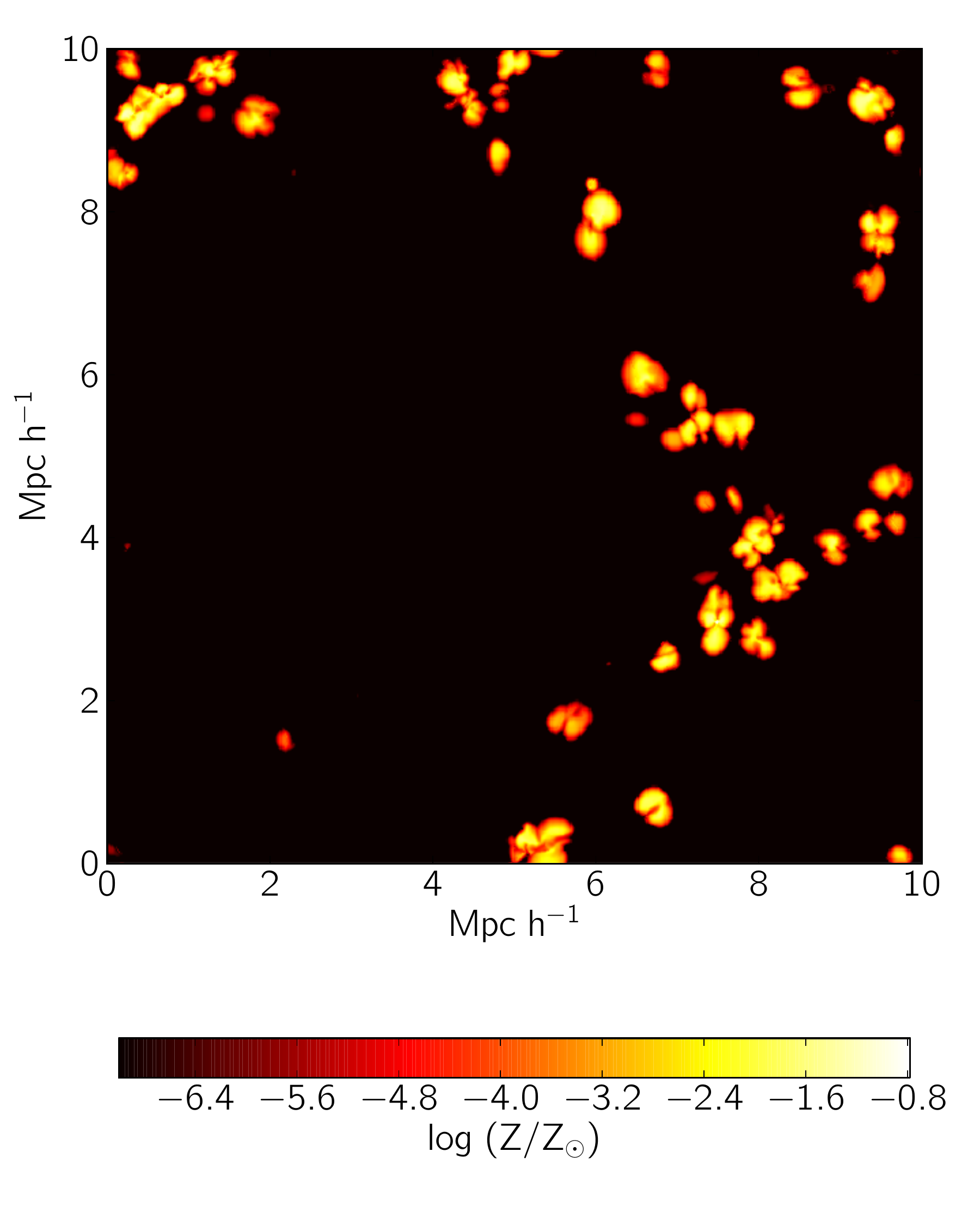}
\caption{Maps of the simulated gas overdensity, $\Delta=\rho/\langle \rho\rangle$ (left), temperature (center), and metallicity (right) at $z=4$ in a slice of thickness $19.53\,h^{-1}$~kpc through the box. \label{mappe}}
\end{figure*}

At all redshifts shown the MDF displays a pronounced double-peak structure. The low metallicity (Pop III) peak is always centered at $Z\simeq 0$, while the Pop II peak grows with cosmic time, contemporarily shifting from $\simeq10^{-2}\zsun $ at $z=9$ to $\simeq 10^{-1} \zsun$ at $z=4$. This evolutionary trend is induced by the monotonic increase of the mean metallicity of star forming regions that we report in Fig. \ref{fig_otf_anal}; as Pop II stars dominate the stellar mass for $4 \leq z \lsim 9$ (see Fig. \ref{fig_sfr_smd}) the peak position closely tracks the value of $\meanz_{\rm SF}$ at a given redshift. It has to be emphasized that since the metallicity is mass weighted, the evolution of $\meanz_{\rm SF}$ in Fig. \ref{fig_otf_anal} is dominated by the most massive star forming regions.

Interestingly, as time proceeds the MDF develops a characteristic low-metallicity tail which resembles the one derived from observations of metal-poor stars in the Milky Way \citep[e.g.][]{Caffau:2011A&A,Yong:2012AJ,Placco:2013ApJ} and Local Group dwarfs \citep[e.g.][]{Starkenburg:2010A&A,Frebel:2010Natur}. Clearly, the persisting non-detection of $Z\lsim10^{-6}\zsun$ stars in current available stellar samples, supports the idea that Pop III were more massive than today forming stars, as already pointed out in \citet{Salvadori:2007MNRAS}. However, whether the local stellar MDF can be considered as universal remains an open problem.

The MDF evolution can be used to characterize the environments in which Pop III stars formed. To this aim, we construct the indicator $R_{\rm P}$ expressing the ratio between the number of Pop III stars formed in extremely metal-poor (but not pristine; therefore these sites have already been enriched by a contiguous star formation episode to some $Z_* < \zcrit$) environments and the total number of Pop III stars formed:
\begin{subequations}
\begin{equation}
R_P=\frac{1}{N_{III}} \int_{Z_*}^{\zcrit} \frac{{\rm d}P}{{\rm d}Z} {\rm d}Z\, ,
\end{equation}
where
\begin{equation}
N_{III}= \int_{0}^{\zcrit} \frac{{\rm d}P}{{\rm d}Z} {\rm d}Z\, .
\end{equation}
\end{subequations}
Assuming $Z_*\equiv10^{-8}\zsun$, from the previous formula we obtain $R_{P}(z=9) = 0.29$ and $R_{P}(z=6)\simeq R_{P}(z=4)= 0.19$. This result implies that Pop III stars are preferentially formed in purely pristine regions, well outside the polluted regions created by nearby/previous star forming galaxies. Only a minor fraction of Pop III is formed in inefficiently enriched sites. This scenario is in agreement with the notion of a ``Pop III wave'' suggested by \citet{Tornatore:2007MNRAS} and confirmed by \citet{Maio:2010MNRAS}, starting from galactic centers and rapidly migrating towards more external regions where pristine gas to sustain their formation can be more easily found. Eventually, the galaxy will run out of unpolluted gas and its Pop III formation comes to a halt.

The $\simeq1\slash3$ decrease of $R_P$ from $z=9$ to $z=6$ indicates that galactic outflows become progressively more efficient in increasing the metallicity of polluted regions above $\zcrit$, thus preventing Pop III to form in regions with very low but non-zero metal content. Finally, the negligible $R_P$ variation between $z=6$ and $z=4$ is a consequence of the saturation of the Pop III SMD toward $10^{6}\msun\,{\rm Mpc}^{-3}$ at $z\simeq5$ (see Fig. \ref{fig_sfr_smd}).

To understand which galaxies are the preferred sites of Pop III star formation we analyze the $\sim1600$ star-forming halos at $z=6$. In the upper panel of Fig. \ref{fig_binned_halom_starm_003} we plot the Pop II and Pop III stellar mass ($M_{\popii}$ and $M_{\popiii}$) vs. halo mass. We define the Pop III mass fraction as $f_{\popiii}=\langle M_{\popiii}\slash (M_{\popii}+M_{\popiii})\rangle$, where the average is performed on the mass bin; $f_{\popiii}$ is shown in the lower panel\footnote{Note that $f_{\popiii}\neq\langle M_{\popiii}\rangle \slash \langle M_{\popii}+M_{\popiii}\rangle$; not keeping this in mind can lead to misinterpretation when comparing the two panels.} of Fig. \ref{fig_binned_halom_starm_003}.

\begin{figure*}
\centering
\includegraphics[width=8.3cm]{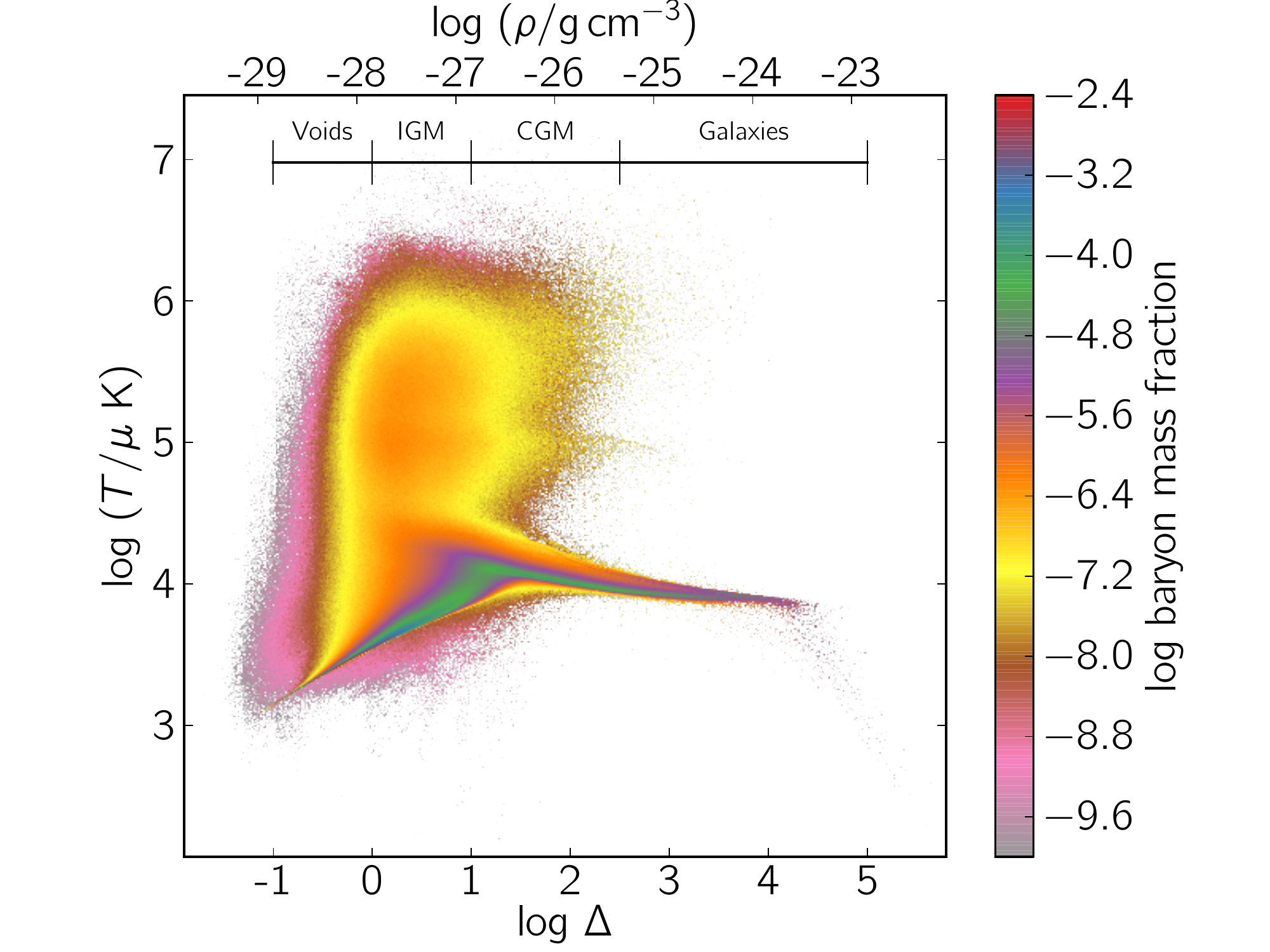}
\includegraphics[width=8.3cm]{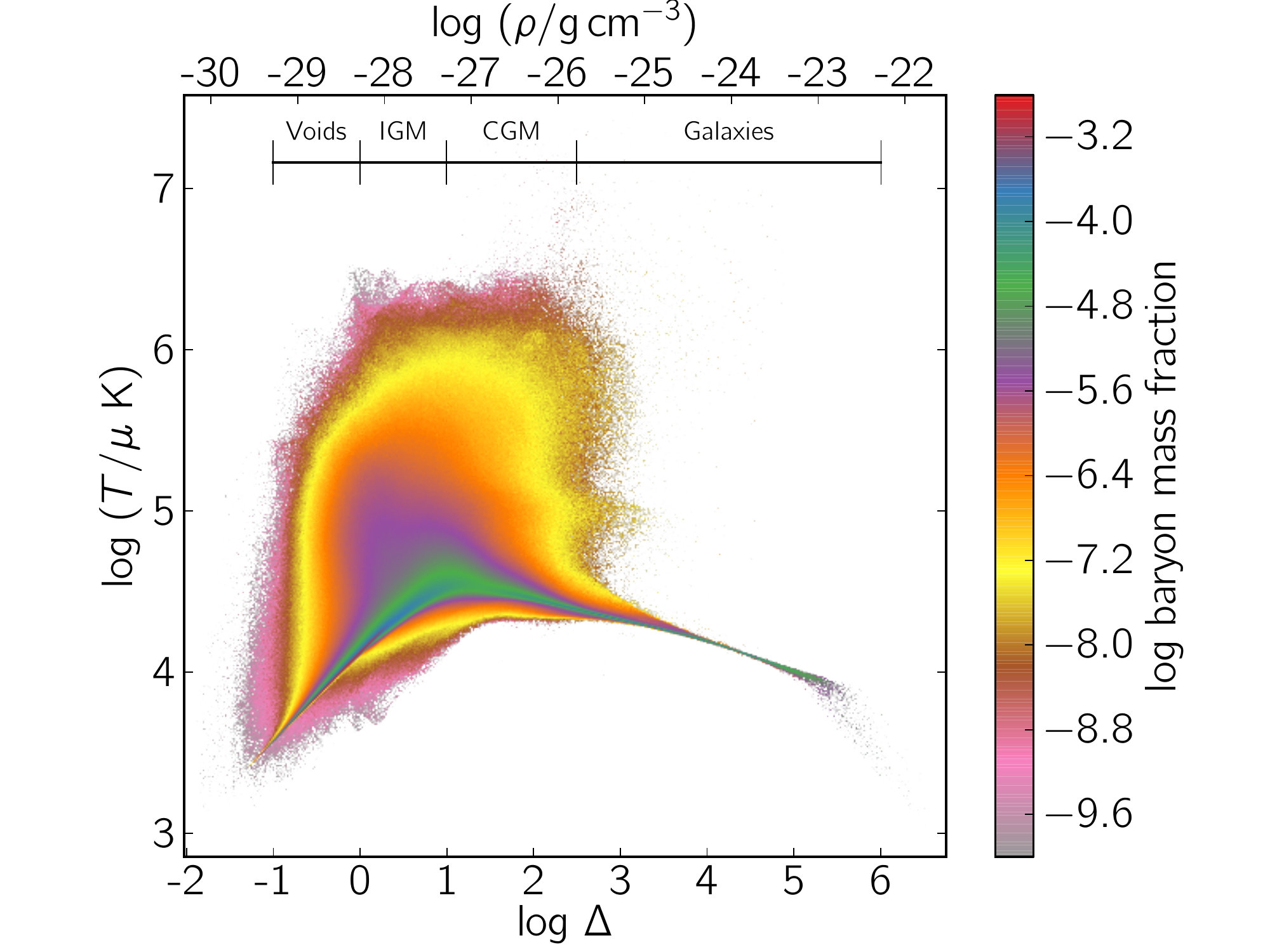}
\caption{Equation of state (EOS) of the baryons at $z=6$ (left) and $z=4$ (right): the colorbar represents the differential mass weighted probability function. Temperature is expressed in molecular weight units; the density is given both in cgs units and in terms of the overdensity $\Delta=\rho/\langle \rho\rangle$. \label{fig_eq_stato}}
\end{figure*}
The Pop III mass fraction monotonically decreases with increasing halo mass, going from $\simeq1$ at $M_{\rm h}\sim 10^{7.5}\msun$ to $\lsim10^{-2}$ for $M_{\rm h}\gsim 10^{9.5}\msun$. This is expected as only small ($M_{\rm h}\lsim10^{8}\msun$) halos can produce Pop III stars, which are then inherited along the hierarchical formation sequence by larger descendants. The very small scattering in the stellar mass seen in the first two mass bins ($10^{7.5}\lsim M_{\rm h}\slash\msun\lsim10^{8}$) results form the fact that these small halos have essentially experienced a single (or a few at most) starburst event. As the amount of metals produced is typically sufficient to enrich the gas to values above $\zcrit$, the subsequent star formation activity will already produce Pop II stars. In conclusion, we suggest that the most suitable halo mass range to observe Pop III stars is $10^{8.4}\lsim M_{\rm h}\slash\msun\lsim10^{8.7}$, since in these galaxies Pop III signature can be distinguished from Pop II, i.e. $f_{\popiii}\simeq0.5$, and at the same time their UV luminosities are sufficiently high to be detected in deep JWST spectroscopic surveys around $z=6$. 
\subsection{Mass-metallicity relation}
Star forming regions can be identified and associated with their parent galaxy using the technique described in Appendix \ref{sec_fof_gal}. We find a one-to-one match of star forming regions and galaxies; this is expected, since the finding algorithm does not allow to resolve substructure on scales $<9.76\,h^{-1}$~kpc. Once the galaxy catalogue at different redshifts has been built, we can relate the gas metallicity of each star forming region, ${Z}_{\rm SF}$, with its total (gas+stars) baryonic mass, ${M}_{\rm SF}$. This relation is shown in Fig. \ref{fig_mass_metal_sf_redshifts} for various redshifts and for a mass bin size log~$M\slash\msun\simeq0.2$. The errorbars represent the r.m.s. scattering within the mass bin.

At $z=9$ (blue line), the metallicity is essentially independent of the stellar mass, achieving an almost constant values of ${Z}_{\rm SF}\simeq10^{-2.7}\zsun$ in the entire mass range, $10^{8}\lsim{M}_{\rm SF}\slash\msun\lsim10^{9.5}$. This surprising behavior is due to the fact that at $z=9$ star forming regions have experienced only one or few bursts. Hence, the ${Z}_{\rm SF}\simeq10^{-2.7}\zsun$ enrichment level is set by the dilution of the heavy elements produced by these early bursts into the surrounding ISM. This point can be clarified by the following simple argument. Suppose that a star forming region has recently formed a mass of stars $M_{\star}$, and consequently a mass of metals $M_Z=Y M_{\star}$. To a first approximation the surrounding interstellar gas mass swept by the SN blast will be $M_g \propto \epsilon_{sn} M_{\rm \star}$. Hence, the expected gas metallicity from a single burst is ${Z}_{\rm SF} = M_Z/M_g \equiv \mathrm{const.}$, and the mass-metallicity relation shows no trend.

As star formation continues and stabilizes on larger scales a more definite trend emerges. At $z=6$ (green line) for example, we see that metallicity increases with stellar mass up to ${\rm M}_{\rm SF}\simeq10^{9.2}$, where the relation starts to flatten. This trend becomes slightly more evident at $z=4$, as it extends to a wider mass range.

On the other hand, for $M_{\rm SF}\gsim 10^{7} \msun$, the shape of the $z=4$ curve (black line) remains almost identical to $z=6$, with little signs of evolution over the time span of 600 Myr elapsed between the two epochs, as already pointed out by a recent study by \citet{Dayal:2013MNRAS}. Also noticeable is the apparent invariance of the flattening scale, which might be related to the ability of systems with $M_{\rm SF}\gsim 10^{9.5} \msun$ to retain most of the produced metals \citep[e.g.][]{Mac_Low:1999ApJ}, in contrast to lower mass systems that are more likely to eject them into the IGM due to their shallower potential wells.

Note that at $z = 4$ the relation extends below $M_{\rm SF}\lsim 10^{7} \msun$: this is due to the presence of satellite galaxies forming in pre-enriched regions of the CGM. Thanks to efficient cooling by metal lines gas in these low-mass systems can now become sufficiently dense to trigger star formation. Similarly to what found at $z = 9$ we see that $M_{\rm SF} \lsim 10^7 \msun$ regions show an almost constant metallicity, $Z_{\rm SF} \sim 10^{-2.7} \zsun$, which is the result of the few bursts of star-formation they experienced. Such a flat metallicity trend resembles the one observed in the faintest Local Group dwarf galaxies \citep[e.g.][]{Kirby:2013arXiv}, and its physical interpretation is in agreement with the findings of \citet{Salvadori:2009MNRAS}.

At $z=4$, we can compare our $\massmetal$ relation with the observed one, inferred from a sample of 40 galaxies at $3\lsim z\lsim 5$ with masses $10^{8}\msun\lsim M_{\star}\lsim 10^{11}\msun$ \citep[e.g.][]{Troncoso:2013arXiv1311}. The simulated high masses ($M_{\rm SF}\gsim10^{9.5}\msun$) star forming regions contain $M_{\star}\simeq10^{8.5}$. Averaging their metallicity we can convert to oxygen abundance by assuming a solar composition \citep[e.g.][]{Asplund:2009ARA&A}: we find $\log({\rm O}\slash H)+12=8.19$, in agreement within 1-$\sigma$ of the values found by \citet[][]{Troncoso:2013arXiv1311}.

\section{Diffuse cosmic gas}\label{sec_global_result}
We start by presenting a qualitative overview of the cosmic gas distribution in Fig. \ref{fig_render_density_04}. The main image shows a rendering of the gas temperature and density fields convolution for the whole simulation box at $z=4$. The rendering has been obtained using the back-front imaging technique presented in Appendix \ref{sec_render}. This technique both allows to clearly identify the typical cosmic web structure made of voids, filaments, knots and clusters, and, thanks to the temperature weighting, the supernova shock structures.

More quantitative information can be gathered from the density, temperature and metallicity maps at $z=4$ (Fig. \ref{mappe}). A comparison between the density and temperature fields allows to identify both active and/or relic star forming regions ($\Delta >\Delta_{\rm th}$) that are characterized by shock-heated hot ($T\, \mu^{-1}\gsim 10^{5}$~K) gas contained in approximately spherical bubbles, of size up to several hundred comoving kpc. The hot gas is enriched in heavy elements up to $\simeq 10^{-1} \zsun$ and many of the bubbles can be identified in both maps. However, we see several bubbles with significant metallicity that contain cooler ($2-5\times 10^4$~K) gas. These bubbles have been produced by earlier stellar populations and had the time to cool via adiabatic expansion and -- to a lesser extent -- radiative cooling. This implies that metallicity, differently from temperature, retains a fair record of the cosmic star formation activity.

An operative classification of the various cosmic environments can be made in terms of the gas overdensity. In Fig. \ref{fig_eq_stato}, we show the gas mass-weighted equation of state (EOS) at $z=6$ and $z=4$. We can define four different environments: (a) the \textit{voids}, i.e. regions with extremely low density ($\Delta\leq 1$); (b) the true \textit{intergalactic medium}, characterized by $1<\Delta\leq 10$; (c) the \textit{circumgalactic medium} (CGM, $10<\Delta\leq 10^{2.5}$), representing the interface between the IGM and galaxies, and (d) high density ($\Delta> 10^{2.5}$) collapsed structures, that for brevity we denote as \textit{galaxies}.

\begin{figure}
\centering
\includegraphics[width=8.3cm]{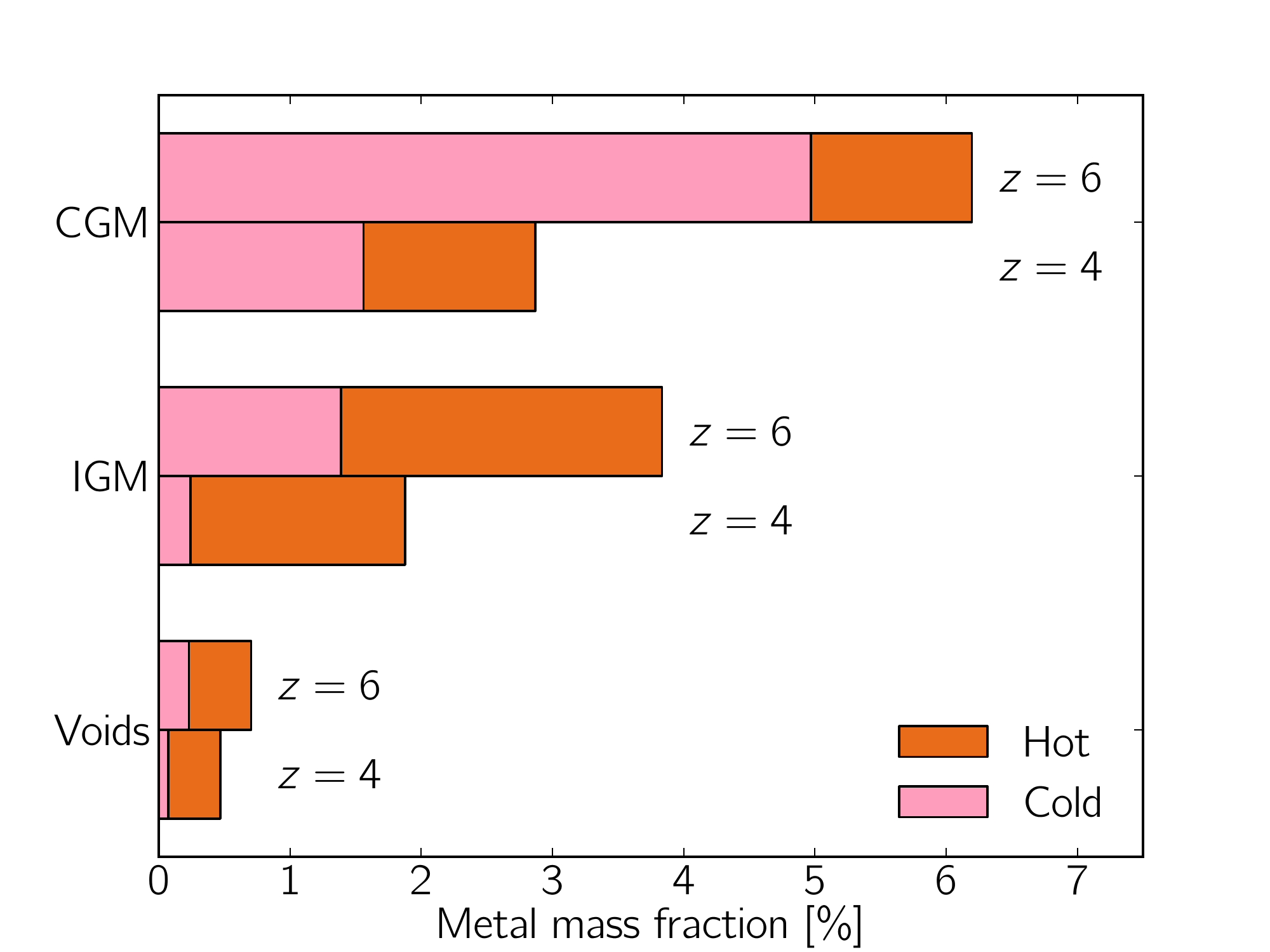}
\caption{Phase distribution of the {\it enriched} intergalactic gas at $z=6$ and $z=4$. For every phase, the relative cold ($T\, \mu^{-1}\le 10^{4.5}$~K) and hot ($T\, \mu^{-1} > 10^{4.5}$~K) parts are shown in pink and orange, respectively.\label{fig_hysto}}
\end{figure}

This classification does not account for the thermal state of the gas. In fact, all the environments but galaxies\label{In reality a small amount of ionized/hot gas might be also present in galaxies due to small-scale energy deposition of supernovae, but our simulations cannot resolve these structures.} are characterized by a range of temperature. For this reason, we will often call ``diffuse phases'' the environments (a), (b), (c). As the gas is heated either by photo-ionization ($T\, \mu^{-1}\lsim 10^{5}$~K) or by shocks ($T\, \mu^{-1}\gsim 10^{5}$~K), to improve our classification we discriminate between gas colder or hotter than $T\, \mu^{-1} =10^{5}$~K, thus allowing to readily get an estimate (see Fig. \ref{fig_hysto}) of the relative importance of such heating mechanisms in the various phases.

Fig. \ref{fig_eq_stato} allows to build a complete census of the cosmic baryons. Most of the baryons reside in the diffuse phases, with galaxies accounting only for a tiny fraction of the total mass steadily increasing from $\simlt 5$\% at $z=6$ to $\simeq 9\%$ at $z=4$. Among diffuse phases, the CGM contains about 15\% of the baryons, the remaining fraction being almost equally divided between voids and the IGM with little variation from $z=6$ to $z=4$.

At both redshifts, there is a continuous transition from voids to the IGM as both components follow a $T \propto\Delta^{\gamma}$ relation, with an adiabatic index $\gamma = 1\slash2$. This relation is well known to arise from the balance between adiabatic expansion cooling and photo-heating \citep[e.g.][]{Theuns:1998MNRAS}. In the CGM, however the EOS flattens as a result of the increasing importance of radiative cooling losses, driven both by a higher density and by a larger metal abundance, as we will see in the next Section. Metal cooling is important also for the shock-heated IGM. Compared to metal-free simulations \citep[e.g.][]{Pallottini:2013_scintillazione}, where typical values of the shock-heated gas temperature can reach $T\mu^{-1}\sim 10^{8}$~K, here metal cooling decreases the bulk temperature of the gas to values $T\mu^{-1}\lsim 10^{6}$~K. Overall this picture is consistent with previous results found by \citet{Rasera:2006} and also by \citet{Cen:2011ApJ} (in particular see their Fig. 20).
\begin{figure*}
\centering
\includegraphics[width=8.3cm]{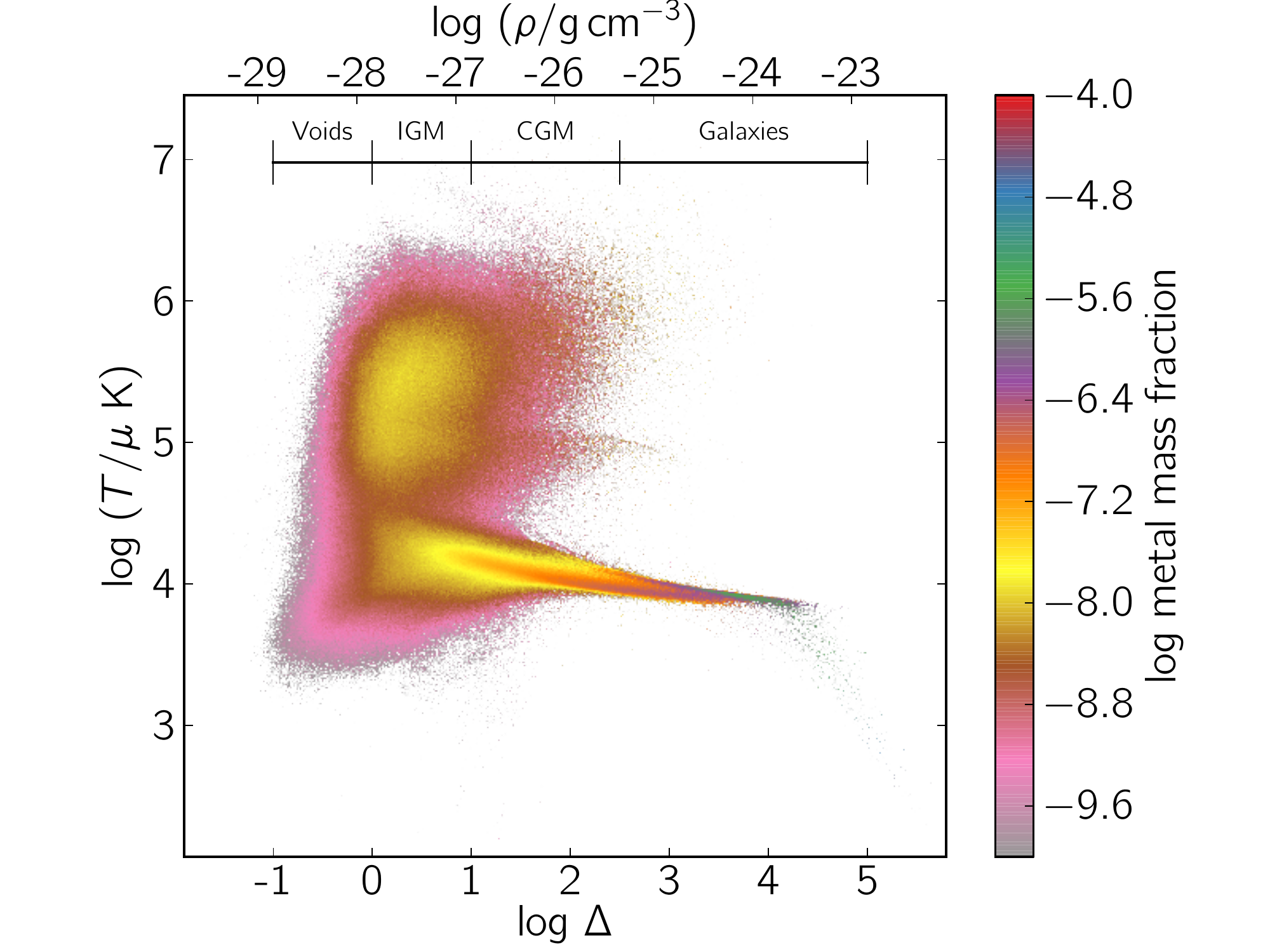}
\includegraphics[width=8.3cm]{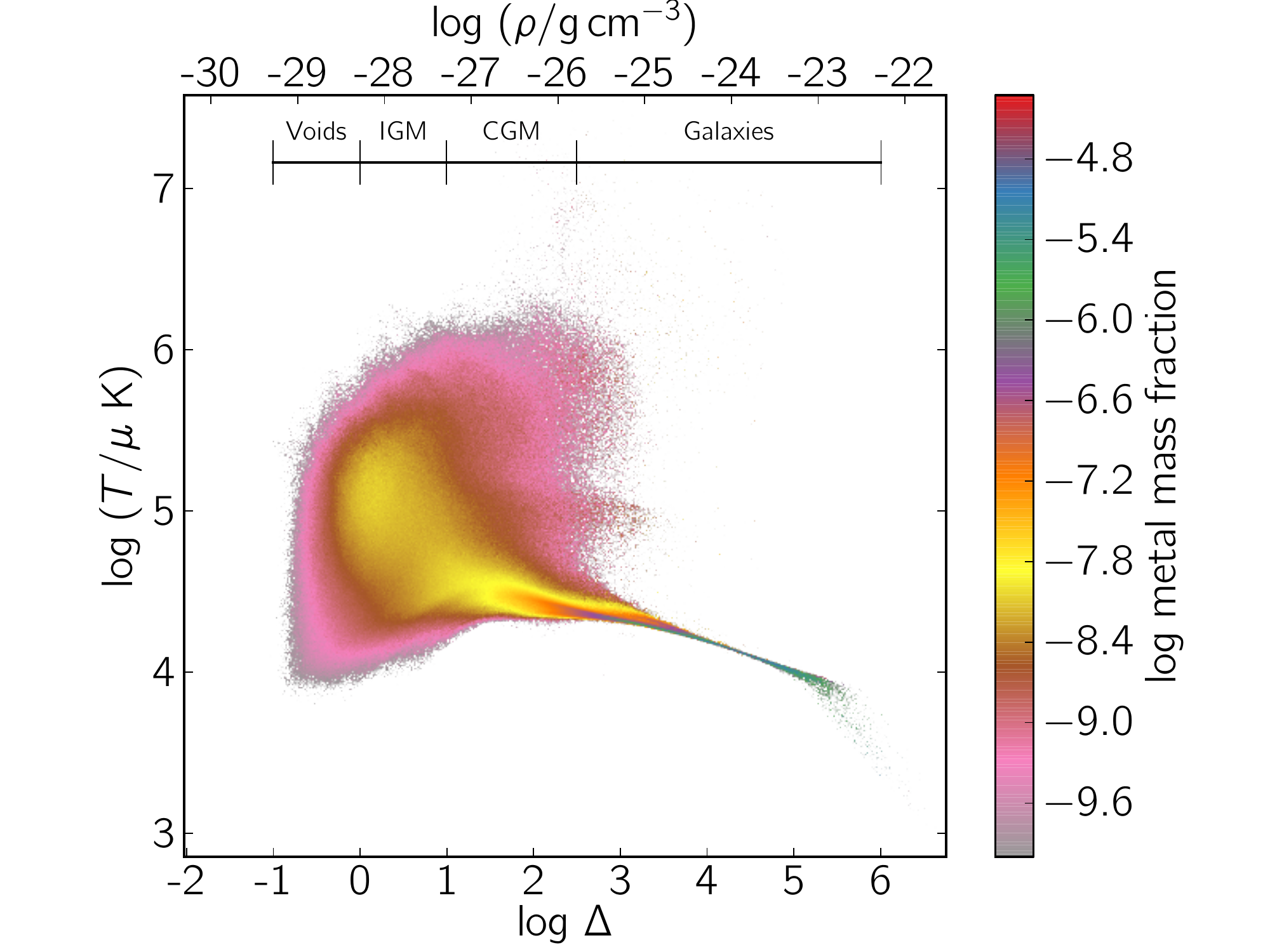}
\caption{Metal weighted EOS at $z=6$ (left) and $z=4$ (right). The notation is the same as in Fig. \ref{fig_eq_stato}.\label{fig_eq_stato_metal}}
\end{figure*}

\begin{figure*}
\centering
\includegraphics[width=8.3cm]{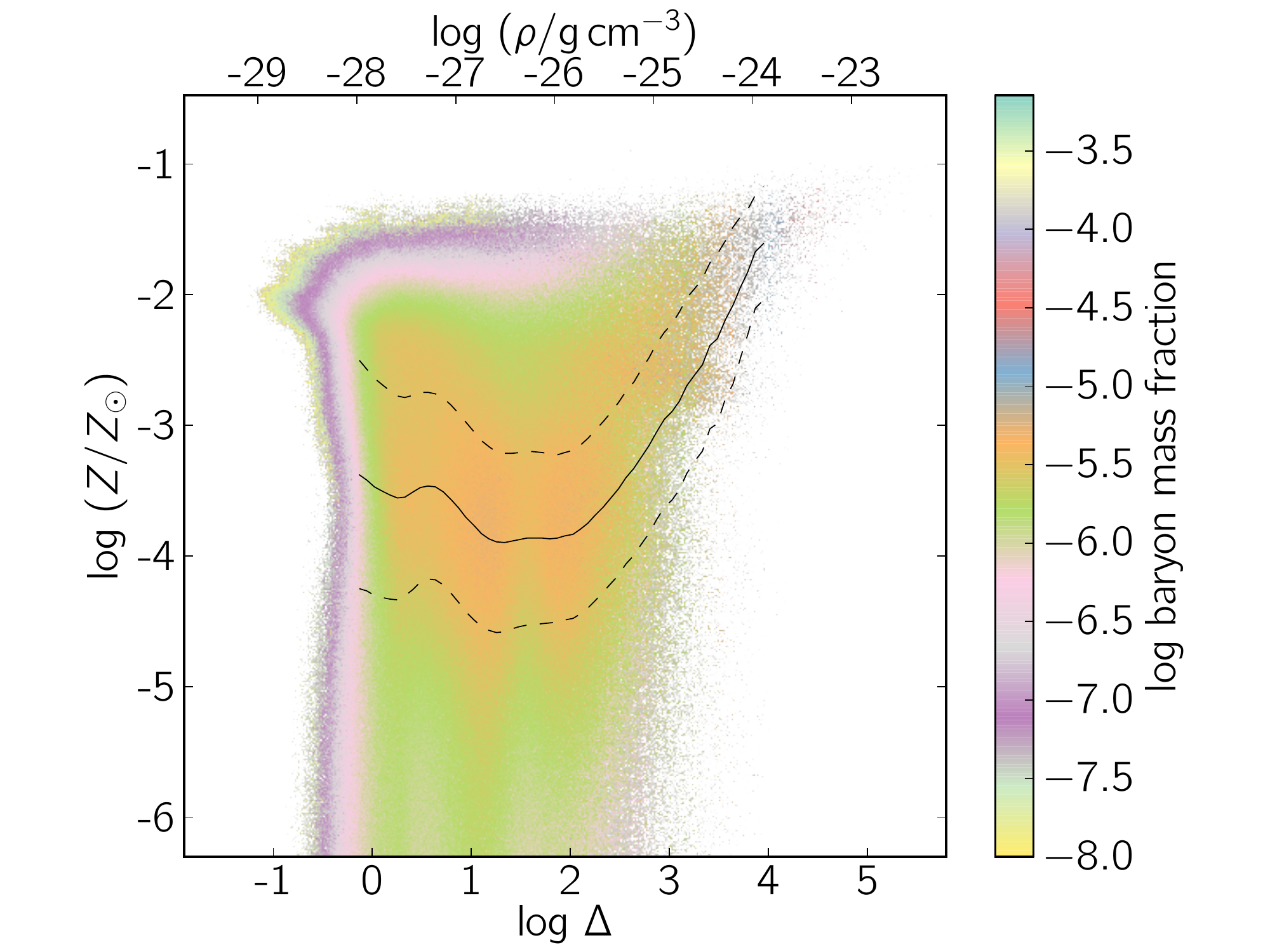}
\includegraphics[width=8.3cm]{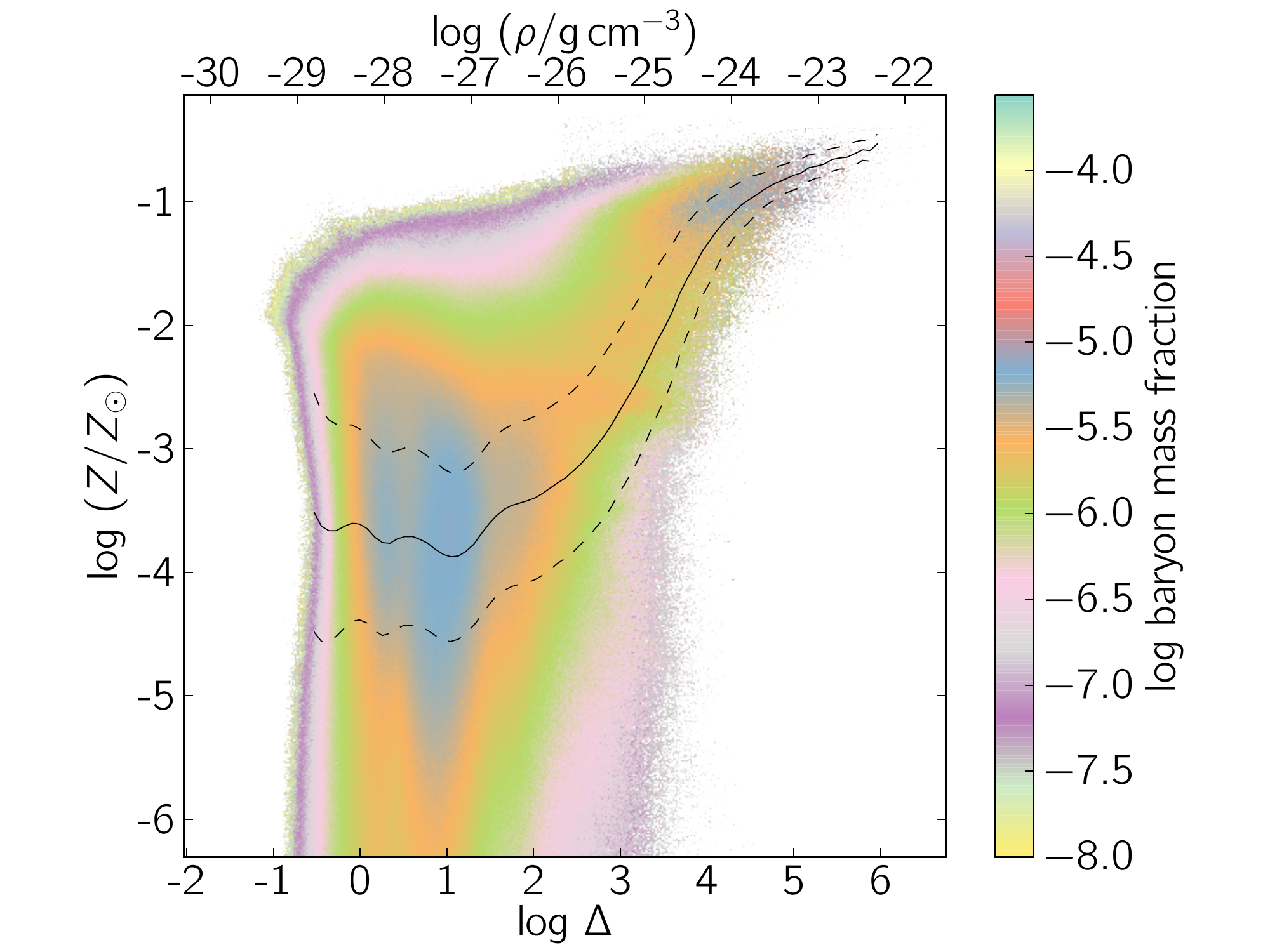}
\caption{Mass weighted probability distribution function (PDF) of the baryons at $z=6$ (left) and $z=4$ (right) in the metallicity-overdensity plane. The solid (dashed) black line is the mean (r.m.s.) metallicity as a function of density.\label{fig_z_stato_04}}
\end{figure*}

\subsection{Metal enrichment}\label{metenr}
Contrary to baryons, which reside predominantly in the IGM, metals are found at any given redshift primarily near their production sites, i.e. in galaxies. However, while at $z=6$ galactic metals make up to about 90\% of the total heavy elements mass, at later epochs ($z=4$) this fraction increases to 95\% as a result of the increased ability of collapsed objects to retain their nucleosynthetic products thanks to their larger potential wells. Among the diffuse components (see Fig. \ref{fig_hysto}), at $z=6$ the CGM is more enriched than the IGM (voids) by a factor 1.6 (8.8), as metals cannot be efficiently transported by winds into distant, low-density regions. Interestingly though, even the most diffuse gas in the voids has been polluted to some extent. We will see later on that by $z=4$ about 1/10 of the cosmic volume has been enriched to a non-zero metallicity. These results are consistent with previous studies \citep[e.g.][]{Oppenheimer:2006MNRAS}.

In Fig. \ref{fig_eq_stato_metal} we plot the metal mass-weighted EOS at $z=6$ (left panel) and $z=4$ (right panel). The temperature structure shows a clear evolution between the two redshifts. The most prominent feature is that the enriched IGM ($\Delta=1-10$) at $z=6$ is characterized by a bimodal temperature distribution, which then merges into a single thermal structure by $z=4$. We interpret this evolution as the result of the early IGM enrichment and heating by winds from low mass galaxies followed by cooling due to radiative losses. Note that the same temperature feature is visible in the baryonic EOS (Fig. \ref{fig_eq_stato}). The relative abundance of cold ($T\, \mu^{-1}\le 10^{4.5}$~K) and hot ($T\, \mu^{-1} > 10^{4.5}$~K) gas in the various diffuse phases is shown in Fig. \ref{fig_hysto}. 

The interesting conclusion is that while in the CGM the majority of metals are found to be cold, in the IGM and in voids densities are too low to allow the enriched gas to efficiently cool. Let us then evaluate the cooling time at $z=6$:
\be
t_c (\Delta)= \frac{3}{2} \frac{k_B T}{n \Lambda(T,Z)} \simeq 4.8 \left(\frac{T}{10^6 \textrm{K}}\right) \Delta^{-1} \textrm{Gyr}, 
\label{tcool} 
\ee
where we have assumed a value for the cooling function $\Lambda=2 \times 10^{-23} \textrm{erg cm}^{3} \rm s^{-1}$ appropriate for a gas with $T\, \mu^{-1}=10^6$~K and $Z=10^{-2} \zsun$. Thus, shock-heated, enriched overdensity with $\Delta>7.6$ will be able to cool during the cosmic time interval between $z=6$ and $z=4$, i.e. 0.63 Gyr, consistently with the results of our simulations.

This simple argument also explains the increasing relative fraction of hot metals with decreasing density of the diffuse phases, shown in Fig. \ref{fig_hysto}. This point has two important implications: (a) a considerable fraction of metals are hidden in a hot phase of the CGM/IGM/voids that is difficult to detect via absorption line experiments, highlighting the long-standing missing metals problem already noted by \citet{Pettini99} and quantified by \citet{Ferrara:2005ApJ}; (b) a small ($\simlt 1$\%) but not negligible fraction of metals managed to reach very rarefied environments as the voids, and in some case also to cool. This implies that these metals must have been injected at sufficiently early epochs that they had the time to cool, i.e. as expected in a pre-enrichment scenario \citep{Madau:2001ApJ}.

The $Z-\Delta$ distribution of the cosmic gas provides additional insights in the metal enrichment process (Fig. \ref{fig_z_stato_04}). At $z=6$ (left panel) baryons are nearly uniformly distributed in $10^{-1}\lsim\Delta\lsim 10^{2.5}$ and the cosmic gas is characterized by a broad range of metallicities ($10^{-6} \lsim {Z}\slash\zsun \lsim 10^{-2}$). Besides containing most of the metals, galaxies ($\Delta\gsim10^{2.5}$) show high metallicities ($10^{-2} \lsim Z\slash\zsun \lsim 10^{-1}$) and a loose $Z-\Delta$ correlation. At $z=4$ (right panel) the distribution evolves and the $Z-\Delta$ correlation at high density becomes tighter and steeper, additionally extending to lower overdensities. Both the IGM and the CGM become preferentially enriched at $10^{-4.5}\lsim Z\slash\zsun\lsim10^{-2.5}$, i.e. around the critical metallicity for the Pop III transition. The overall shape of the distributions at different $z$ agree well with those found by \citet{Gnedin:1997}, \citet{Oppenheimer:2012MNRAS} and with the observed IGM metallicities \citep[e.g.][]{Meiksin:2009RvMP}. 

\begin{figure}
\centering
\includegraphics[width=8.3cm]{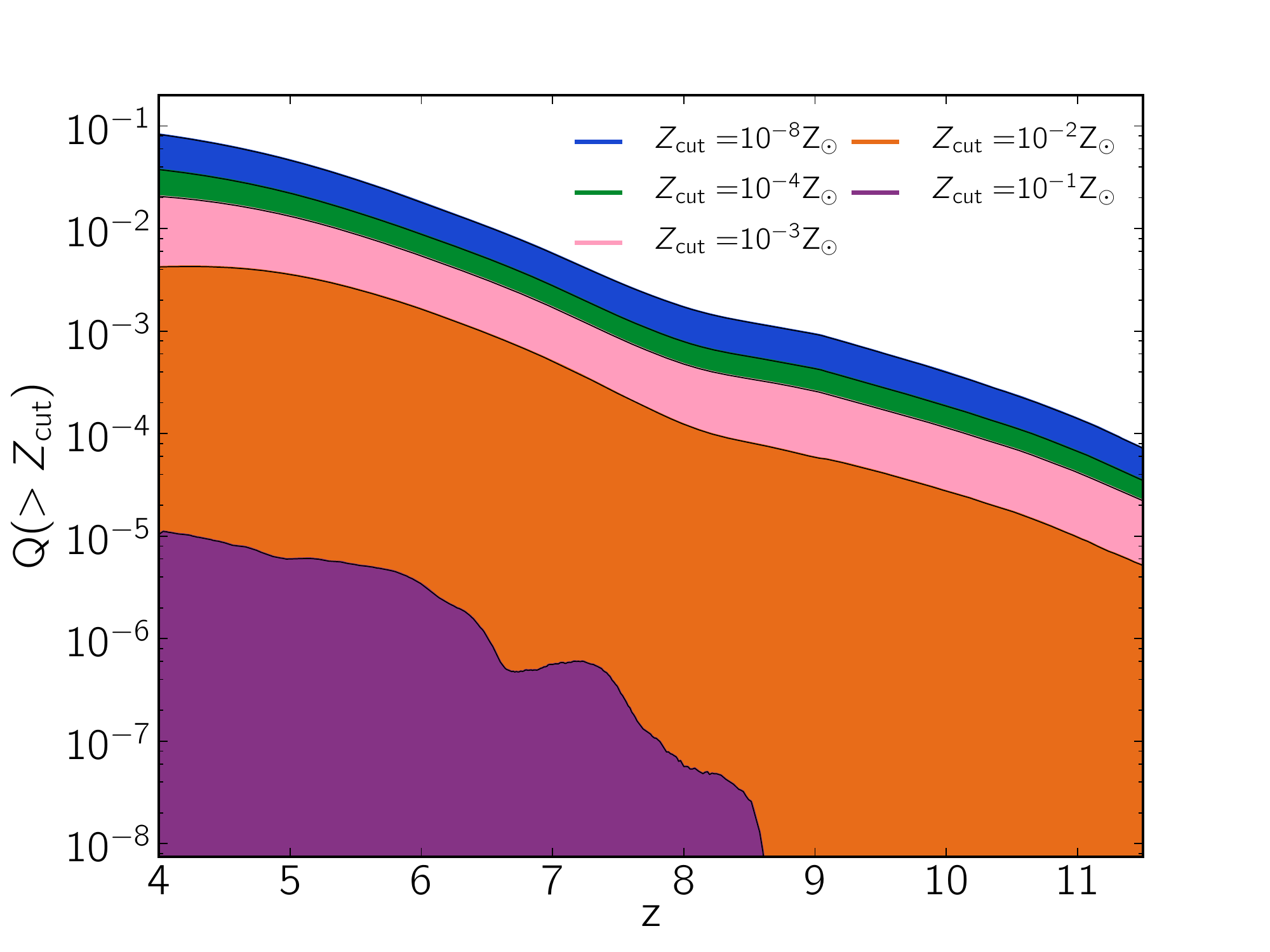}
\caption{Redshift evolution of the metal volume filling factor, $Q(>\zcut)$, for different values of the metallicity cut, $\zcut$. See eq. \ref{eq_def_q} for the definition.
\label{filling}
} 
\end{figure}

As time evolves, metals are not only produced at an increasing rate but they are transported by winds away from the production sites. To see this it is useful to derive the fraction of the cosmic volume, $V$, filled with heavy elements at a metallicity larger than a given value, $\zcut$. Formally, this can be written as 
\begin{equation}\label{eq_def_q}
Q(>\zcut)=\frac{1}{V} \int \Theta(Z-\zcut)\mbox{d}V.
\end{equation}
The behavior of $Q$ for different values of $\zcut$ has been traced on-the-fly in the simulation. The result is shown in Fig. \ref{filling}, which highlights interesting features of the enrichment process.

As we have already seen (Fig. \ref{fig_otf_anal}), the typical metallicity of star forming regions at $z\simeq4$ is $\simeq 10^{-1} \zsun$ and corresponds to the lowermost curve in Fig. \ref{filling}. Even at the lowest redshifts, star forming regions fill a very small fraction ($\simeq 10^{-5}$) of the volume; however, these sites represent the metal production factories, out of which metals are ejected and distributed by outflows in the CGM and IGM. As a matter of fact, the non-monotonic behavior at $z\lsim 6.5$ of $Q(>10^{-1}\zsun)$ is related to the increasing ability of galaxies to retain their metals as they become on average more massive (see Fig. \ref{fig_mass_metal_sf_redshifts}).

For lower values of $\zcut$, $Q(>\zcut)$ increases rapidly: already the region encompassing a mean metallicity $Z>10^{-2}\, \zsun$ fills a volume $\gsim10^3$ times larger (corresponding to a physical scale $\gsim 10$ times larger than the star forming regions). This region corresponds roughly to the CGM, i.e. the transition region surrounding the galaxy that is strongly influenced by the energy and mass input from the latter.

Finally, the $10^{-8}\leq\zcut\slash\zsun\leq 10^{-3}$ range corresponds to the IGM (see Fig. \ref{fig_z_stato_04}), where the metal abundance is largely diluted by intergalactic hydrogen. Note that the $Q(>\zcut)$ curves for $\zcut$ in the IGM range show little variation, indicating that the IGM metallicity is relatively constant at a value $10^{-3.5} \zsun$. However, only a fraction $< 10$\% of the cosmic volume has been ever polluted by heavy elements.

Overall, the filling factor evolution is consistent with the one found by \citet{Johnson:2013MNRAS} for $6\leq z\lsim 10$. Additionally, we found an evolution for $10^{-3}\leq\zcut\slash\zsun\leq10^{-2}$ similar to previous works \citep[i.e.][]{Cen:2005, Oppenheimer:2006MNRAS, Oppenheimer:2009MNRAS}. However, we find $Q(\geq10^{-1}\zsun)$ is roughly $\sim10$ times smaller than the corresponding one quoted in \citet{Oppenheimer:2009MNRAS}. This is somehow expected, since high metallicity regions correlate with density peaks (see Fig. \ref{fig_z_stato_04}); therefore such discrepancy likely arises from variations in the feedback prescriptions \citep[e.g.][]{aquila:2012MNRAS} and intrinsic differences between AMR and SPH \citep[e.g.][]{AGORA:2013arXiv,Power:2013arXiv}. The filling factor of metals has important implications for the relative evolution of Pop III stars and the local transition to a Pop II star formation mode that we will discuss in Sec. \ref{sec_test}. 

\subsection{The circumgalactic medium}\label{sec_spatialanal}
We devote this Section to some additional points concerning the CGM. We restrict the analysis to $z=6$ because of the importance of this epoch for reionization \citep[e.g.][]{Pentericci:2011ApJ,Schroeder:2013MNRAS} and because it represents the current limiting redshift for QSO absorption line statistical studies \citep[e.g.][]{DOdorico:2013MNRAS}. The CGM plays a key role in galaxy evolution as it represents the interface between galaxies and the IGM; moreover, ought to its relatively large overdensity and metallicity, it is more readily observed and may serve as a laboratory to study supernova feedback. Supernova winds carve large, hot, metal enriched bubbles in the CGM surrounding their host galaxies, as it is visually represented in Fig. \ref{mappe}.

It is then natural to ask whether relations exist between the properties of these bubbles and their parent galaxy. To answer this question, we first identify a galaxy and the associated metal bubble with the method described in Appendix \ref{sec_fof_gal}, based on a metallicity threshold criterion, i.e. the gas inside a bubble must have $Z > Z_{\rm th} = 10^{-7}\zsun$. Choosing a different ${Z}_{\rm th}$ would somewhat change the inferred bubble properties; however, the correlations we are going to discuss are insensitive to normalization constants.
\begin{figure*}
\centering
\includegraphics[width=8.3cm]{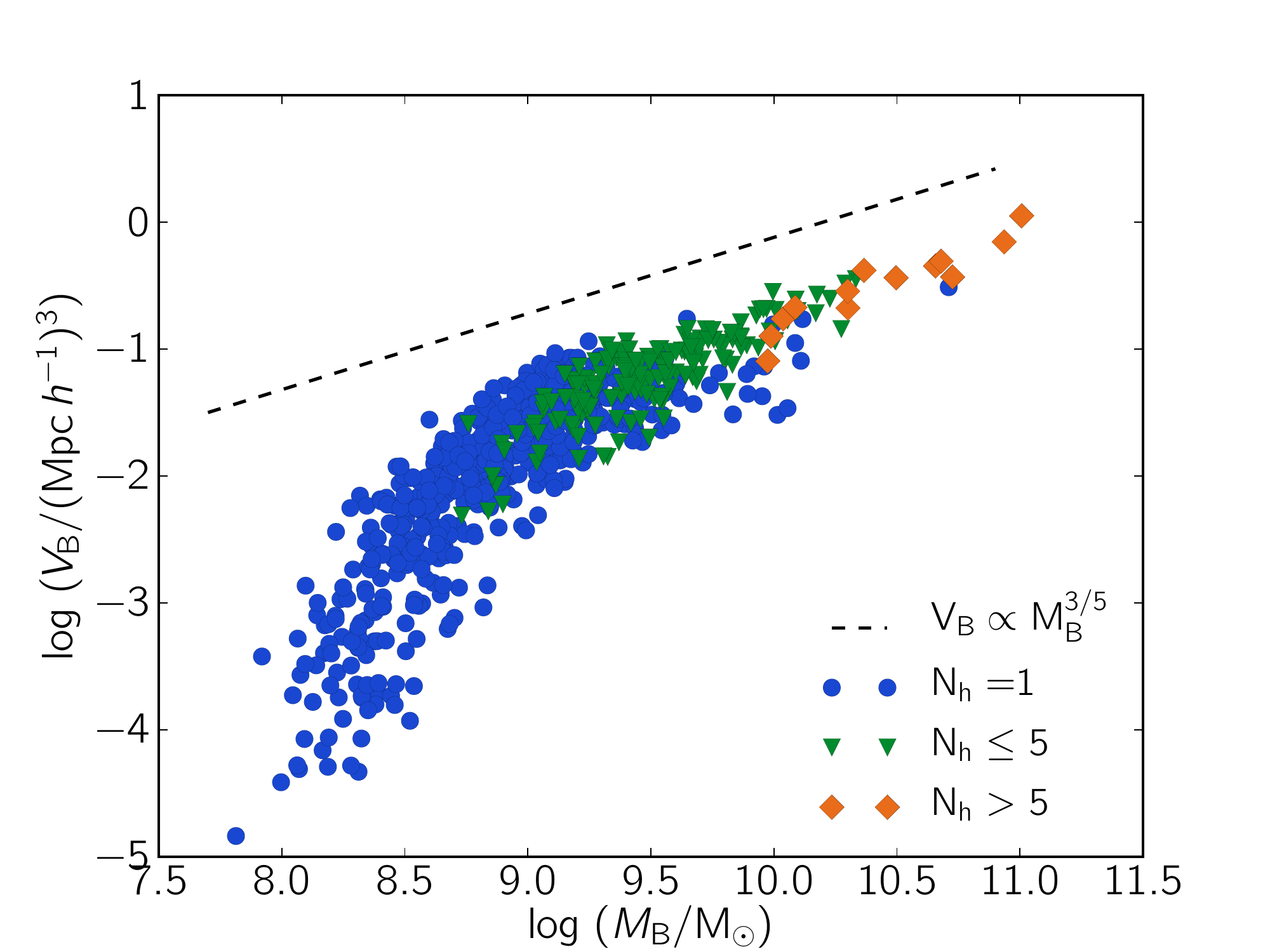}
\includegraphics[width=8.3cm]{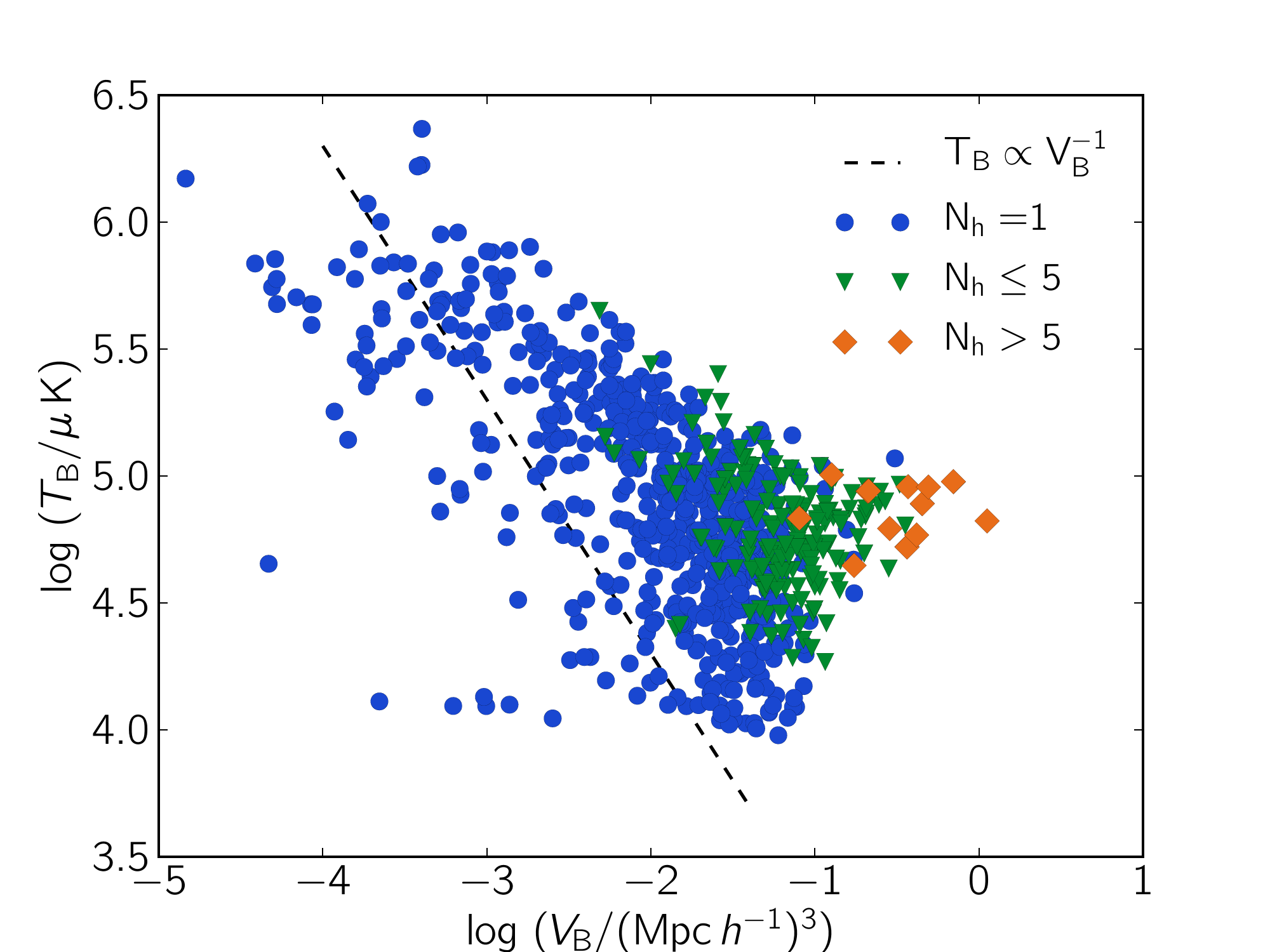}\\
\includegraphics[width=8.3cm]{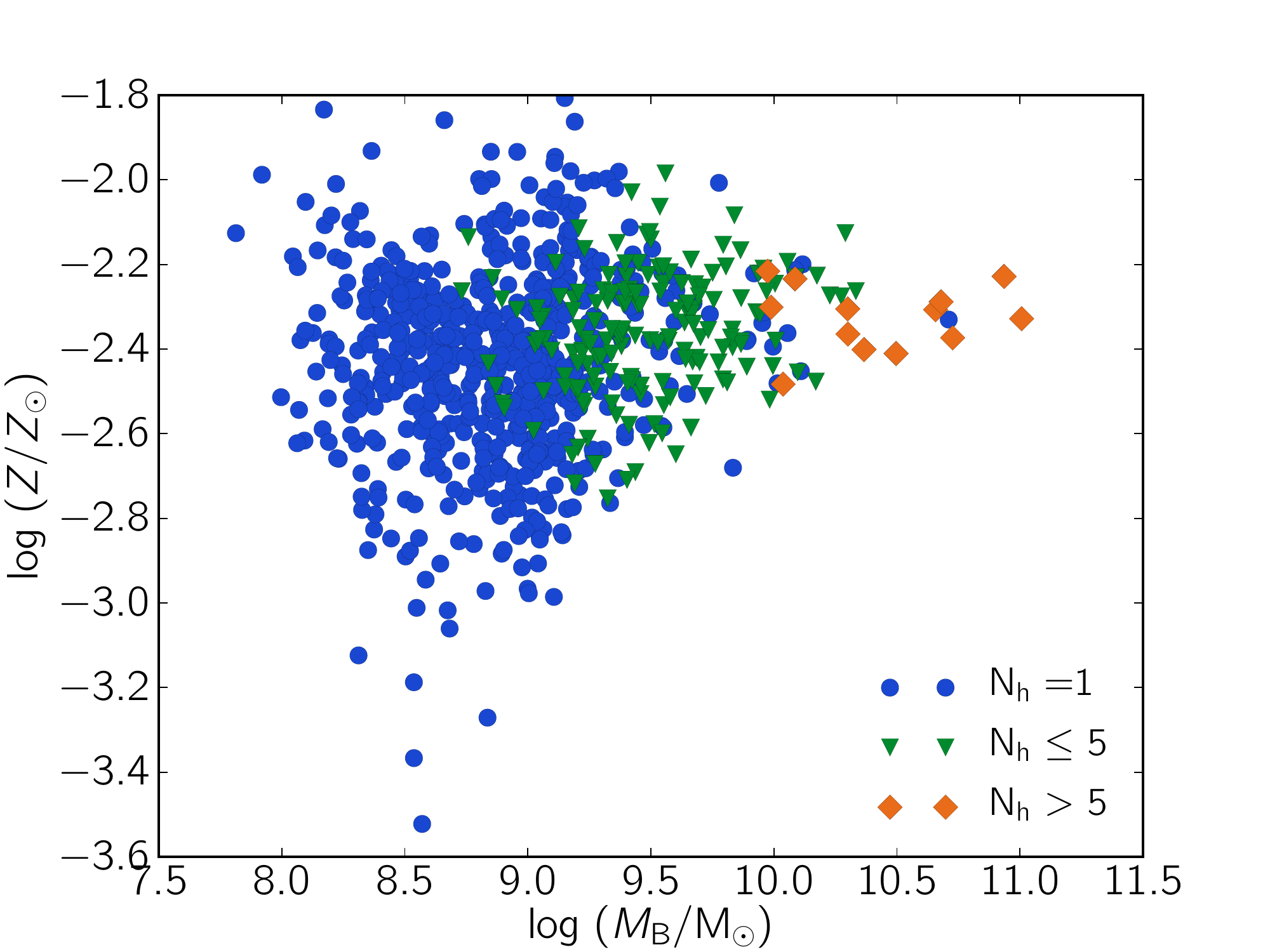}
\includegraphics[width=8.3cm]{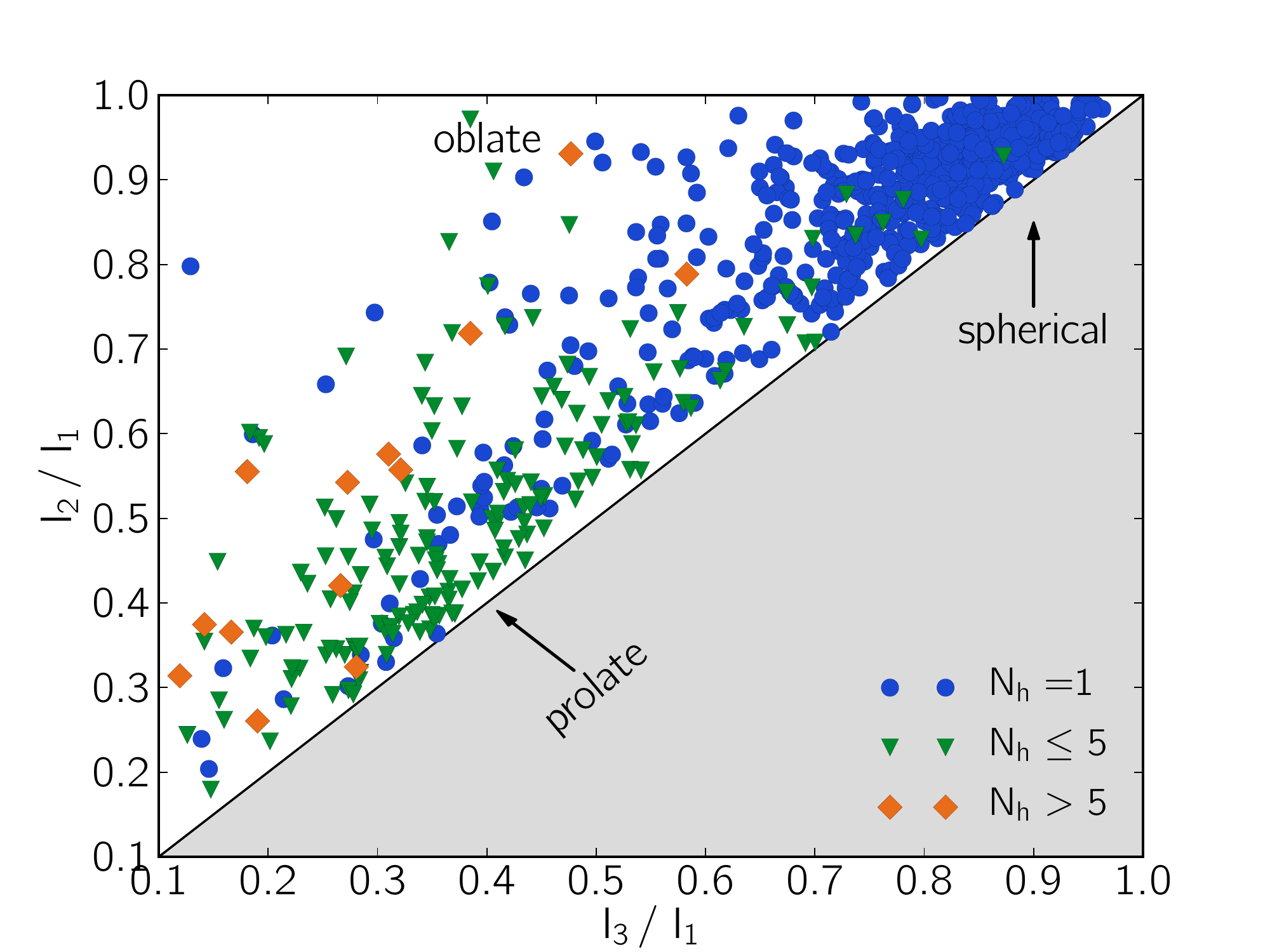}
\caption{Physical and geometrical properties of metal bubbles at $z=6$. \textit{Upper-left panel:} bubble volume vs. enclosed gas mass; \textit{Upper-right:} mean bubble temperature vs. volume; \textit{Lower-left:} mean bubble metallicity vs. enclosed gas mass; \textit{Lower-right:} bubble shape parameters. The number of halos ($N_{\rm h}$) inside the bubble is also indicated. See Appendix \ref{sec_fof_gal} for the details on the definitions. Dashed black lines are the analytical relations inferred from the Sedov-Taylor blast solution (see text).\label{fig_bolle_characteristic}}
\end{figure*}
As a first step we classify metal bubbles on the basis of the number of galaxies present in their interiors: in Fig. \ref{fig_bolle_characteristic} we use different symbols to differentiate bubbles according to the number of halos they contain, i.e. $N_{\rm h}=1$ (blue circle), $1<N_{\rm h}\leq5$ (green triangle), and $N_{\rm h}>5$ (orange squares). Broadly speaking, one can think of the $N_{\rm h}$ parameter as an indicator of merging activity experienced by the bubble. The detailed definitions of the physical quantities characterizing the bubbles are given in Appendix \ref{sec_fof_gal}.

As we will see in the following, to a first order metal bubble properties can be very well described by the Sedov-Taylor adiabatic blast solution. It is then useful to recall its functional form:
\be
V_B(t) = \zeta \left(\frac{E}{\rho_{e}}\right)^{3/5} t^{6/5}
\label{eq_ST}
\ee
where $V_B$ is the bubble volume, $E =\sum_i^{N_{\rm h}} E_i$ is the total SN energy produced by the $N_{\rm h}$ halos inside the bubble, $\rho_{e}$ is the mean density of the environment, $t$ is the bubble age and $\zeta$ is a dimensionless factor of order unity. In the simplest case, $E_i$ is proportional to the stellar mass formed and hence, for a fixed star formation efficiency (see eq. \ref{eq_mimic_kennicutt}), to the baryonic mass of the i-th star forming region, $M_{{\rm SF},i}$. It follows that $E\propto \sum_i^{N_{\rm h}} M_{{\rm SF},i} \propto M_B$, having further assumed a universal radial density profile within each bubble. In addition, as the post-shock gas temperature is $T_B \propto v_B^2$, where $v_B$ is the expansion velocity of the bubble, we also find, using eq. \ref{eq_ST}
\be
T_B \propto\left(\frac{E}{\rho_{e}}\right) V_B^{-1}.
\ee
We can directly compare the predictions from these simple formulae with the simulation results. The upper-left panel of Fig. \ref{fig_bolle_characteristic} shows that the volume of bubbles generally increases with the mass of the galaxies they contain; however, the largest ones result for the coherent action of $\ge 2$ galaxies powering them. The $V_B-M_B$ relations follows nicely the $M_B^{3/5}$ analytical form, but smaller bubbles which are still in the initial phases of their evolution deviate from this simple law and have smaller volumes with respect to the mass of the galaxies they contain. The Sedov-Taylor relation accounts well also for the gas temperature within bubbles (upper-right panel) providing a good fit to the slope of singly-powered bubbles, while those with $N_{\rm h}\ge 2$ tend to have hotter bubbles as a result of the larger energy input and replenishment rate of hot gas. Small ($V_{B}\lsim 10^{-2}({\rm Mpc}\,h^{-1})^{3}$) bubbles are typically hotter and reach $T_{B}\mu^{-1}\gsim 10^{5}$~K. Viceversa, in bubbles with $V_{B}\sim 10^{-1}({\rm Mpc}\,h^{-1})^{3} $ the temperatures can be as low as $T_{B}\mu^{-1}\sim 10^{4}$~K, i.e. they contain metals that had the time to cool and are currently purely advected with the expansion.

We find no clear correlation between $Z$ and $M_{B}$ (lower-left panel). This degeneracy mainly depends on the temporal evolution and the geometry. The elapsed time from the last SN explosion characterizes the metallicity spread, as injected metals are diluted into the surrounding medium \citep{Madau:2001ApJ}. Additionally, the relative outflow-galaxy geometry affects the pollution. As shown by \citet{Recchi:2013A&A} using 2D chemo-dynamical simulations, the surrounding gas distribution can change the enriched gas metallicity by a factor $\sim10$. This is particularly relevant at high masses ($M_{B}\gsim10^{9}\msun$) where the bubbles tend to merge, thus shape effects are dominant. The absence of correlation could have been also expected from the smooth distribution of the CGM/IGM ($\Delta\lsim10^{2.5}$) in the $Z-\Delta$ plane (left panel of Fig. \ref{fig_z_stato_04}).

The geometry of the bubbles is influenced by the topology of the cosmic web. The shape of the bubbles can be described in terms of the eigenvalues $I_i$ of the inertia tensor, where $I_{1}\geq I_{2} \geq I_{3}$. The ratios of the principal axis are used as index of sphericity ($I_{3}\slash I_{1}$), prolateness ($I_{3}\slash I_{2}$) and oblateness ($I_{2}\slash I_{1}$). The geometry of the bubbles can be analyzed in the sphericity-oblateness plane, i.e. $I_{3}\slash I_{1}$ vs. $I_{2}\slash I_{1}$, shown in the lower-right panel and obtained directly from the actual shape of bubbles in the simulation output. Almost $40\%$ of the bubbles are in the spherically symmetric region ($I_{3}\slash I_{1}\simeq1$); most of them ($\sim90\%$) correspond to bubble around isolated galaxies. Another $\sim30\%$ of the bubbles conserves at least a cylindrical symmetry and populates the $I_{3} \slash I_{1}\simeq I_{2} \slash I_{1}$ diagonal (prolate) and the $I_{2} \slash I_{1}\simeq 1$ stripe (oblate). These region contains $\sim70\%$ bubbles which have experienced few ($1<N_{\rm h}\leq5$) merging events. Because of the elongated shapes, the merging must have occurred along filaments of the cosmic web. The rest ($\sim35\%$) have lost any kind of symmetry, and almost all ($\sim90\%$) bubbles with large $N_{\rm h}$ values are located in this region. They correspond to metal polluted bubbles stretching along filaments and linking various knots of the cosmic density field.

\section{Effects of Pop III IMF variations}\label{sec_test}
To understand the interplay between cosmic metal enrichment and the Pop III - Pop II transition leading to a progressive disappearance of Pop III stars, we have performed an additional set of simulations. The suite is composed of three runs, solely differing in terms of the values adopted for the Pop III stellar parameters ($R$, $Y$, $\epsilon_{\sn}$), as summarized in Tab. \ref{tabella_yield_popiii}. The single runs are named after the Pop III type selected, i.e. SALP, FHN and PISN.

These simulations, similarly to the fiducial case examined here so far, evolve a 10~Mpc~$h^{-1}$ box; however, to limit the computational cost they are made with $256^{3}$ DM particles. Consequently, the mass (spatial) resolution is lower by a factor $8$ ($2$). The reduced resolution must be compensated with a more efficient star formation, and the best fit parameters of the subgrid prescription are now $t_{\star}=1.5$~Gyr and $\eta_{\sn}=0.3$. Again, the free parameters are fixed by matching the SALP run to the SFR observations \citep{Bouwens:2012ApJ} at low ($z\leq7$) redshift. Note that these parameters are then kept fixed for all the Pop III star choices.

\begin{figure}
\centering
\includegraphics[width=8.3cm]{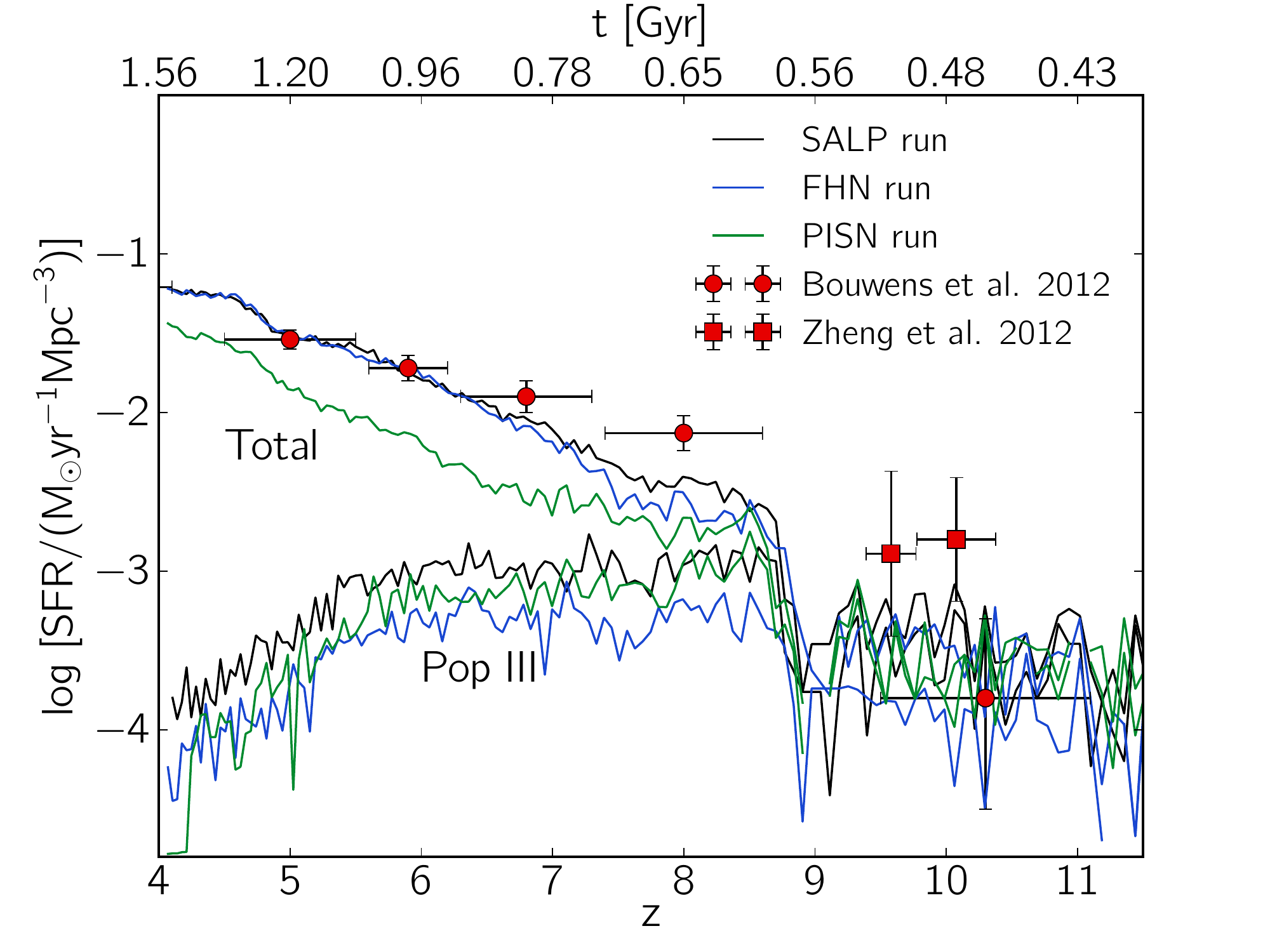}
\caption{Simulation suite result of cosmic SFRs as a function of redshift (age of the Universe). Different colors correspond to different runs and distinguish the total and Pop III only SFR. See Tab. \ref{tabella_yield_popiii} for the reference values of the adopted assumptions for the Pop III.\label{fig_confronto_stellar_model}}
\end{figure}

Fig. \ref{fig_confronto_stellar_model} shows the resulting SFR for the three runs. By comparing with the $512^{3}$ fiducial run presented in Fig. \ref{fig_sfr_smd}, we notice that the lower resolution affects the SFR at $z \simgt 8$. However for the comparison purposes of different Pop III properties this should not represent a major problem, and we also refrain from draw conclusions from $z\simgt 8-9$. The lower resolution causes the total SFR to be dominated by Pop III stars up to $z\sim 9$, as feedback is artificially more effective.

Note that there is weak dependence of the redshift at which galaxies become (on average) able to sustain a steady star formation process (see Sec. \ref{sec_gal_enrichment}) on resolution. For the fiducial simulation this occurs at $z\simeq8$, while in the rest of the suite it happens at $z\simeq 7.5$. Such epoch roughly coincides with the start of Pop III formation quenching. The two effects are obviously linked as follows from the discussion in the present Section.

\begin{figure*}
\centering
\includegraphics[width=5.6cm]{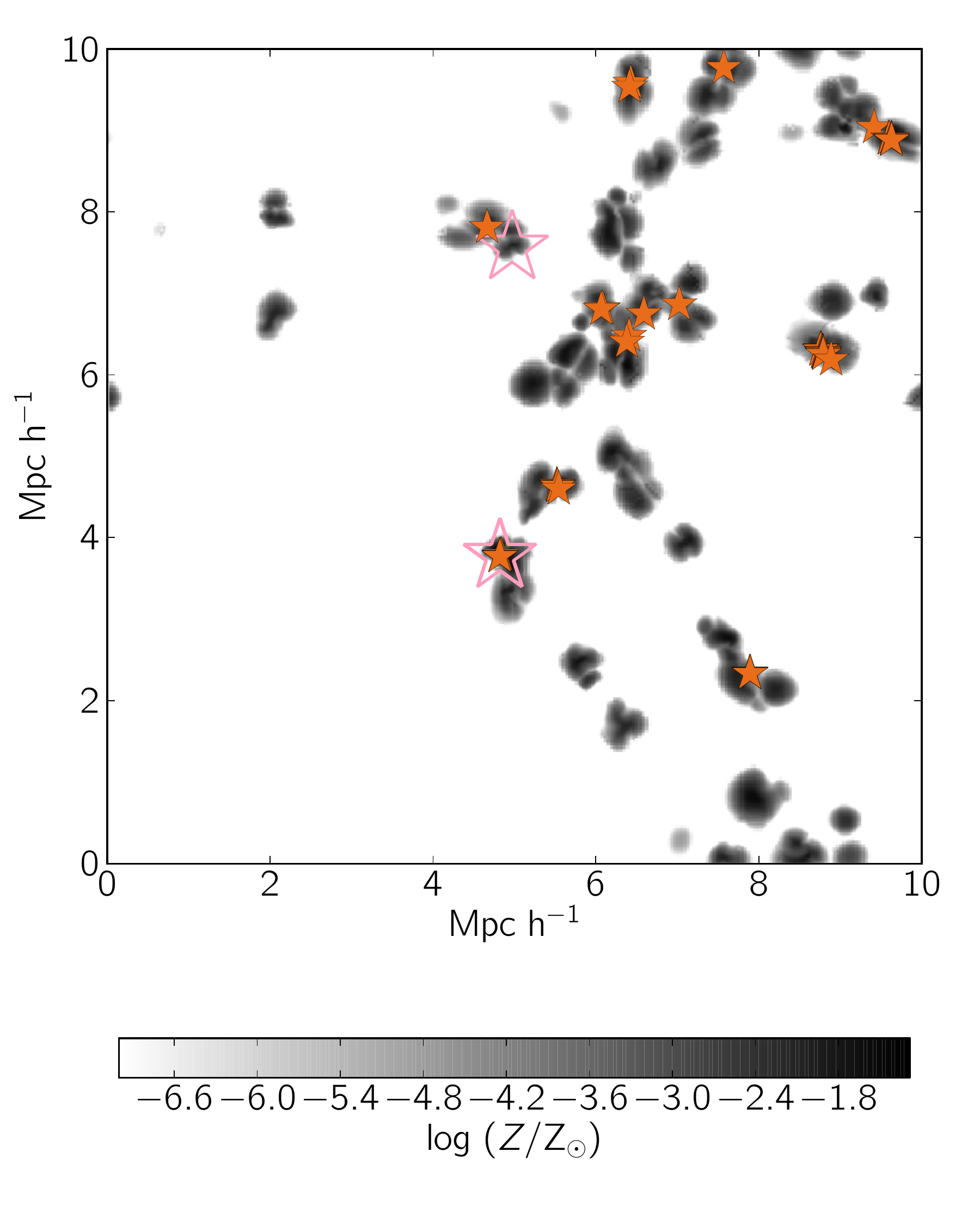}
\includegraphics[width=5.6cm]{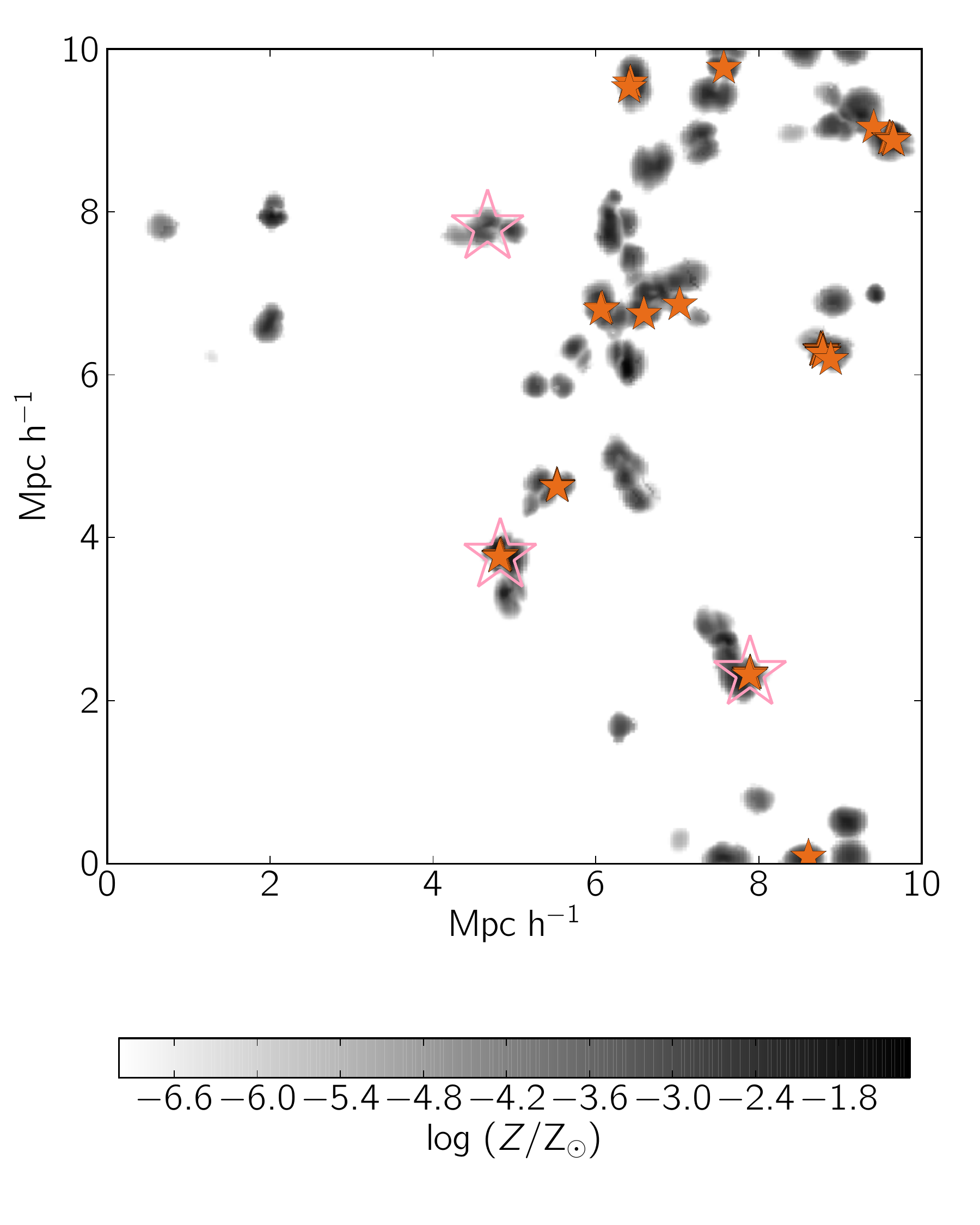}
\includegraphics[width=5.6cm]{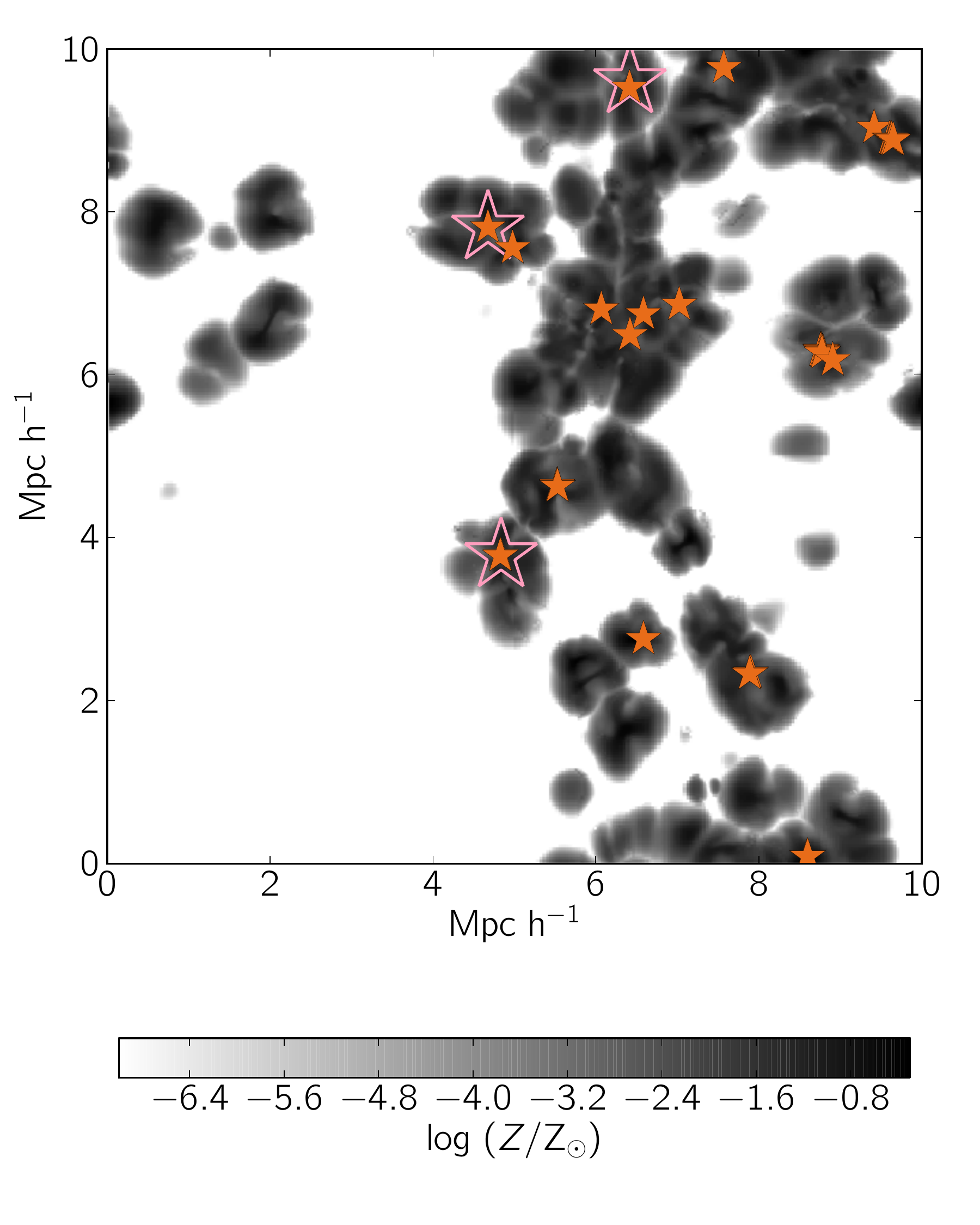}
\caption{Metallicity maps at $z=4$ for the SALP (left), FHN (middle), PISN (right) runs. Filled orange small (empty pink large) stars indicate the positions of Pop II (Pop III). The slice thickness is $39.06\,h^{-1}$~kpc.\label{fig_cfr_mappe_stelle_metal}}
\end{figure*}

Let us start by considering the evolution of Pop III SFR. Rather surprisingly, different prescriptions make essentially no differences on Pop III star formation history, apart from a slightly faster drop of PISN SFR below $z=6$. This more rapid fading can be understood as a result of the $\simeq 10 \times$ higher PISN energy input and metal yield resulting in a more widespread pollution of the volume above $\zcrit$. This can be visually appreciated from Fig. \ref{fig_cfr_mappe_stelle_metal}, featuring metallicity maps at $z=4$ for the different runs. 

Independently of $z$, the filling factor of regions enriched above the critical metallicity, $Q(>\zcrit)$, in the PISN run is $\simeq 3$ times larger than for the SALP$\slash$FHN runs; in addition, the maximum metallicity is about 10 times higher. Based on this evidence one would expect a much more pronounced suppression of Pop III stars in the PISN case, which however is not observed, apart from the above mentioned small relative drop at $z<6$. The results instead point towards a different interpretation. The key point is that enriching to $Z>\zcrit$ levels low density regions in the periphery of galaxies or the CGM does not produce a major effect in the SFR history of Pop III stars because the bulk of active Pop III formation sites is localized in newly forming halos far from the ones already hosting star formation. This is consistent with the previously discussed outcome of the fiducial run (see Fig. \ref{fig_metalpdf_star} and corresponding text) that Pop III stars are preferentially formed in a metal-free (rather than with $Z\ne 0$ but below $\zcrit$) environment; using semi-analytic models, \citet{Crosby:2013} showed that for $z\lsim10$ Pop III formation takes place in pristine regions separated by sufficiently large distances.

To better substantiate the last statement we point out that the Pop III quenching strength is related to the ratio between the typical size of metal bubbles, $\langle R_B \rangle=\langle V_{B}^{1/3} \rangle$ (see Sec. \ref{metenr} and Appendix \ref{sec_fof_gal}), and $r_{\rm SF}$, the correlation length of star forming regions\footnote{The correlation length is defined as the scale at which the two point correlation function of star forming regions $\xi_{\rm SF}(r_{\rm SF})=1$ \citep[e.g.][]{Reed:2009MNRAS}.}. In our simulations, we find that $r_{\rm SF}\simeq 2\, {\rm Mpc}\,h^{-1}$ at both $z=6$ and $z=4$. At the same time, the mean size of the bubbles goes from $\langle R_B \rangle\simeq 0.3 \, {\rm Mpc}\,h^{-1}$ at $z=6$ to $\simeq 0.5 \, {\rm Mpc}\,h^{-1}$ at $z=4$, thus quenching PopIII star formation. In fact, we can see from Fig. \ref{fig_sfr_smd} that the fiducial model predicts ${\rm SFR}_{\popiii}(z=6)\simeq 10^{1.3}~{\rm SFR}_{\popiii}(z=4)$.

In conclusion, the similar evolutionary trend of a flat Pop III SFR, persisting at a level of about $10^{-3}\msun$~yr$^{-1}{\rm Mpc}^{-3}$, up to $z=6$ and followed by a rapid drop thereafter, appears to be a solid feature of our model and to be independent of the details of Pop III IMF.

This physical interpretation implies that chemical feedback might be artificially enhanced in simulations when the mean pollution radius $\langle R_B \rangle$ becomes comparable to the simulation box size. We further discuss how resolution and box-size effects affect this issue in Appendix \ref{sec_mass_res}. Note that additional uncertainties might come from the treatment of radiative feedback. A proper investigation of resolution and box size effects would require a suite of simulations with increasing box size and fixed resolution and a convergence study with fixed box-size and increasing mass resolution. We defer this study to a future work.

The effects of Pop III IMF variations have instead a larger impact on Pop II (and hence total) SFR. From Fig. \ref{fig_confronto_stellar_model} we see that, as for the Pop III case, the differences between a SALP or FHN IMF are minor: using Fig. \ref{fig_cfr_mappe_stelle_metal} they can be quantified by $\langle|\log Z_{\rm FHN}/\zsun -\log Z_{\rm SALP}/\zsun|/ \log Z_{\rm SALP}/\zsun \rangle\simeq0.01$, where the average is calculated on the slice. This difference is produced by the yields of Pop III SALP bs. FHN (Tab. \ref{tabella_yield_popiii}) and from the stochasticity of star formation prescription (eq. \ref{eqs_local_sto_sfr}).

On the other hand, differences are noticeable if a PISN IMF is assumed. In this case, the total SFR is depressed by a factor $\simeq 5$ for $z\gsim 5$. Because of the higher $\epsilon_{\sn}$ value, a pair-instability supernova can reach and disrupt a nearby potential star formation site, as also noted by \citet[][]{Greif:2010ApJ}. Instead, less energetic hyper- or supernovae can only pollute their immediate surroundings, failing to reach other more distant and denser environments. Under the assumption of similar local star formation time scales ($t_{\star}$) for both Pop II and Pop III, the PISN scenario is difficult to be reconciled with the observed global SFR history, as feedback from these stars is probably too effective.

\section{Synthetic spectra}\label{sec_spettri}

We compute mock QSO absorption spectra along several l.o.s. drawn through the simulated box. The details of the adopted technique to compute the $\HI$ optical depth are given in \citet{Gallerani:2006MNRAS}. Here, we also consider metal absorption lines due to ionic species such as $\SIV$, $\CIV$, $\SII$, $\CII$ and $\OI$. For these species, we compute the Doppler parameter according to the following expression $b_i=\large \sqrt{2k_{B}T\slash m_i}$, where $m_{i}$ is the mass of the $i$-th species. In Tab. \ref{lambda_ostrenght}, we report the wavelengths and oscillator strengths \citep[][]{Prochaska:2004ApJ} adopted for the considered species.

In order to compute the number density of different ionic species, we have built a grid of model calculations using {\tt CLOUDY} \citep[][]{cloudy:1998PASP} version 10. We consider a plane parallel slab of gas in pressure equilibrium, illuminated by a Haardt-Madau ionizing background \citep[][]{Haardt:1996, Haardt:2012}. The intensity of the ionizing field at 1 Ryd has been normalized so that the photoionization rate, $\Gamma_{\rm HI}$ (units: ${\rm s^{-1}}$), matches the values predicted by the two reionization models presented in \citet{Gallerani:2008MNRASa,Gallerani:2008MNRASb}, namely an Early Reionization Model (ERM) and a Late Reionization Model (LRM), predicting a reionization redshift of $z_{\rm rei} \simeq7$ (ERM) and $z_{\rm rei}\simeq6$ (LRM), respectively.

\begin{figure*}
\centering
\includegraphics[width=6.5cm]{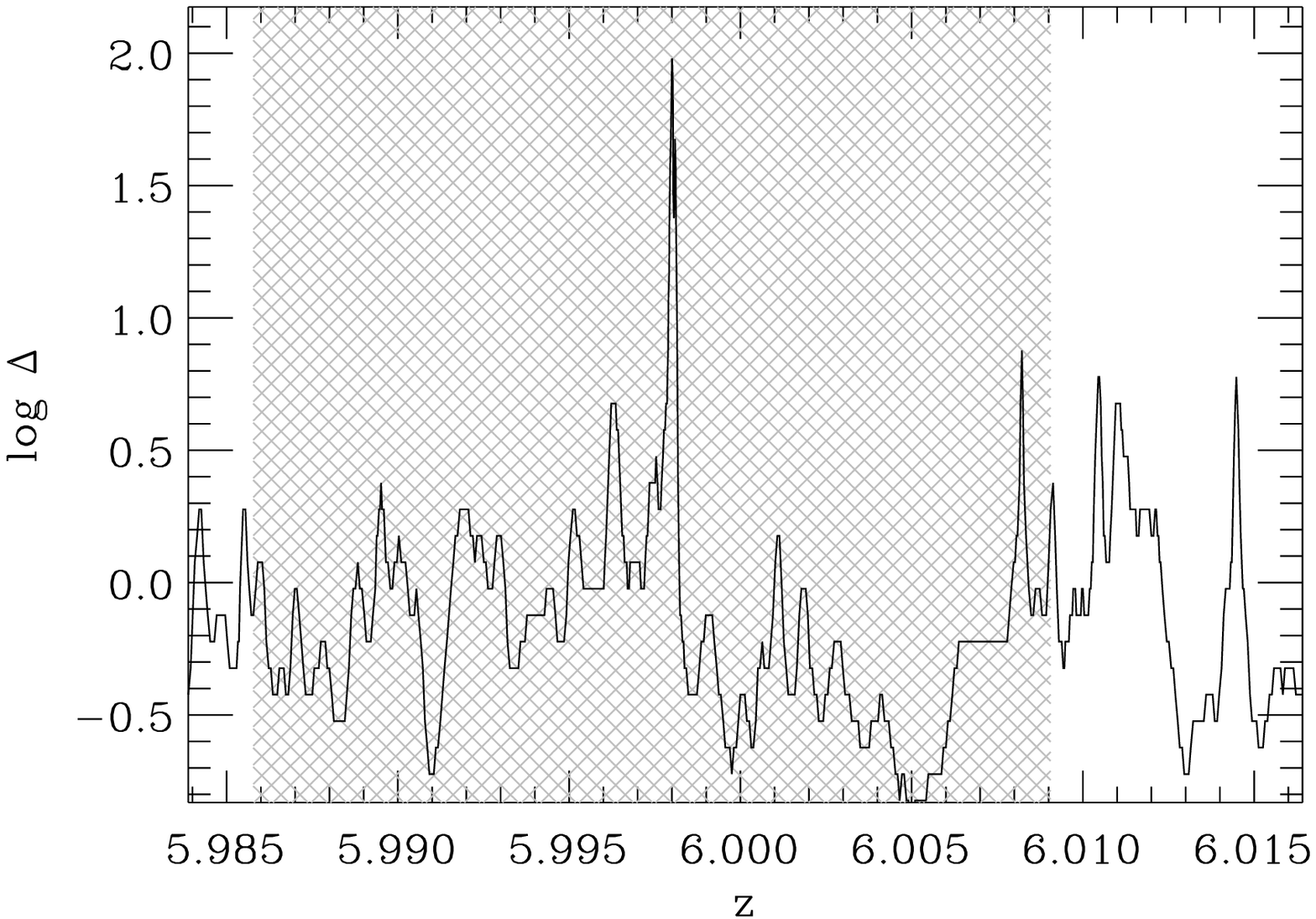}
\includegraphics[width=6.5cm]{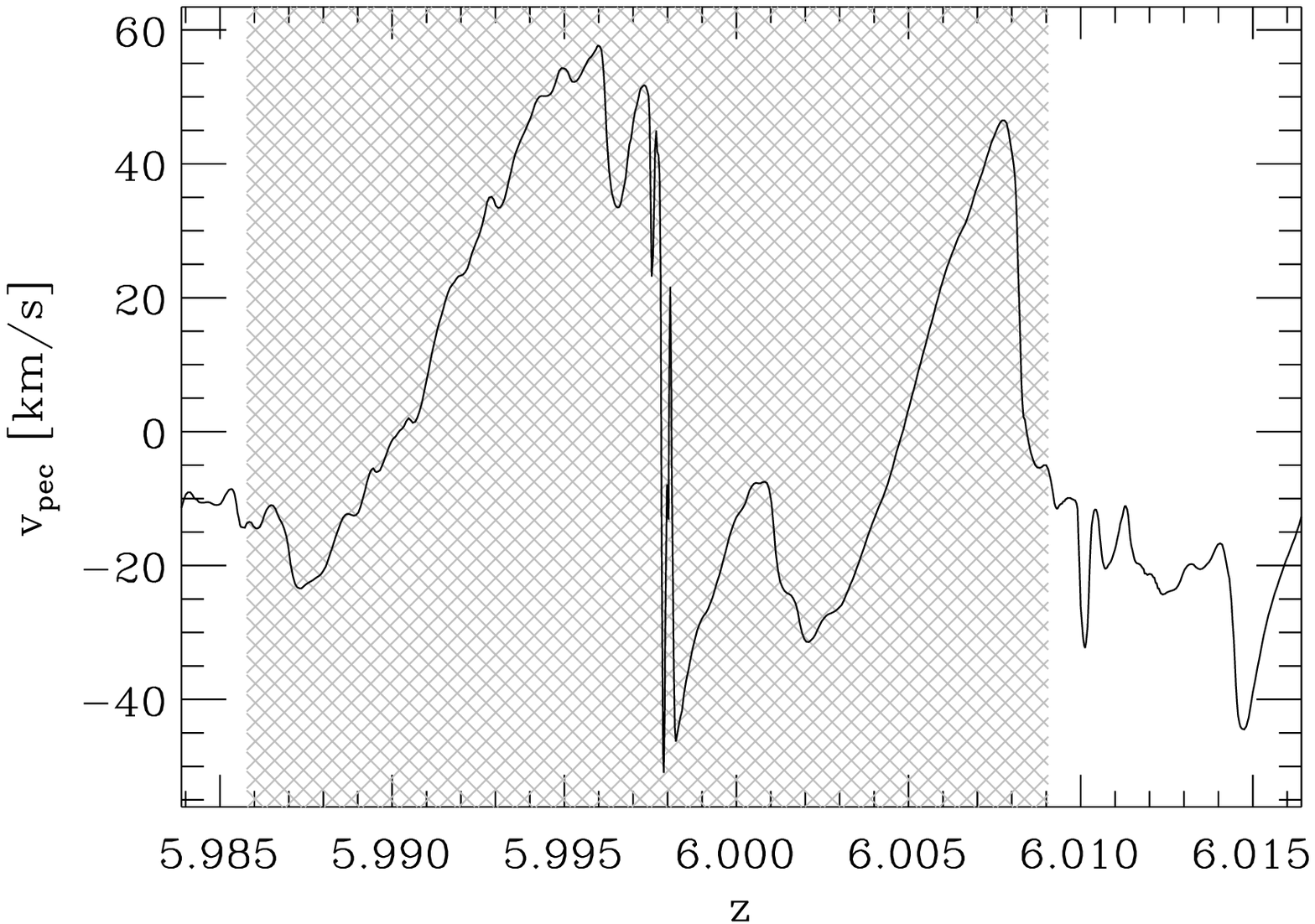}\\
\includegraphics[width=6.5cm]{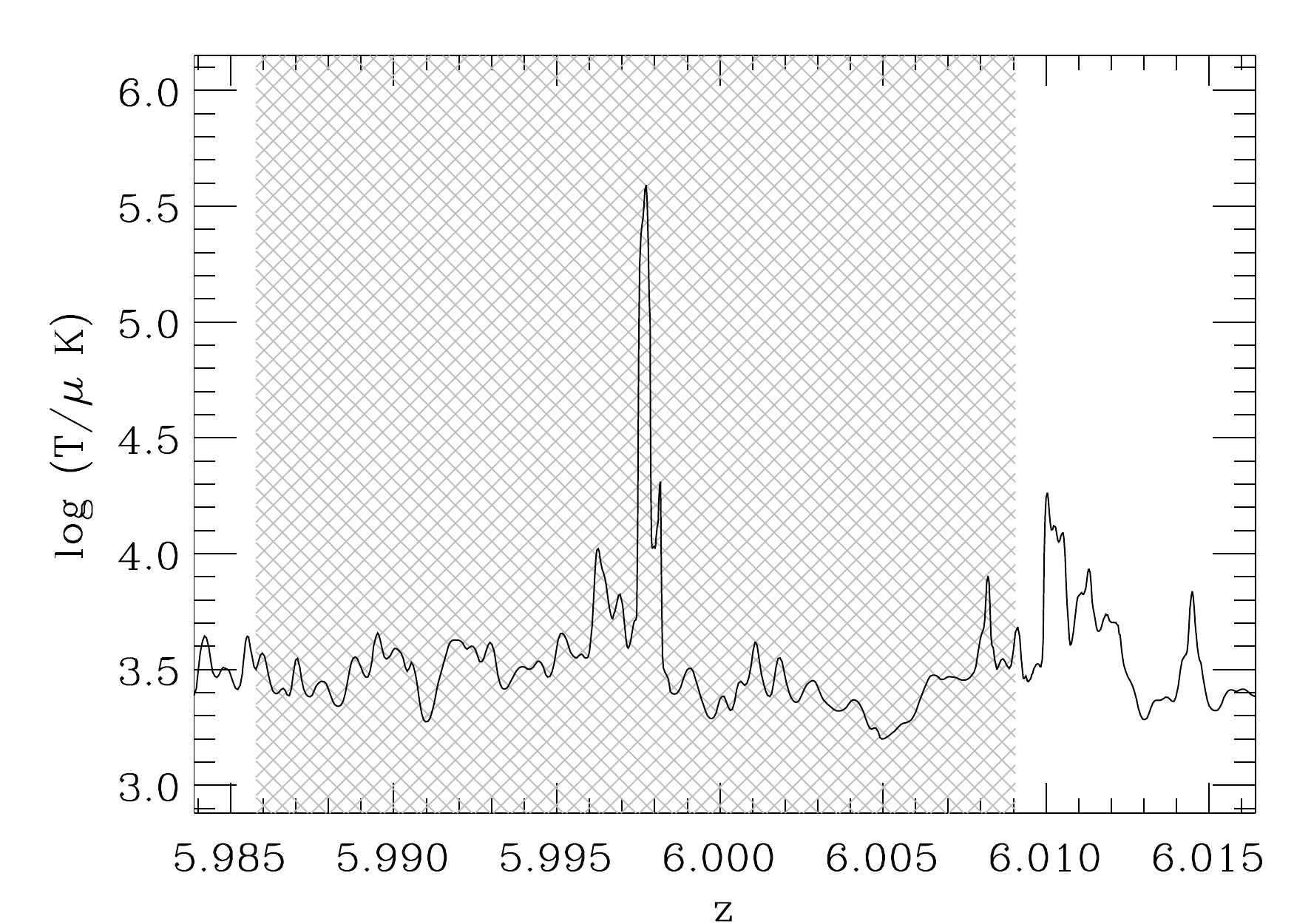}
\includegraphics[width=6.5cm]{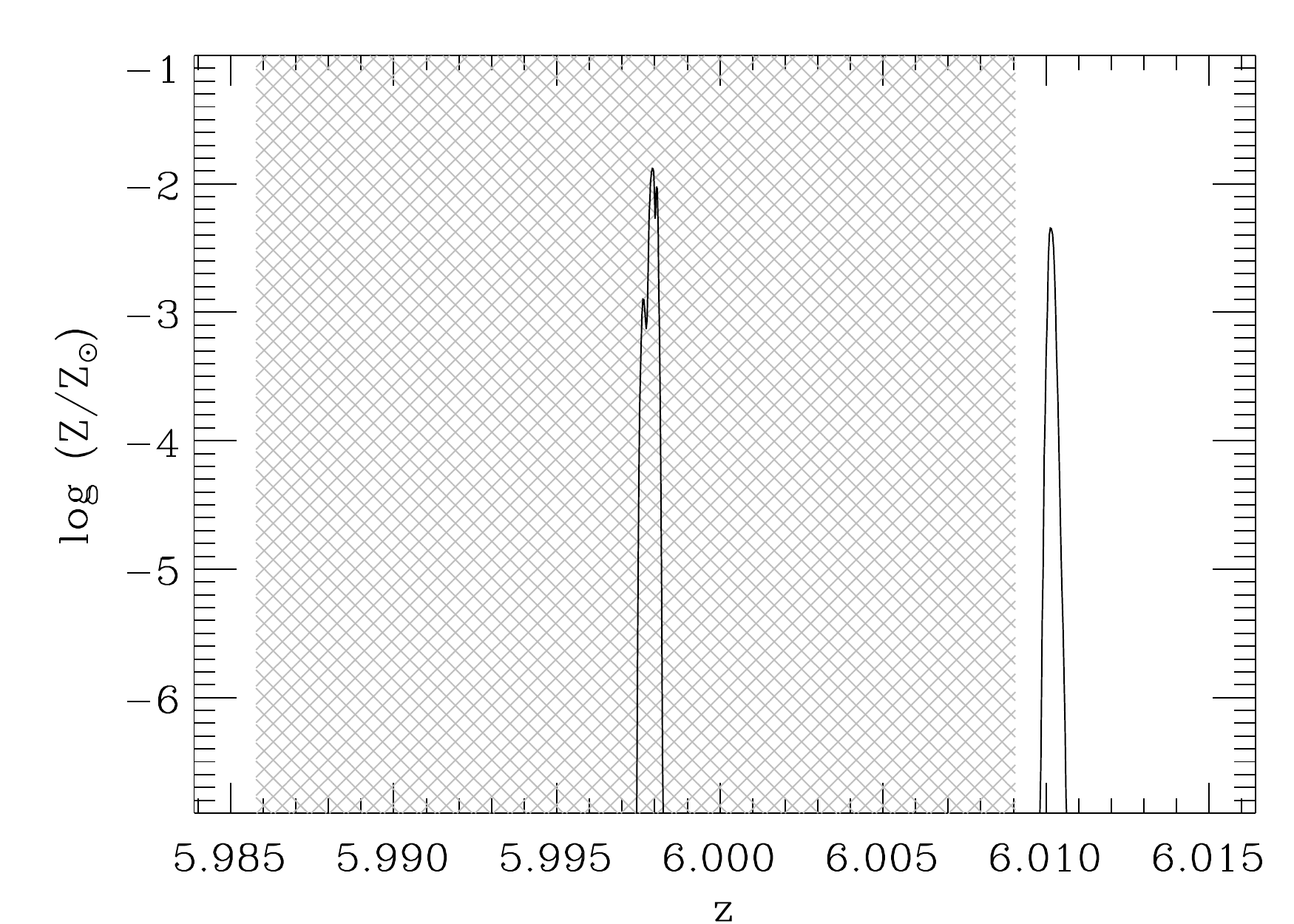}
\caption{Distribution of the gas density (upper-left panel), peculiar velocity (upper-right panel), temperature (lower-left panel), metallicity (lower-right panel) along a random l.o.s. drawn through the simulated box at $z=6$.\label{fig_proplos}}
\end{figure*}
\begin{figure*}
\centering
\includegraphics[width=15.3cm]{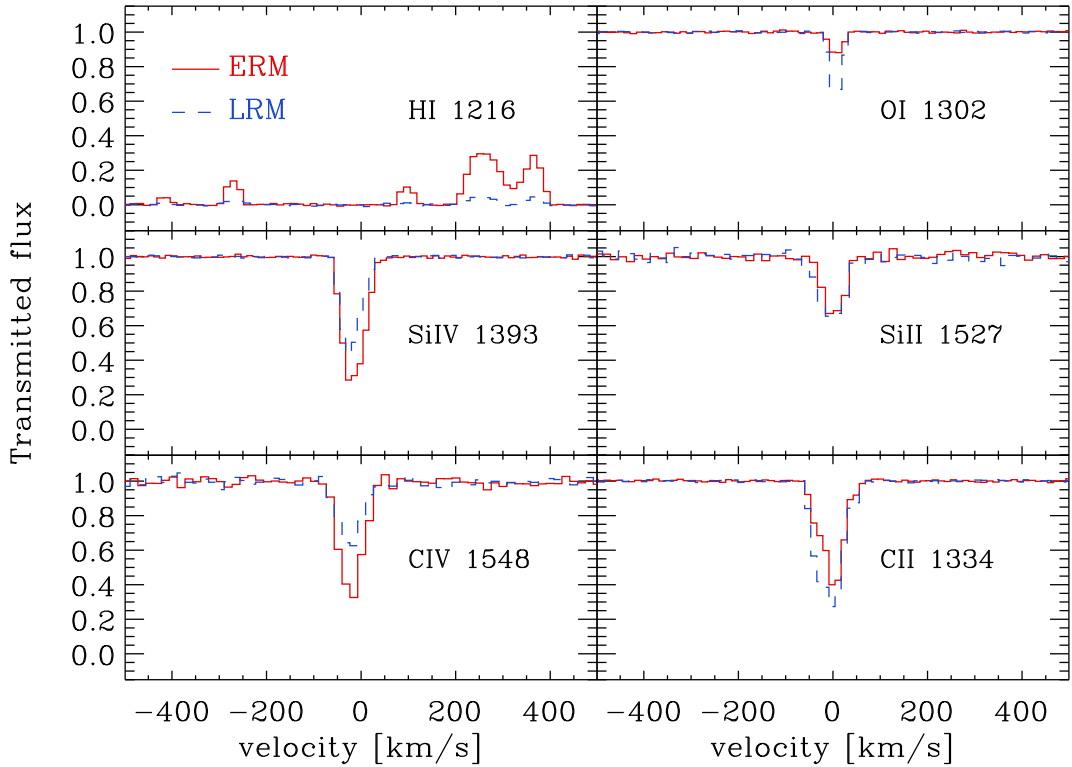}
\caption{Synthetic spectra extracted from the l.o.s. drawn through the simulated box with physical properties shown in Fig. \ref{fig_proplos}. The spectra are calculated using an Early Reionization Model (ERM, red solid line) and a Late Reionization Model (LRM, blue dashed line) for Ly$\alpha$ forest (upper-left panel) and metal absorption lines ($\OI$, upper-right panel; $\SIV$ and $\SII$, middle-left and middle-right panel; $\CIV$ and $\CII$, lower-left and lower-right panel, respectively). See the text for the definitions.
\label{spectra}}
\end{figure*}

\begin{figure*}
\centering
\includegraphics[width=18cm]{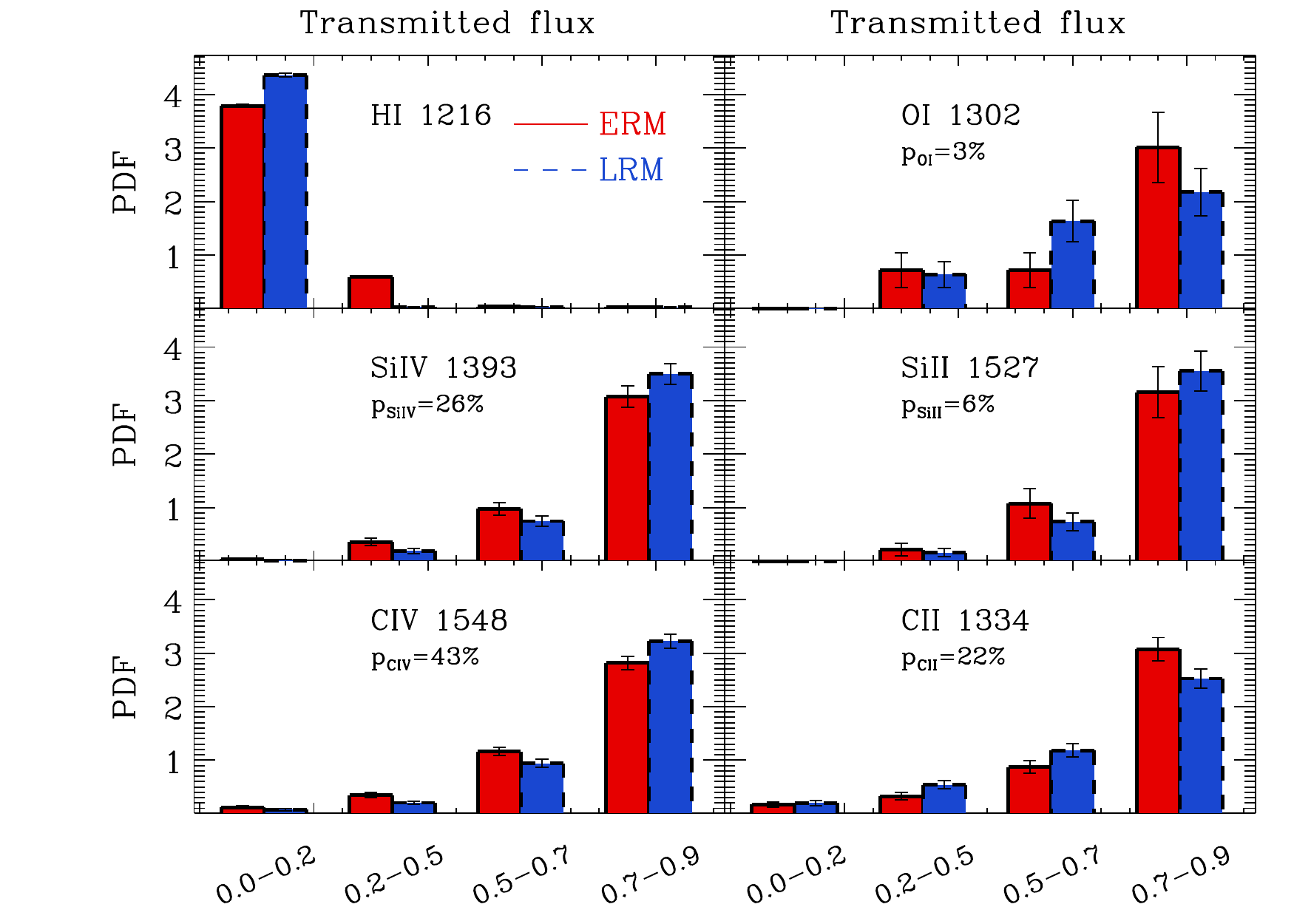}
\caption{PDFs of the transmitted flux for the species considered in Fig. \ref{spectra}. The solid line (red shaded region) represents our predicted PDF in the case of the ERM, while the dashed line (blue shaded region) denotes LRM results. For each bin, we report the corresponding poissonian error bars. For each ion, we indicate the probability of encountering its corresponding absorption line in the l.o.s. sample, as defined in the text. \label{pdf_trans}}
\end{figure*}

We point out that the ionizing background values predicted both by the ERM ($\log\Gamma_{\rm HI}=-12.46$) and LRM ($\log\Gamma_{\rm HI}=-12.80$) are consistent within 1-$\sigma$ with the measurement of the UVB at $z=6$ through the flux decrement technique \citep[$\log\Gamma_{\rm HI}=-12.74\pm 0.30$,][]{Wyithe:2011MNRAS}; moreover, the ERM (LRM) is consistent within 2.1-$\sigma$ (0.3$\,$-$\sigma$) with the results obtained through the proximity effect technique \citep[$\log\Gamma_{\rm HI}=-12.84\pm 0.18$,][]{Calverley:2011MNRAS}. Each {\tt CLOUDY} model is characterized by a gas metallicity, density and a constant temperature. As for the chemical composition, we adopt the solar elemental abundance ratios provided in {\tt CLOUDY}. Calculations are stopped when the depth of the slab reaches our cell resolution.

We interpolate {\tt CLOUDY} results with the gas properties predicted by the simulation. In Fig. \ref{fig_proplos}, we show the distribution along a random l.o.s. of several quantities, namely: gas density (upper-left panel), peculiar velocity (upper-right panel), gas temperature (lower-left panel), and metallicity (lower-right panel).

Although a detailed, quantitative comparison with observations is beyond the aim of the present paper, we include observational artifacts in our simulated spectra, following \citet[][]{DOdorico:2013MNRAS}, a work based on X-shooter spectra. For species with absorption features in the wavelength range $\lambda_i<1440 $~A ($\lambda_i>1440 $~A), since the absorption systems of interest at $z=6$ are redshifted in the VIS (NIR) region of the X-Shooter spectrum, we smooth the synthetic spectra to a resolution $R=8800$ ($R=5600$), we add to each pixel a Gaussian random deviate, yielding a signal-to-noise ratio $S/N=50$ ($S/N=10$), and we finally rebin the simulated transmitted flux in channels of width 0.4~A (0.6~A).

An example of the simulated spectra obtained through this procedure is shown in Fig. \ref{spectra}, where the velocity interval corresponds to the redshift range marked through gray hatched regions in Fig. \ref{fig_proplos}. Metal absorption lines show up preferentially in correspondence of metal-rich ($Z\gsim10^{-2}~\zsun$), IGM/CGM regions ($\Delta \gsim1-10$), characterized by mean temperatures $T\mu^{-1}\simeq10^{4.5}~{\rm K}$, as the ones occurring at $z\simeq 5.997$ in our sample l.o.s.. As expected, absorption lines due to atoms at high ionization levels (e.g. $\CIV$ and $\SIV$) are more pronounced with respect to low-ionization absorption lines (e.g. $\SII$ and $\OI$) in the case of the ERM, which predicts a higher ionizing background at the redshift of the absorption system. The above relations are reversed in the LRM case.

To get further insight, for each species and reionization model we compute $N_{\rm los}^{\rm tot}=300$ synthetic spectra as the ones shown in Fig. \ref{spectra} and we analyze them in terms of the PDF of the transmitted flux ($F_{\rm tr}$). The result is shown in Fig. \ref{pdf_trans}. We divide the transmitted flux in four bins, and for each of them we show the PDF values predicted for the ERM (solid line, red shaded region), LRM (dashed line, blue shaded region), with the corresponding poissonian error bars. For metal absorption lines, we restrict our analysis to those spectral regions characterized by $F_{\rm tr}<0.9$. For each species considered, we also report the probability to find at least one absorption line per l.o.s., $p_i=N_{\rm los}^{\rm abs,i}/N_{\rm los}^{\rm tot}$, where $N_{\rm los}^{\rm abs,i}$ is the number of l.o.s. in which we find at least one absorber of the $i$-th species.

Fig. \ref{pdf_trans} confirms that, in the case of the Ly$\alpha$ forest, the PDF analysis does not provide a fair diagnostics of the IGM ionization level at $z\sim 6$ \citep{Fan:2006ARA&A,Gallerani:2006MNRAS,Gallerani:2008MNRASa}. In fact, from the upper-left panel, it is clear that in both reionization models, most of the pixels are characterized by $F_{\rm tr}\simeq 0.1$, meaning that this statistics is basically dominated by noise.

The same figure shows that, instead, the PDF obtained from metal absorption lines contains a wealth of information, and may provide strong constraints on the ionizing background intensity. It is also interesting to note that the presence of strong $\OI$ absorbers ($F_{\rm tr}\simeq 0.3$) does not exclude the possibility that the IGM/CGM is ionized, since these absorption systems are also found in the ERM, which predicts a small neutral hydrogen fraction ($x_{\rm HI}\simeq 10^{-4}$) at $z\simeq 6$. Observations of $\OI$ absorption systems at these redshifts are useful tools for understanding the metal enrichment and cosmic reionization processes \citep{Oh:2002MNRAS,Finlator:2013MNRAS,Keating:2013arXiv}.

Finally, as a consistency check, we compute the $\Omega_{\rm CIV}$ predicted by our simulation at $z=6$. The cosmic density parameter relative to the $i$-th species can be written as \citep[e.g.][]{DOdorico:2013MNRAS}
\begin{subequations}\label{eq_calc_omega}
\be
\Omega_{i}(z)=\frac{H_{0}m_{i}}{c \rho_{c}} \int N_{i} f_{i} {\rm d}N_{i}\, ,
\ee
where $\rho_{c}$ is the critical density and $f_{i}(N_{i},z)$ is the PDF of the $i$-th column density $N_{i}$. For a discrete set of absorbers indexed by $j$, the integral can be approximated as \citep[i.e.][]{Storrie-Lombardi:1996MNRAS}
\be
\Omega_{i}(z)\simeq \frac{H_{0}m_{i}}{c \rho_{c}}  \frac{\sum_{j} N_{i}^{j}}{ \Delta X}\, ,
\ee
where $\Delta X$ is the cosmological path length of the l.o.s. that can be computed using the following relation:
\be
{\rm d} X= (1+z)^{2} \left(\Omega_{\Lambda}+\Omega_{m}(1+z)^{3}\right)^{-1\slash2} {\rm d}z\, .
\ee
Additionally, the statistical error, $\delta\Omega_{i}$, is given by
\be
\frac{\delta\Omega_{i}}{\Omega_{i}}=\frac{\sqrt{\sum_{j} (N_{i}^{j})^{2}}}{\sum_{j} N_{i}^{j}}\,.
\ee
\end{subequations}

Using eq.s \ref{eq_calc_omega}, $N_{\rm los}^{\rm tot}=300$, and taking into account the $\CIV$ column density provided by {\tt CLOUDY}, we find $\Omega_{\rm CIV}^{\rm ERM} =(3.8\pm 0.1)\times 10^{-8}$ and $\Omega_{\rm CIV}^{\rm LRM} =(3.1\pm 0.1)\times 10^{-8}$. We also compute $\Omega_{\rm CIV}$ only considering those systems which produce observable absorption lines ($F_{\rm tr} < 0.9$), and which are characterized by $10^{13.4}<N_{\rm CIV}/\rm cm^{-2} <10^{15}$. In this case, we find $\Omega_{\rm CIV}^{\rm ERM} =(2.9\pm 0.3)\times 10^{-8}$ and $\Omega_{\rm CIV}^{\rm LRM} =(2.3\pm 0.3)\times 10^{-8}$. Considering the total produced metals ($\Omega^{\rm SFH}_{Z}$ in Sec. \ref{sec_gal_enrichment}) and assuming solar abundances, it results $\Omega_{\rm CIV}\slash\Omega^{\rm SFH}_{\rm C}\simeq1.6\times10^{-2}$. In other words, we predict that through $\CIV$ absorption line experiments it is possible to probe $\simeq$ 2\% of the total carbon present in the IGM/CGM. 

We note that the $\Omega_{\rm CIV}$ values resulting from our calculations are a factor $\simeq 3-4$ greater then the ones found by \citet{DOdorico:2013MNRAS}. This discrepancy is not surprising, since we are not properly comparing simulations and observations. While our calculations take into account the actual $\CIV$ distribution in the simulated box, the observed $\Omega_{\rm CIV}$ is inferred from absorption lines spectra through a Voigt profile fitting procedure. Moreover, we restate that our simulations do not account for radiative transfer effects. This is a crucial point, since the ionizing flux which determines the ionization level of atomic species can be dominated by the presence of local sources, rather than the background. In particular, this is relevant for the CGM, which is closer to galaxies and is responsible of strong absorption features.

Constraining the ionization level of the IGM at $z\sim 6$ through metal absorption lines requires both the proper inclusion of radiative transfer effects and an extended statistical analysis (e.g. equivalent width and column density distribution, etc) of the synthetic spectra. We defer to a future study a comprehensive comparison between observations and absorption spectra extracted from our simulations.

\begin{table}
\centering
\resizebox{\linewidth}{!}{
    \begin{tabular}{|c|c|c|c|c|c}
            \hline
	    &$\SIV$ & $\CIV$& $\SII$&$\CII$&$\OI$\\
            \hline\hline
            $\lambda_{i}$&1393.7550&1548.1950 &1526.7066&1334.5323& 1302.1685\\
            $f_{i}$ &0.5280& 0.1908&0.127&0.1278&0.0488\\
           \hline
   \end{tabular}
}
\caption{Rest frame wavelengths $\lambda_{i}$ and oscillator strengths $f_{i}$ \citep[][]{Prochaska:2004ApJ} for the transition considered in the simulated QSO spectra with metal absorption lines.\label{lambda_ostrenght} 
}
\end{table}

\section{Conclusions}\label{sec_conclusioni}
We have studied cosmic metal enrichment via a suite of $\Lambda$CDM hydrodynamical simulations using a customized version of the adaptive mesh refinement code {\tt RAMSES} to evolve a $(10$~Mpc~$h^{-1})^{3}$ volume up to $z=4$ with $512^{3}$ dark matter (DM) particles, a corresponding number of coarse grid cells and allowing for 4 additional levels of refinement. The subgrid prescription for star formation is based on a local density threshold criterion ($\Delta>\Delta_{\rm th}$) and on a critical metallicity criterion ($\zcrit = 10^{-4}\zsun$), allowing us to follow the transition from Pop III to Pop II stars. To assess the impact of variations in the unknown Pop III IMF we have investigated three different choices: (a) a standard Larson-Salpeter IMF (SALP), (b) a $\delta$-function describing faint hypernovae (FHN), and (c) a top-heavy IMF allowing for pair-instability supernovae (PISN). We account for thermal feedback from supernovae and implemented a metal-dependent parameterization of stellar yields and return fractions.

This set-up enables the resolution of DM halos masses of $10^{7.5}\msun$ with $\simeq 100$ particles and to build a statistically significant sample of galaxies at all redshifts of interest. The two free parameters of our subgrid model (star formation timescale and supernova coupling efficiency) have been fixed by reproducing the observed cosmic star formation rate \citep[SFR,][]{Bouwens:2012ApJ,Zheng:2012Natur} and stellar mass densities \citep[SMD,][]{Gonzalez:2011} at $4\leq z \lsim 10$.

By constructing halo catalogues and identifying the associated stars and star forming regions ($\Delta>\Delta_{\rm th}$), it has been possible to analyze the evolution of metal enrichment on galactic scales at two representative redshifts, $z=6$ and $z=4$. Galaxies account for $\lsim 9\%$ of the baryonic mass; the complementary fraction resides in the diffuse medium, which we have classified according to the environmental overdensity into: (a)~\textit{voids}, i.e. regions with extremely low density ($\Delta\leq 1$), (b) the true \textit{intergalactic medium} (IGM, $1<\Delta\leq 10$) and (c) the \textit{circumgalactic medium} (CGM, $10<\Delta\leq 10^{2.5}$), representing the interface between the IGM and galaxies. 

We have computed synthetic spectra of metal absorption lines through the simulated box at $z=6$. The number density of different ionic species are calculated in post-processing with {\tt CLOUDY} and by considering two physically motivated and observationally constrained reionization models, i.e. an Early Reionization Model (ERM, $\log(\Gamma_{\rm HI}/{\rm s^{-1}})=-12.46$ at $z=6$) and a Late Reionization Model (LRM, $\log(\Gamma_{\rm HI}/{\rm s^{-1}})=-12.80$).

We have tried to analyze separately the metal enrichment properties of galaxies and diffuse medium for sake of clarity, but obviously the intimate connection between these two components makes it impossible to separate their description completely. Readers mostly interested in galaxies/stars (diffuse gas) can directly refer to Sec. \ref{sec_gal_enrichment} (Sec. \ref{sec_global_result}); those specifically interested in Pop III stars should also find Sec. \ref{sec_test} relevant. The summary of the main results given below is organized in points attempting to keep these distinctions.
\begin{itemize}
\item[\bf 1.] Between $z=9$ and $z=6$ a galactic mass-metallicity relation is established. For star forming regions of mass $M_{\rm SF}\gsim 10^{7} \msun$, such relation shows little evolution from $z=6$ to $z=4$. In particular, at $z=4$, galaxies hosting a stellar mass $M_{\star}\simeq10^{8.5}\msun$ show a mean oxygen abundance of $\log({\rm O}\slash H)+12=8.19$, consistent with observations \citep[][]{Troncoso:2013arXiv1311}.

\item[\bf 2.] At $z = 4$ such relation extends to $M_{\rm SF}\lsim 10^{7} \msun$: these are satellite galaxies forming whose star formation has been enabled by the progressive enrichment of the diffuse gas out of which they form. For $10^{6}\lsim M_{\rm SF}\slash\msun\lsim 10^{7}$ the metallicity trend is flat and resembles the one observed in the faintest Local Group dwarf galaxies \citep[e.g.][]{Kirby:2013arXiv}.

\item[\bf 3.] The total amount of heavy elements produced by star formation rises from $\Omega^{\rm SFH}_{Z} = 1.52 \times 10^{-6}$ at $z=6$ to $8.05 \times 10^{-6}$ at $z=4$. Metals in galaxies make up to $\simeq 0.89$ of such budget at $z=6$; this fraction increases to $\simeq 0.95$ at $z=4$. At $z=6$ ($z=4$) the remaining metals are distributed in the three diffuse phases, CGM/IGM/voids, with the following mass fractions: $0.06/0.04/ 0.01$ ($0.03/0.02/ 0.01$).

\item[\bf 4.] In all the diffuse phases a considerable fraction of metals is in a warm/hot ($T\, \mu^{-1}>10^{4.5}K$) state. In particular, a small but not negligible mass fraction ($\simeq0.003$) of metals in voids shows $T\, \mu^{-1}\leq10^{4.5}K$. This implies that these metals must have been injected at sufficiently early epochs that they had the time to cool as expected in a pre-enrichment scenario.

\item[\bf 5.] Analogously to the mass-metallicity relation for star forming regions, at $z=4$ a density-metallicity ($\Delta-Z$) relation is in place for the diffuse phases. Independently of $\Delta$, the IGM/voids show an uniform distribution around $Z\sim10^{-3.5}\zsun$, while in the CGM $Z$ steeply rises with density up to~$\simeq10^{-2}\zsun$.

\item[\bf 6.] The geometry of metal bubbles is influenced by the topology of the cosmic web. At $z=6$, $\sim40\%$ are spherically symmetric and are mostly found around isolated galaxies; $30\%$ show instead a cylindrical shape which mainly results from merging of bubbles aligned along filaments. 

\item[\bf 7.] The cosmic Pop III star formation history is almost insensitive to the chosen Pop III IMF. Pop III stars are preferentially formed in pockets of pristine ($Z=0$) gas, well outside polluted regions created by nearby/previous star formation episodes. This supports the ``Pop III wave'' scenario suggested by \citet{Tornatore:2007MNRAS} and confirmed by \citet{Maio:2010MNRAS}.

\item[\bf 8.] In the PISN case, the Pop II SFR is suppressed by a factor of $\simeq5$ with respect to the SALP/FHN cases. Because of the higher energy deposition, a pair-instability SN can reach and disrupt a nearby potential star formation site, quenching Pop II formation. Assuming the same star formation timescales for Pop II and Pop III, a PISN scenario is difficult to be reconciled with the observed SFR history, as the feedback from these stars is probably too effective.

\item[\bf 9.] Metal absorption line spectra extracted from our simulations at $z\sim 6$ contain a greater wealth of information with respect to the Ly$\alpha$ forest. Given the prevailing thermodynamical/ionization conditions of the enriched gas, $\CIV$ absorption line experiments can only probe up to $\simeq$ 2\% of the total carbon present in the IGM/CGM. However, metal absorption lines are very effective tools to study reionization.

\item[\bf 10.] The occurrence of low-ionization metal systems (e.g. $\OI$ and $\CII$) in $z\sim 6$ quasar (gamma-ray burst) absorption spectra does not exclude the possibility that the IGM/CGM is on average highly ionized at these epochs. In fact, such systems, although with a lower incidence than in a Late Reionization Model, are also detectable in the Early Reionization Model, which predicts a lower $\HI$ fraction ($x_{\rm HI}\sim 10^{-4}$) at $z\simeq 6$.
\end{itemize}

In the future, we will perform a more extended statistical analysis of the synthetic spectra, in terms of the equivalent width and column density distributions. This study will enable a direct comparison with recent high-$z$ observations \citep{DOdorico:2013MNRAS} and will allow to constrain cosmic reionization models. Since the ionization level of metal atoms is sensitive to the proximity effect of ionizing sources, it will be crucial to take into account radiative transfer effects. Therefore, we plan to couple our simulation with the new version of the radiative transfer code {\tt CRASH3} \citep{Graziani:2013MNRAS}. This code allows to include an arbitrary number of point sources and to reprocess the ionizing radiation through an inhomogeneous distribution of metal-enriched gas, therefore representing a perfect tool for our planned research.

As a final remark, we have highlighted the potential problem that chemical feedback might be artificially enhanced in a simulation when the box size becomes smaller or comparable to the pollution radius $\langle R_B \rangle$. Although the box size and resolution have a significant impact on the determination of Pop III cosmic SFR, additional uncertainties come from the treatment of radiative feedback. A proper demonstration would involve a suite of simulations with increasing box size and fixed resolution and a convergence study with fixed box-size and increasing mass resolution. The situation may be worth a closer scrutiny, which we defer to future work.

\section*{Acknowledgments}
SG thanks INAF for support through an International Post-Doctoral Fellowship. SS acknowledges support from Netherlands Organization for Scientific Research, VENI grant 639.041.233. This work has been partially supported by the PRIN-INAF 2010 grant "The 1 Billion Year Old Universe: Probing Primordial Galaxies and the Intergalactic Medium at the Edge of Reionization``. {\tt RAMSES} is governed by the CeCILL license under French law and was written by R. Teyssier, to whom we are very grateful. We thank the anonymous referee for helpful comments.

\bibliographystyle{mn2e}
\bibliography{metal_enrichment}
\appendix

\section{Resolution effects}\label{sec_mass_res}

As stated in the Introduction, the nature of Pop III stars is still under debate, and there is a lack of consensus on their formation properties and subsequent evolution. Although this  paper is not specifically focused on Pop III stars, it is necessary to address numerical effects that could possibly affect Pop III evolution.

Using definitions given in Sec. \ref{sec_sfr_prescription}, each Pop III formation event on average spawns a star particle with a mass $M_{s}=m_{\star}\langle N\rangle$. Using eq. \ref{eq_mean_starpt}, by assuming $\langle N\rangle>0$ and that one stellar particle is sufficient to pollute its surrounding environment, we can write
\be\label{eq_massres}
M_{s}\propto (\Delta x)^{3} \Delta t/t_{\star}\simeq M_{\rm res}^{4/3}/t_{\star}\, ,
\ee
where $M_{\rm res}$ is the mass resolution of the AMR simulation, and we have implicitly assumed a Lagrangian mass threshold-based refinement criterion. In eq. \ref{eq_massres} $M_s$ is limited from below, since we expect PopIII to form in clouds of mass $\sim 10^{2}-10^{3} \msun$ \citep{Bromm:2002ApJ,Yoshida:2006ApJ,Greif:2012MNRAS,Hosokawa:2012ApJ,Meece:2013arXiv}; however, for the present estimate, we can neglect this point. Throughout the paper we have shown that each halo can host at most one Pop III formation event. Assuming no external pollution, an upper limit for the Pop III SFR can be approximated by
\be
{\rm SFR}_{\popiii}\lsim M_{s} n(M>M_{h}^{\rm min})/ t_{\star}\, ,
\ee
where $n(M>M_{\rm h}^{\rm min})$ is the number of halos with mass larger than $M_{\rm h}^{\rm min}$, the minimum mass that can host star formation. For low mass halos we can approximate $n(M>M_{\rm h}^{\rm min})\sim 1/M_{\rm h}^{\rm min}$ \citep{Press:1974,sheth:1999MNRAS}, thus
\be\label{eq_sfr_mass}
{\rm SFR}_{\popiii}\propto \frac{M_{\rm res}^{4/3}}{M_{h\rm }^{\rm min} t_{\star}^{2} }  \, .
\ee
Numerically, $M_{\rm h}^{\rm min}$ is the mass of halos resolved by a suitable minimum number of particles, usually taken to be $M_{\rm h}^{\rm min}\simeq 10^{2.5}M_{\rm res}$ \citep{Christensen:2010ApJ}. From the physical point of view, $M_{\rm h}^{\rm min}$ is determined by feedbacks, star formation criteria and the presence of a LW background \citep[e.g.][]{Wise:2012ApJ,Johnson:2013MNRAS,xu:2013arXiv}. \citet{Johnson:2013MNRAS} show that the LW background can induce differences in ${\rm SFR}_{\popiii}$ up to a factor $\simeq 8$ during the early stages of star formation ($z\gsim10$). Then the scaling provided by eq. \ref{eq_sfr_mass} reproduces well such result once the proper values of $M_{\rm h}^{\rm min}$ in the \citet{Johnson:2013MNRAS} simulation are inserted for the case with or without LW background. 

There is a caveat regarding eq. \ref{eq_sfr_mass}. In our formalism $t_{\star}$ depends on $M_{\rm res}$, and is calibrated by reproducing the observed SFR/SMD; thus, without a dedicated simulation suite, the $t_{\star}(M_{\rm res})$ functional dependence is uncertain. Moreover, differences in both models and implementations \citep{aquila:2012MNRAS,AGORA:2013arXiv} might hinder the effectiveness of the estimate when making a cross-code comparison.

For our fiducial simulation at $z=6$ the Pop III SFR is $10^{-2.7}\msun$~yr$^{-1}{\rm Mpc}^{-3}$. Using eq. \ref{eq_sfr_mass} and considering a slowly varying $t_{\star}$, our result is compatible with the values reported by \citet{Johnson:2013MNRAS} (SFR$_{\popiii}\simeq10^{-4}\msun$~yr$^{-1}{\rm Mpc}^{-3}$) and \citet{Wise:2012ApJ} (SFR$_{\popiii}\simeq10^{-4.5}\msun$~yr$^{-1}{\rm Mpc}^{-3}$) at the same redshift.

We remind that we have not taken into account external enrichment. The estimate thus holds up to $z\gsim5$, where in our simulation Pop III star formation is definitively quenched (see Sec. \ref{sec_halostar}). Such quenching redshift is almost independent of resolution (see Sec. \ref{sec_test}). 

In principle, the pollution efficiency depends on the ratio between the galaxy correlation length, $r_{\rm SF}$, and the metal bubble size $\langle R_B \rangle$ (see Sec. \ref{sec_test} for the definitions). A rough approximation for $r_{\rm SF}$ is given by the autocorrelation scale of DM halos of mass $\gsim M_{h}^{\rm min}$. Since $M_{h}^{\rm min}\geq10^{2.5}M_{\rm res}$, the quantity can be considered almost independent from resolution \citep{Reed:2009MNRAS,Guo:2014MNRAS}. On the other hand $\langle R_B \rangle$ can be estimated using the Sedov-Taylor approximation (eq. \ref{eq_ST}) with $\langle R_B \rangle\propto \langle (\eta_{\sn} M_{\star})^{1/5} \rangle$, where $M_{\star}$ is the stellar mass per halo of mass $M_{\rm h}$. The total stellar mass is calibrated with observations (see Fig. \ref{fig_sfr_smd}) and for $z\lsim6$, in each halo with $M_{\rm h}\gsim 10 M_{h}^{\rm min}$, the stellar mass is dominated by Pop II (see Fig. \ref{fig_binned_halom_starm_003}). Thus $\langle R_B \rangle$ is expected to be weakly resolution dependent as $\langle R_B \rangle\propto (\eta_{sn})^{1/5}$.

As a corollary, when in a simulation $\langle R_B \rangle$ becomes comparable with the box size, chemical feedback might be artificially enhanced. However, a robust test of this idea would involve a simulation suite with increasing box size and fixed resolution and a convergence study with fixed box-size and increasing mass resolution. While this is outside the purpose of the current work, it would be interesting to analyze the problem in the future.

As a final note, in the current paper we have assumed the same $t_{\star}$ for Pop III and Pop II. Considering two distinct star formation timescales would introduce an extra degree of freedom in the model. The natural way to fix the introduced Pop III time scale would be fitting cosmic SFR$_{\popiii}$ observations; however, these are not currently available. 

\section{Structure identification}\label{sec_fof_gal}

In the present simulation structure identification is achieved in post processing via a Friend-Of-Friend (FOF) algorithm \citep[e.g.][]{Davis:1985ApJ}. While the FOF is readily able to identify particle groups, the method must be modified when dealing with cell based structure, such as the baryons of {\tt RAMSES}.

In the simulation, DM and stars are traced by particles. Considering them altogether, the FOF is able to construct the complete halo catalogue and the stellar content associated with each DM group. It is important to highlight that the chosen FOF linking length is calculated by accounting only DM particles. Note that in principle the DM-stellar association can be achieved first by founding with the FOF the DM halo catalogue, then associate each star to a particular halo if the relative distance\footnote{Distances are calculated by properly accounting for the periodic boundary conditions of the box.} is less then the halo virial radius.

Additionally, in the simulation we have identified baryonic groups, such as star forming regions ($\Delta>\Delta_{\rm th}$) and metal bubbles ($Z>Z_{\rm th}\equiv10^{-7}\zsun$). This is achieved by slightly changing the FOF algorithm to let it operate on cells rather then particles. Firstly, from the whole baryons we extract only those cells satisfying the relevant threshold criterion, i.e. density based for star forming regions or metallicity based for bubbles. The we let the FOF link the baryons by treating them as particles. It is important to state that the FOF linking length is equal to half the coarse grid size of the simulation. This choice allows the reconstruction of the proper catalogue of structure, but deny substructure identification.

To describe the properties of the baryonic regions found, we can adopt definitions similar to the one used to characterize DM halos \citep[e.g.][]{Springel:2004IAUS,deSouza:2013MNRAS}. The position of a region is given by the location of the density peak, while its "radius`` is defined as $R\equiv V^{1\slash 3}$, where $V$ is the volume occupied by the selected region. As the baryons are extracted from an AMR code, there is no ambiguity in the volume definition. The shape can be described in terms of the eigenvalues $I_i$ of the inertia tensor, where $I_{1}\geq I_{2} \geq I_{3}$. The ratios of the principal axis are used as index of sphericity ($I_{3}\slash I_{1}$), prolateness ($I_{3}\slash I_{2}$) and oblateness ($I_{2}\slash I_{1}$). The other physical quantities, such as the temperature $T_{\rm SF}$ and metallicity $Z_{\rm SF}$, are calculated as mass weighted mean on the cells.

The method for baryon groups identification have been tested on metal bubbles. We have constructed the metal bubble catalogue at various redshifts and for different values of the metallicity threshold, $Z_{\rm th}$. We have checked that the total volume occupied by the identified bubbles gives a filling factor equal to $Q(Z_{\rm th})$ at the selected redshifts (see Fig.~\ref{fig_otf_anal}).

Throughout the paper we have associated DM halos and stars with baryonic regions. This is achieved using a distance based criterion. Each baryonic region is linked with every DM halos whose position of the center of mass is inside the boundary of the region. We indicate with $N_{\rm h}$ the number of DM halos associated with each region. Broadly speaking, for metal bubbles $N_{\rm h}$ can be regarded as an index indicating degree of merging experienced during the evolution.

\section{Rendering technique}\label{sec_render}

Rendering is an efficient and widely used support tool for a cosmological simulation, since it allows an immediate visual qualitative representation of the data. A large amount of 3-D volume rendering algorithms are already present and have been specifically implemented for SPH \citep{Price:2007PASA,Dolag:2008NJPh}, AMR \citep{Turk:2011arXiv,Labadens:2012arXiv} and moving mesh \citep{Vogelsberger:2013arXiv} cosmological codes.

The majority of the available methods are based on a raycasting approach, which is best suited to create images from data obtained from simulations with SPH-type data structure. For the present paper we implement a rendering technique aimed at exploiting the intrinsic AMR nature of the data structure. As a matter of fact, the presence of refinement levels naturally allows the generation of high-definition images. The method is based on a voxel representation, and currently the code is still under development.

Let $\mathbf{u}$ denote the bidimensional coordinate of the image to be created and $\mathbf{x}$ the spatial coordinate of the data to be rendered. Let $\hat{n}$ denote the direction of the l.o.s. of the observer. As the data is made of AMR cubic cells \citep[e.g.][]{Labadens:2012ASPC,Labadens:2012arXiv}, the projection $\mathbf{u}(\mathbf{x})$ from the real space to the image plane depends both on the cell position, orientation respect to $\hat{n}$ and the field of view of the observer. Note that the image resolution is taken to be equal to the finest level of refinement resolved in the simulation; this implies that $\mathbf{u}(\mathbf{x})$ depends on the size of the cell.

The basic idea is that the cells can be rendered similarly to the marching cube technique \citep[i.e.][]{Lorensen:87}, thus, in principle, the part of the projection matrix depending on relative orientations can be calculated a priori. Although for now, the algorithm allows only fixed $\hat{n}$ face-on oriented respect to the cells.

To calculate the intensity $I$ of the image we make use of a back-to-front emission-absorption rendering \citep[e.g.][]{Kaehler:2006}. After sorting the data by $\hat{n}$, at every pixel $I$ is updated via a transfer like equation
\begin{equation}\label{eq_raycasting}
 \mbox{d}I\left(\mathbf{u}\right)=\left(E_{\rm rt}(\mathbf{u}(\mathbf{x}))-A_{\rm rt}(\mathbf{u}(\mathbf{x}))I\left(\mathbf{u}\right)\right)\rho(\mathbf{u}(\mathbf{x}))\mbox{d}n
\end{equation}
where $E_{\rm rt}$ and $A_{\rm rt}$ are respectively the emission and absorption coefficients, while $\rho$ is the density field. Like in \citet{Dolag:2008NJPh} we set $E_{\rm rt}=A_{\rm rt}$ in order to obtain appealing images without artifact effects. We let the coefficient dependents of the characteristic of the cell at $\mathbf{x}$ by selecting {\it generalized isosurface}, in order to have a smoother and less noisy final image. The rendering of the physical field $q$ is calculated by selecting $i=1,\dots,n$ isosurface implicitly defined by calculating the emission coefficient as
\begin{equation}
E_{\rm rt}\left(\mathbf{x}\right)\propto\max_{i}\left\{ \exp\left[K\left(\frac{q\left(\mathbf{x}\right)-c_{i}}{ h_{i}} \right)\right]\right\}
\end{equation}
where $K$ is a kernel smoothing function, $h_{i}$ the bandwidth and $c_{i}$ the center of the $i$-th isosurface. The normalization for $E_{\rm rt}$ is chosen in order to avoid $I$ saturation. By using different kernels and a varying the isosurfaces, it is possible to obtain different visual effects, which can be best suited for the rendering of different physical quantities.

In particular, Fig. \ref{fig_render_density_04} is calculated by choosing 6 temperature isosurfaces. These are defined by a B-spline kernel function, with centers $c_{i}$ equispaced in log~$T$, with constant bandwidths satisfying $c_{i}-c_{i-1}=h_{i}$. The balancing in the level selection ensure a good dynamical range for the temperature, thus the output is a visual appealing image that gives a representation of the temperature field convolved with the density structure. Note that the convolution is obtained by definition, since in eq. \ref{eq_raycasting} the density has the role of an optical depth.

As said the image resolution is taken is linked to the finest level of refinement resolved in the simulation. This over-sampling of the image \citep[e.g.][]{Crow:1977APC} avoid most of the aliasing problems that occurs in rendering a 3D-voxel mesh. On the other hand, the technique augment the image processing time \citep[e.g.][]{Labadens:2012arXiv}. However, since the imaging algorithm can be massively parallelized, this does not represent a relevant issue.

Note that it is possible to have additional antialias by directly smoothing the image pixels with their neighbors \citep{Vogelsberger:2013arXiv} or --equivalently-- by convolving the final image with a proper filter \citep{Labadens:2012arXiv}. However, these technique degrade the image resolution and should not be required after the full implementation of the projection method.
\bsp

\label{lastpage}
\end{document}

%% file: pacchetti.tex
\usepackage[english]{babel}
\usepackage{amsmath,amssymb}
\usepackage{graphicx, subfig}
\usepackage[normalem]{ulem}

\usepackage{color}
\usepackage{mathtools}
\usepackage{epstopdf}

%% file: definizioni.tex
\def\be{\begin{equation}} 
\def\ee{\end{equation}} 
\def\ba{\begin{eqnarray}} 
\def\ea{\end{eqnarray}}

\def\msun{{\Msun}}

\def\HH{${\rm {H_2}}\,\,$}

\def\HI{\hbox{H~$\scriptstyle\rm I\ $}}

\def\HeII{\hbox{He~$\scriptstyle\rm II\ $}} 
 
\def\CIV{\hbox{C~$\scriptstyle\rm IV\ $}} 
\def\SIV{\hbox{Si~$\scriptstyle\rm IV\ $}}

\def\gsim{\lower.5ex\hbox{\gtsima}} 
\def\lsim{\lower.5ex\hbox{\ltsima}} \def\gtsima{$\; \buildrel > \over 
\sim \;$} \def\ltsima{$\; \buildrel < \over \sim \;$} \def\prosima{$\; 
\buildrel \propto \over \sim \;$} \def\gsim{\lower.5ex\hbox{\gtsima}} 
\def\lsim{\lower.5ex\hbox{\ltsima}} 
\def\simgt{\lower.5ex\hbox{\gtsima}} 
\def\simlt{\lower.5ex\hbox{\ltsima}} 
\def\simpr{\lower.5ex\hbox{\prosima}} \def\la{\lsim} \def\ga{\gsim} 
  
 \def\gtsima{$\; \buildrel > \over \sim \;$} 
\def\ltsima{$\; \buildrel < \over \sim \;$} 
\def\gsim{\lower.5ex\hbox{\gtsima}} 
\def\lsim{\lower.5ex\hbox{\ltsima}} 
\def\simgt{\lower.5ex\hbox{\gtsima}} 
\def\simlt{\lower.5ex\hbox{\ltsima}} 
\def\simpr{\lower.5ex\hbox{\prosima}} 
\def\la{\lsim} 
\def\ga{\gsim}

\def\msun{\,{\rm \Msun}}

\def\E3{{\cal E}_{\rm g}^{III}}

\def\r12{r_{1/2}} 
\def\x12{x_{1/2}} 
\def\v12{v_{1/2}}